%% file: main.tex
\documentclass[10pt,letterpaper]{article}
\usepackage[margin=1in]{geometry}

\usepackage{mathtools}
\usepackage{amsmath,amssymb}

\usepackage{changepage}

\usepackage{textcomp,marvosym}

\usepackage[natbib=true, style=vancouver,sorting=none]{biblatex}
\AtEveryBibitem{\clearlist{language}}
\usepackage{nameref,hyperref}

\usepackage[right]{lineno}

\usepackage[nopatch=eqnum]{microtype}
\DisableLigatures[f]{encoding = *, family = * }

\usepackage[dvipsnames]{xcolor}

\usepackage{array}

\usepackage[thinc]{esdiff}

\newcolumntype{+}{!{\vrule width 2pt}}

\newlength\savedwidth

\raggedright
\usepackage[aboveskip=1pt,labelfont=bf,labelsep=period,justification=raggedright,singlelinecheck=off]{caption}

\makeatletter
\renewcommand{\@biblabel}[1]{\quad#1.}
\makeatother

\usepackage{lastpage,fancyhdr,graphicx}
\usepackage{epstopdf}
\fancyheadoffset[L]{2.25in}
\usepackage{tcolorbox}
\newtcolorbox{summarybox}{colback=gray!20,
boxrule=0pt,arc=0pt,boxsep=10pt,left=7pt,right=7pt,leftrule=0pt}

\usepackage{soul}
\usepackage{cancel}

\begin{document}

\vspace*{0.2in}

\begin{flushleft}
{\Large
\textbf\newline{Deciphering regulatory architectures from synthetic single-cell expression patterns} 
}
\newline
\\
Rosalind Wenshan Pan\textsuperscript{1*},
Tom Röschinger\textsuperscript{1},
Kian Faizi\textsuperscript{1},
Hernan Garcia\textsuperscript{2,3,4,5},
Rob Phillips\textsuperscript{1,6*}
\\
\bigskip
\textsuperscript{1}Division of Biology and Biological Engineering, California Institute of Technology, Pasadena, CA; 
\textsuperscript{2}Biophysics Graduate Group, University of California, Berkeley, CA;
\textsuperscript{3}Department of Physics, University of California, Berkeley, CA;
\textsuperscript{4}Department of Molecular and Cell Biology, University of California, Berkeley, CA;
\textsuperscript{5}Institute for Quantitative Biosciences-QB3, University of California, Berkeley, CA;
\textsuperscript{6}Department of Physics, California Institute of Technology, Pasadena, CA
\bigskip

* Correspondence: rosalind@caltech.edu and phillips@pboc.caltech.edu

\end{flushleft}
\section*{Abstract}
For the vast majority of genes in sequenced genomes, there is limited understanding of how they are regulated. Without such knowledge,  it is not possible to perform a quantitative theory-experiment dialogue on how such genes give rise to physiological and evolutionary adaptation.  One category of high-throughput experiments used to understand the sequence-phenotype relationship of the transcriptome is massively parallel reporter assays (MPRAs). However, to improve the versatility and scalability of MPRA pipelines, we need a ``theory of the experiment'' to help us better understand the impact of various biological and experimental parameters on the interpretation of experimental data. These parameters include binding site copy number, where a large number of specific binding sites may titrate away transcription factors, as well as the presence of overlapping binding sites, which may affect analysis of the degree of mutual dependence between mutations in the regulatory region and expression levels. To that end, in this paper we create tens of thousands of synthetic single-cell gene expression outputs using both equilibrium and out-of-equilibrium models.  These models make it possible to imitate the summary statistics (information footprints and expression shift matrices) used to characterize the output of MPRAs and from this summary statistic to infer the underlying regulatory architecture. Specifically, we use a more refined implementation of the so-called thermodynamic models in which the binding energies of each sequence variant are derived from energy matrices. 
Our simulations reveal important effects of the parameters on MPRA data and we demonstrate our ability to optimize MPRA experimental designs with the goal of generating thermodynamic models of the transcriptome with base-pair specificity.  Further, this approach makes it possible to carefully examine the mapping between mutations in binding sites and their corresponding expression profiles, a tool useful not only for better designing MPRAs, but also for exploring regulatory evolution.

\bigskip

\begin{summarybox}
\section*{Author summary}


With the rapid advancement of sequencing technology, there has been an exponential increase in the amount of data on the genomic sequences of diverse organisms. Nevertheless, deciphering the sequence-phenotype mapping of the genomic data remains a formidable task, especially when dealing with non-coding sequences such as the promoter. In current databases, annotations on transcription factor binding sites are sorely lacking, which creates a challenge for developing a systematic theory of transcriptional regulation. To address this gap in knowledge, high-throughput methods such as massively parallel reporter assays (MPRAs) have been employed to decipher the regulatory genome. In this work, we make use of thermodynamic models to computationally simulate MPRAs in the context of transcriptional regulation and produce thousands of synthetic MPRA datasets. We examine how well typical experimental and data analysis procedures of MPRAs are able to recover common regulatory architectures under different sets of experimental and biological parameters. By establishing a dialogue between high-throughput experiments and a physical theory of transcription, our efforts serve to both improve current experimental procedures and enhancing our broader understanding of the sequence-function landscape of regulatory sequences.
\end{summarybox}



\begin{refsection}
\section*{Introduction}
With the widespread emergence of sequencing technology, we have seen an explosion of genomic data in recent years. However, data on transcriptional regulation remains  far behind. Even for organisms as widely studied as \textit{E. coli}, many promoters lack annotations on the transcription factor binding sites that underlie transcriptional regulation. Moreover, existing binding site annotations are largely without experimental validation for functional activity, as a large proportion are determined through DNA-protein interaction assays such as ChIP-Seq~\cite{Bartlett2017-dj, Trouillon2023-sb, Gao2021-lz, Mundade2014-xd} or computational prediction~\cite{Bulyk2003-xf}. This fundamental gap in knowledge poses a major obstacle for us to understand the spatial and temporal control of cellular activity, as well as how cells and organisms respond both physiologically and evolutionarily to environmental signals. 

One strategy to understand the regulatory genome is by conducting massively parallel reporter assays (MPRAs), where the regulatory activities of a library of sequences are measured simultaneously via a reporter. The library of sequences may be genomic fragments~\cite{Muerdter2015-hg} or sequence variants containing mutations relative to the wild-type regulatory sequence~\cite{Kinney2019-nt}. There are two main ways to measure regulatory activities in MPRAs. The first approach uses fluorescence-activated cell sorting to sort cells into bins based on the expression levels of a fluorescent reporter gene~\cite{Kinney2010-ue}. Subsequently, deep sequencing is utilized to determine which sequence variant is sorted into which bin. The second approach uses RNA-sequencing (RNA-Seq) to measure the counts of barcodes associated with each sequence variant as a quantitative read-out for expression levels. The two approaches have been used in both prokaryotic~\cite{Urtecho2019-po, Urtecho2020-bk, Han2023-ks, Vvedenskaya2018-qg} and eukaryotic systems~\cite{De_Boer2020-rd, Zheng2023-yl, Klein2020-bu} to study diverse genomic elements including promoters and enhancers.

In particular, our group has developed Reg-Seq~\cite{Ireland2020-bp}, an RNA-Seq-based MPRA that was used to successfully decipher the regulatory architecture of 100 promoters in \textit{E. coli}, with the hope now to complete the regulatory annotation of entire bacterial genomes. Mutations in regulatory elements lead to reduced transcription factor binding, which may result in measurable changes in expression. Therefore, the key strategy to annotate transcription factor binding sites based on MPRA data that we focus on is to identify sites where mutations have a high impact on expression levels. To do this, one approach is to calculate the mutual information between base identity and expression levels at each site. A so-called information footprint can then be generated by plotting the mutual information at each position along the promoter. Positions with high mutual information are identified as putative transcription factor binding sites.

In this paper, we develop a computational pipeline that simulates the RNA-Seq-based MPRA pipeline. Specifically, we make use of equilibrium statistical mechanics to build synthetic datasets that simulate the experimental MPRA data and examine how various parameters affect the output of MPRAs. Conventionally, thermodynamic models are not sequence specific. The binding energies are usually phenomenological parameters that are fit once and for all.  Here, by way of contrast, the idea is to leverage the experimentally determined or synthetically engineered energy matrices that allow us to consider arbitrary binding site sequences and to compute their corresponding level of expression.

With this computational pipeline, we examine tens of thousands of unique promoters and hence tens of thousands of unique implementations of the sequence-specific thermodynamic models. These sequences are then converted into the two primary summary statistics used to analyze the experimental data, namely, information footprints and expression shift matrices.  Given that the experimental implementations of these MPRAs entail tens of millions of unique DNA constructs, the computational pipeline gives us the opportunity to systematically and rigorously analyze the connection between key parameters. These parameters include experimental parameters, such as the rate of mutation used to generate the sequence variants, as well as biological parameters such as transcription factor copy number, where a large number of specific binding sites may titrate away transcription factors. This computational pipeline will help us to optimize MPRA experimental design with the goal of accurately annotating transcription factor binding sites in regulatory elements, while revealing the limits of MPRA experiments in elucidating complex regulatory architectures. Additionally, the insights gained from our simulation platform will enable further dialogue between theory and experiments in the field of transcription including efforts to understand how mutations in the evolutionary context give rise to altered gene expression profiles and resulting organismal fitness.

Our use of thermodynamic models in this pipeline is motivated by several principal considerations. First, because of their simplicity, these models have served and continue to serve as a powerful null model when considering signaling, regulation, and physiology.  Their application runs the gamut from the oxygen binding properties of hemoglobin~\cite{Hill1910-hu, Hill1913-ml, Adair1925-tx, Monod1965-ij, Pauling1935-pt}, to the functioning of membrane-bound receptors in chemotaxis and quorum sensing~\cite{Keymer2006-ck, Tu2008-kd, Mello2003-ho, Swem2008-rv}, and to the binding of transcription factors at their target DNA sequences~\cite{Buchler2003-yt, Kuhlman2007-qj, Hammar2014-fp, Martin2008-tv, Ackers1982-ax, Shea1985-js, McGhee1974-jq, Vilar2003-rj, Vilar2023-ux}. Second, for our purposes, the thermodynamic models form an internally consistent closed theoretical system in which we can generate tens of thousands of ``single-cell'' expression profiles and use the same tools that we use to evaluate real MPRA data to evaluate these synthetic datasets, thus permitting a rigorous means for understanding such real data. That said, despite their many and varied successes, thermodynamic models deserve continued intense scrutiny since many processes of the central dogma involve energy consumption and thus may involve steady state probabilities that while having the form of ratios of polynomials, are no longer of the Boltzmann form~\cite{Hammar2014-fp, Kreamer2015-gy, Eck2020-ma, Meijsing2009-mc, Garcia2012-js}. These shortcomings of thermodynamic models are further discussed in~\ref{S1_Appendix} of our Supplemental Information. To address these shortcomings, we also provide a preliminary study of how broken detailed balance might change the interpretation of summary statistics such as information footprints.


The remainder of this paper is organized as follows. In Sec~\ref{sec:pipeline}, we introduce our procedure to construct and analyze synthetic datasets for both promoters regulated by a single transcription factor and promoters regulated by a combination of multiple transcription factors. In Sec~\ref{sec:mutation} and Sec~\ref{sec:library-size}, we discuss the choice of parameters related to the construction of the mutant library, including the rate of mutation, mutational biases, and library size. After setting up the computational pipeline, we perturb biological parameters and examine how these perturbations affect our interpretation of MPRA summary statistics. The parameters that we will explore include the free energy of transcription factor binding (Sec~\ref{sec:free-energy}), the regulatory logic of the promoter (Sec~\ref{sec:logic}), the copy number of the transcription factor binding sites (Sec~\ref{sec:plasmid}), and the concentration of the inducers (Sec~\ref{sec:inducer}). Next, we explore factors that may affect signal-to-noise ratio in the information footprints. These factors include stochastic fluctuations of transcription factor copy number (Sec~\ref{sec:extrinsic}), non-specific binding events along the promoter (Sec~\ref{sec:spurious}), as well as the presence of overlapping binding sites (Sec~\ref{sec:overlapping}). Additionally, in Sec~\ref{sec:nonequilibrium}, we generalize our pipeline and consider the cases of transcriptional regulation where detailed balance may be broken. Finally, we discuss the insights generated from our computational pipeline in relation to future efforts to decipher regulatory architectures in diverse genomes.

\section*{Results}

\subsection{Mapping sequence specificity and expression levels}

\subsubsection{Computational pipeline for deciphering regulatory architectures from first principles} \label{sec:pipeline}

In MPRA pipelines, the goal is to make the connection between regulatory sequences, transcription factor binding events, and expression levels. In Reg-Seq for example, the authors start with a library of sequence variants for an unannotated promoter, each of which contains a random set of mutations relative to the wild type sequence. Then, RNA-Seq is used to measure the expression levels of a reporter gene directly downstream of each promoter. By calculating the mutual information between mutations and the measured expression levels, the regulatory architecture of the promoter can be inferred. Finally, Bayesian models and thermodynamic models can be built using statistical mechanics to infer the interaction energies between transcription factors and their binding sites in absolute $k_BT$ units at a base-by-base resolution~\cite{Ireland2020-bp}.

\begin{figure}[!h]
\centering
\includegraphics[width=\textwidth]{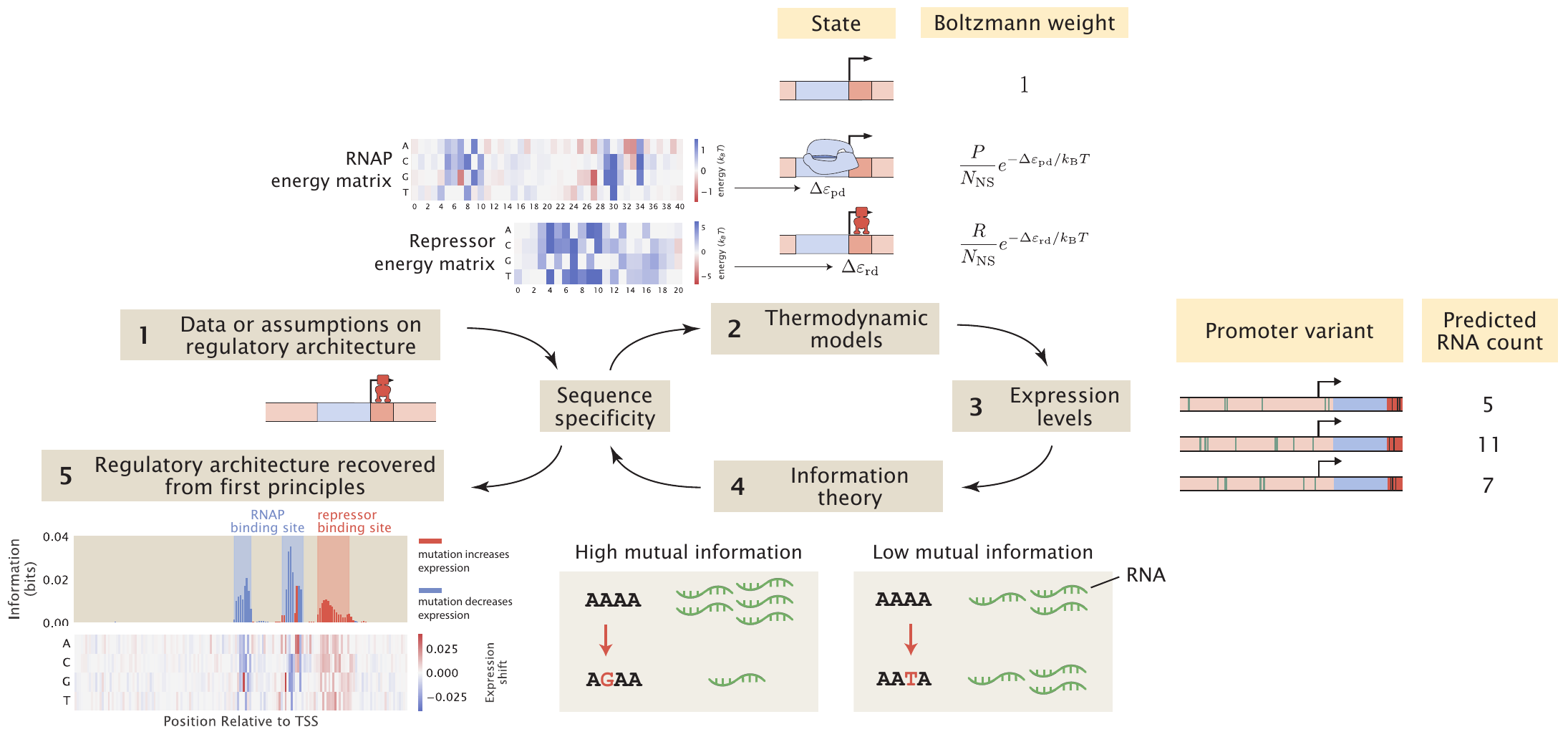}
\caption{{\bf A computational pipeline for deciphering regulatory architectures from first principles.}
Given (1) knowledge or assumptions about the regulatory architecture of a promoter, we make use of (2) thermodynamic models to construct a states-and-weights diagram, which contains information about all possible states of binding and the associated Boltzmann weights. Here, in the states-and-weights diagram, $P$ is the copy number of RNAP, $R$ is the copy number of the repressor, $N_\mathrm{NS}$ is the number of non-binding sites, $\Delta \varepsilon_{\mathrm{pd}}$ and $\Delta \varepsilon_{\mathrm{rd}}$ represent the binding energies of RNAP and the repressors at their specific binding sites relative to the non-specific background, respectively. Using these states-and-weights diagrams as well as the energy matrices, which are normalized to show the change in binding energies for any mutation along the promoter compared to the wild-type sequence, we can (3) predict the expression levels for each of the promoter variants in a mutant library. To recover the regulatory architecture, we (4) calculate the mutual information between the predicted expression levels and mutations at each position along the promoter according to Eq~\ref{eq:MI}. In particular, there is high mutual information if a mutation leads to a large change in expression and there is low mutual information if a mutation does not lead to a significant change in expression. The mutual information at each position is plotted in an information footprint, where the height of the peaks corresponds to the magnitude of mutual information, and the peaks are colored based on the sign of expression shift, defined in Eq~\ref{eq:exp_shift}. Given the assumption that the positions with high mutual information are likely to be RNAP and transcription factor binding sites, we (5) recover the regulatory architecture of the promoter. The base-specific effects of mutations on expression levels can also be seen from expression shift matrices, which is calculated using Eq~\ref{eq:exshift_matrix}, where the difference between the expression levels of sequences carrying a specific mutation at a given position and the average expression level across all mutant sequences is computed.}
\label{fig1}
\end{figure}

Our computational MPRA pipeline involves similar steps, but instead of starting from experimental measurements of expression levels, we use thermodynamic models to predict expression levels given the sequences of the promoter variants and the corresponding interaction energies, as schematized in Fig \ref{fig1}. Through this process, we generate synthetic datasets of expression levels that are in the same format as the datasets that we obtain via RNA-Seq. Subsequently, we analyze the synthetic datasets in the same way as we would analyze an experimental dataset. Importantly, we can perturb various experimental and biological parameters of interest within this pipeline and examine how changing these parameters affect our ability to discover unknown transcription factor binding sites through MPRAs.

We first demonstrate our computational pipeline using a promoter with the simple repression regulatory architecture, i.e. the gene is under the regulation of a single repressor. Specifically, we use the promoter sequence of LacZYA. We assume that it is transcribed by the $\sigma^{70}$ RNAP and only regulated by the LacI repressor, which binds to the O1 operator within the LacZYA promoter. As we have derived in \ref{S2_Appendix}, for a gene with the simple repression regulatory architecture, the probability of RNAP being bound~\cite{Bintu2005-ib, Bintu2005-oz} is given by
\begin{align}
\label{eq:pbound}
    p_{\mathrm{bound}} = \frac{\frac{P}{N_\mathrm{NS}}e^{-\beta \Delta \varepsilon_\mathrm{pd}}}{1 + \frac{P}{N_\mathrm{NS}}e^{-\beta \Delta \varepsilon_\mathrm{pd}} + \frac{R}{N_\mathrm{NS}}e^{-\beta \Delta \varepsilon_\mathrm{rd}}}.
\end{align}
$\beta = \frac{1}{k_BT}$ where $k_B$ is Boltzmann's constant and $T$ is temperature. As we can see, the parameters that we need are the copy number of RNAP ($P$), the copy number of repressor ($R$), the number of non-binding sites ($N_\mathrm{NS}$), and the binding energies for RNAP and the repressors ($\Delta \varepsilon_\mathrm{pd}$ and $\Delta \varepsilon_\mathrm{rd}$), respectively. We begin by assuming that $P$ and $R$ are constant and $N_\mathrm{NS}$ is the total number of base pairs in the \textit{E. coli} genome. On the other hand, the values for $\Delta \varepsilon_\mathrm{pd}$ and $\Delta \varepsilon_\mathrm{rd}$ depend on the sequence of the promoter variant.

\begin{figure}[!h]
\centering
\includegraphics[width=0.8\textwidth]{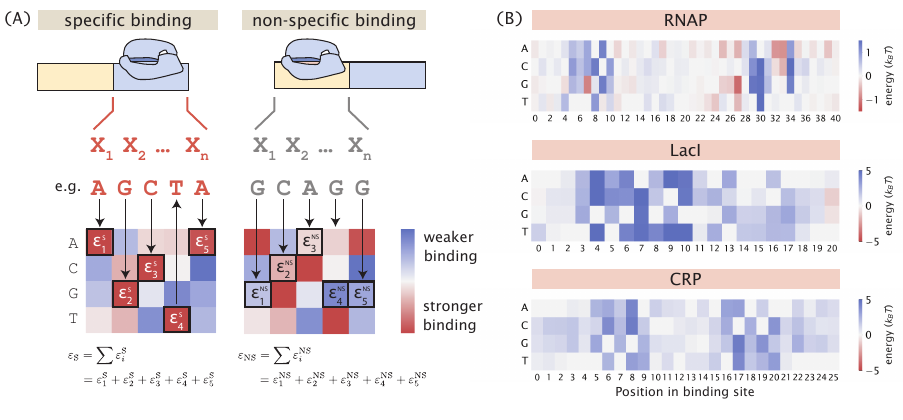}
\caption{{\bf Mapping binding site sequences to binding energies using energy matrices.}
(A)~Given the assumption that binding energies are additive, we can use an energy matrix to determine how much energy each base along the binding site contributes and compute the total binding energy by taking the sum of the binding energies contributed by each position. The total binding energy can be used to compute the Boltzmann weight for each of the states, which is then used to calculate the probability of RNAP being bound. (B)~Experimentally measured energy matrices of RNAP~\cite{Brewster2012-ds}, the LacI repressor~\cite{Barnes2019-dx}, and the CRP activator.~\cite{Kinney2010-ue}}
\label{fig2}
\end{figure}

We calculate the binding energies by mapping the sequences of the promoter variants to the energy matrices of the RNAP and the repressor, as shown in Fig~\ref{fig2}(A). Specifically, we assume that binding energies are additive. This means that given a sequence of length $l$, the total binding energy $\Delta \varepsilon$ can be written as
\begin{align}
    \Delta \varepsilon = \sum_{i=1}^l \varepsilon_{i,b_i},
\end{align}
where $\varepsilon_{i,b_i}$ is the binding energy corresponding to base identity $b_i$ at position $i$ according to the energy matrix. Here, we use the energy matrices of the RNAP and LacI that were previously experimentally determined using Sort-Seq~\cite{Brewster2012-ds, Barnes2019-dx}, as shown in Fig~\ref{fig2}(B). Unless otherwise specified, these energy matrices are used to build all synthetic datasets in the remainder of this paper. It should be acknowledged that the additive model does not take into account epistasis effects~\cite{Bulyk2002-yr, Zhao2012-uw}. It may be beneficial to include higher-order interaction energy terms in future simulations of MPRA pipelines.

After computing the sequence-specific binding energies, we can then substitute the relevant energy terms into Eq~\ref{eq:pbound} and calculate the probability of RNAP being bound. To connect the probability of RNAP being bound to expression levels, we make use of the occupancy hypothesis, which states that the rate of mRNA production is proportional to the probability of RNAP occupancy at the promoter \cite{Phillips2019-ld}. The rate of change in mRNA copy number is given by the difference between the rates of mRNA production and degradation. In general, there can be multiple transcriptionally active states, each with its own transcription rate. For example, for a promoter that is regulated by an activator, there are two transcriptionally active states. One is where only RNAP is bound to the promoter and one is where both RNAP and the activator are bound to the promoter. Each of these two states could have a different rate of mRNA production. With this, we define an average rate of mRNA production, which is given by the sum of each state's production rate, weighted by the probability of the state $\langle r \rangle = \sum_i r_i p_{\mathrm{bound}, i}$. Hence, the rate of change of mRNA copy number is given by
\begin{align} \label{eq:dmdt}
    \diff{m}{t} = \sum_i r_i p_{\mathrm{bound}, i} - \gamma m,
\end{align}
where for transcriptionally active state $i$, $r_i$ is the rate of transcription, $p_{\mathrm{bound}, i}$ is the probability of RNAP occupancy in state $i$, $m$ is the copy number of mRNAs, and $\gamma$ is the rate of mRNA degradation. Therefore, the steady-state level of mRNA is given by
\begin{align}
    m^* = \frac{1}{\gamma} \sum_i r_i p_{\mathrm{bound}, i}.
\end{align}
For simplicity, we assume that each transcriptionally active state has the same rate of mRNA production, $r$. Therefore,
\begin{align} \label{eq:mrna-level}
    m^* = \alpha \sum_i p_{\mathrm{bound}, i},\ \mathrm{where}\ \alpha=\frac{r}{\gamma}.
\end{align}
Using the above expression, we can calculate the expected RNA count for each of the promoter variants in our library. Assuming that $r$ and $\gamma$ do not depend on the mRNA sequence, the total probability of RNAP being bound, given by $p_{\mathrm{bound}} = \sum p_{\mathrm{bound}, i}$, is scaled by the same constant to produce the mRNA count of each promoter variant. Therefore, the choice of $\alpha$ does not affect our downstream calculations involving the probability distribution of expression levels. Depending on the sequencing depth, the mRNA count for each promoter variant is typically on the order of $10$ to $10^3$ \cite{Ireland2020-bp}. Here, we take the geometric mean and set $\alpha$ to $10^2$ to ensure that the mRNA count is on a realistic scale.

Up until this point, we have constructed a synthetic RNA-Seq dataset containing the predicted expression levels of each sequence variant in a mutant library. MPRA data is often described using several interesting summary statistics. Using thousands of synthetically derived mRNA counts, we can compute such summary statistics and ask how both biological and experimental parameters change them. These summary statistics can then be used to infer the underlying regulatory architecture from these large-scale synthetic datasets. To do this, we calculate the mutual information between mutations and expression levels. The mutual information at position $i$ is given by
\begin{align} \label{eq:MI}
    I_i = \sum_{b} \sum_{\mu} \Pr{_i}(b, \mu) \log_2{\left(\frac{\Pr{_i}(b, \mu)}{\Pr{_i}(b)\Pr(\mu)}\right)},
\end{align}
where $b$ represents base identity, $\mu$ represents expression level, $\Pr{_i}(b)$ is the marginal probability distribution of mutations at position $i$, $\Pr(\mu)$ is the marginal probability distribution of expression levels across all promoter variants, and $\Pr{_i}(b, \mu)$ is the joint probability distribution between expression levels and mutations at position $i$. In general, $b$ can be any of the four nucleotides, i.e. $b \in \{\mathrm{A}, \mathrm{C}, \mathrm{G}, \mathrm{T}\}$. This means that $\Pr{_i}(b)$ is obtained by computing the frequency of each base per position. Alternatively, a more coarse grained approach can be taken, where the only distinction is between the wild-type base and mutation, in which case $b$ is defined as 
\begin{align}
    b = 
    \begin{cases}
        0, & \text{if the base is mutated}, \\
        1, & \text{if the base is wild type}.
    \end{cases}
\end{align}
As shown in~\ref{S4_Appendix}, using the coarse-grained definition of $b$ improves the signal-to-noise ratio of the information footprint as reducing the number of states reduces articifical noise outside of the specific binding sites. Therefore, we use this definition for our subsequent analysis.

On the other hand, to represent expression levels as a probability distribution, we group sequences in each range of expression levels into discrete bins and compute the probabilities that a given promoter variant is found in each bin. As shown in~\ref{S4_Appendix}, we found that increasing the number of bins leads to a lower signal-to-noise ratio in the information footprints because the additional bins contribute to artificial noise. Therefore, we choose to use only two bins with the mean expression level as the threshold between them. This means that $\mu$ can take the values of
\begin{align}
    \mu = 
    \begin{cases}
        0, & \text{if expression is lower than mean expression} \\
        1, & \text{if expression is higher than mean expression}.
    \end{cases}
\end{align}
In~\ref{S5_Appendix}, we derive the information footprint for a constitutive promoter analytically and demonstrate that in the absence of noise, mutual information is expected to be 0 outside of the specific binding sites and non-zero at a specific binding site.

To decipher the regulatory architecture of a promoter, another important piece of information is the direction in which a mutation changes expression. This can be determined by calculating the expression shift, which measures the change in expression when there is a mutation at a given position~\cite{Belliveau2018-xp}. Suppose there are $n$ promoter variants in our library, then the expression shift $\Delta s_l$ at position $l$ is given by
\begin{align} \label{eq:exp_shift}
    \Delta s_l = \frac{1}{n} \sum^n_{i=1} \xi_{i,l} \left(c_i - \langle c \rangle \right),\ \mathrm{where}\ \langle c \rangle = \frac{1}{n} \sum^n_{i=1} c_i,
\end{align}
where $c_i$ represents the RNA count of the $i$-th promoter variant, $\xi_{i,l} = 0$ if the base at position $l$ in the $i$-th promoter variant is wild type, and $\xi_{i,l} = 1$ if the base is mutated. If the expression shift is positive, it indicates that mutations lead to an increase in expression and the site is likely to be bound by a repressor. On the other hand, a negative expression shift indicates that mutations lead to a decrease in expression, and therefore the site is likely to be bound by RNAP or an activator.

\begin{figure}[!h]
\centering
\includegraphics[width=0.8\textwidth]{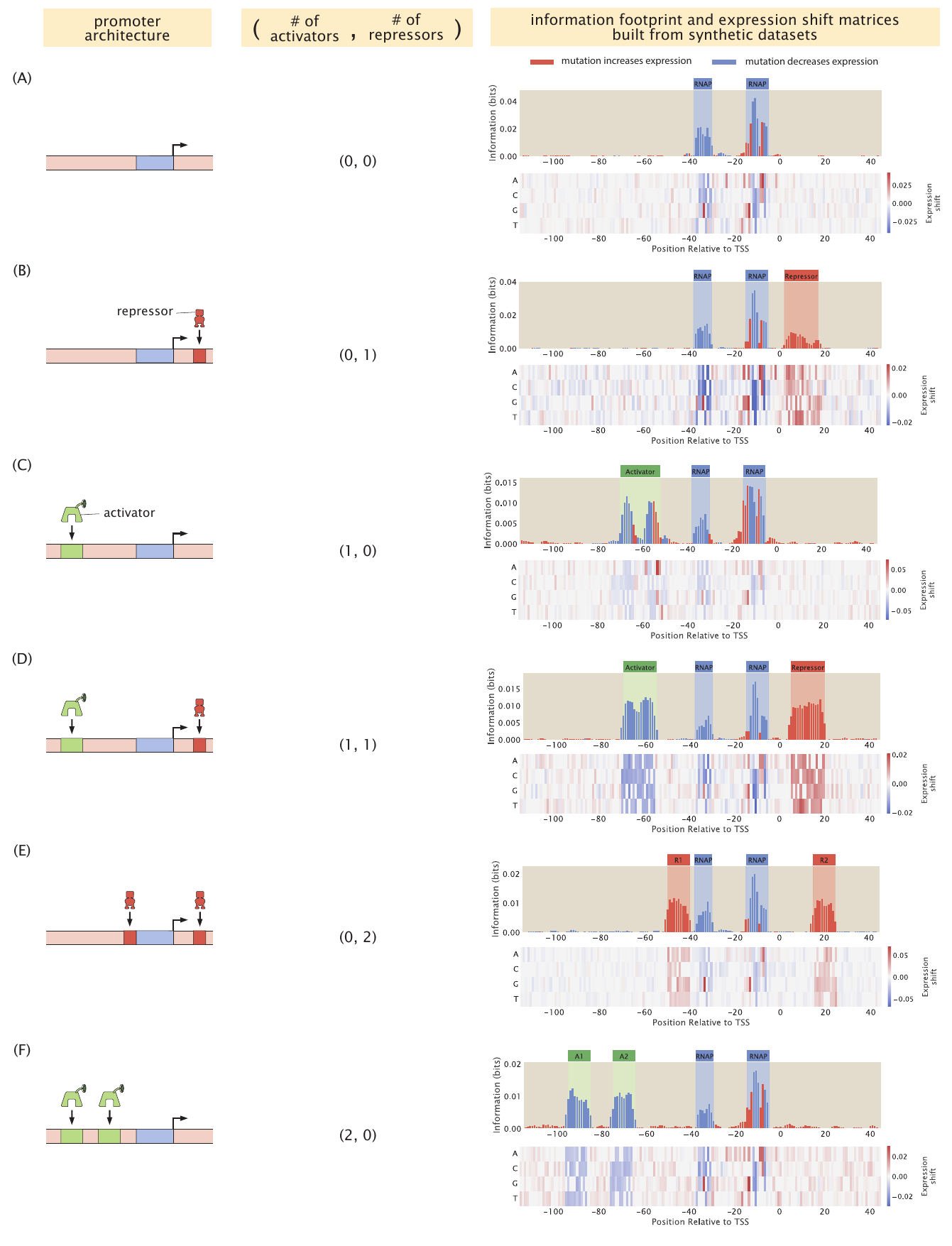}
\caption{{\bf Building information footprints and expression shift matrices based on synthetic datasets of different regulatory architectures.}
We describe each of the regulatory architectures using the notation (A,R), where A refers to the number of activator binding sites and R refers to the number of repressor binding sites. The corresponding information footprints and expression shift matrices built from synthetic datasets are shown on the right. The architectures shown in panels (A)-(F) are a constitutive promoter, simple repression, simple activation, repression-activation, double repression, and double activation, respectively. For panels (A)-(C), we use energy matrices of RNAP, LacI, and CRP shown in Fig~\ref{fig2}(B). For panels (D)-(F), we continue to use the experimentally measured energy matrix for RNAP; the energy matrices for the repressors and the activators are constructed by hand, where the interaction energies at the wild type bases are set to $0\ k_BT$ and the interaction energies at the mutant bases are set to $1\ k_BT$.}
\label{fig3}
\end{figure}

By calculating the mutual information and expression shift at each base position along the promoter, we can plot an information footprint for a promoter with the simple repression regulatory architecture, as shown in Fig~\ref{fig3}(B). There are two peaks with negative expression shifts near the -10 and -35 positions, which correspond to the canonical RNAP binding sites. There is another peak immediately downstream from the transcription start site with a positive expression shift, which corresponds to the binding site of the LacI repressor. Taken together, we have demonstrated that by calculating mutual information, we are able to recover binding sites from our synthetic dataset on expression levels.

To have a more precise understanding on how much each mutation to each possible base identity $b \in \mathrm{A, C, G, T}$ changes expression levels, we can extend Eq~\ref{eq:exp_shift} to calculate an expression shift matrix. Specifically, the value in the expression shift matrix at position $l$ corresponding to base $b$ is given by
\begin{align} \label{eq:exshift_matrix}
    \Delta s_{b,l} = \begin{dcases}
        \frac{1}{n} \sum^n_{i=1} \xi_{b,i,l} \left(\frac{c_i}{\langle c \rangle} - 1\right),\ \mathrm{where}\ \langle c \rangle = \frac{1}{n} \sum^n_{i=1} c_i,  & \text{if $b$ is mutated} \\
        0,  & \text{if $b$ is wild type}
    \end{dcases}  
\end{align}
where $\xi_{b,i,l} = 1$ if the base at position $l$ in the $i$-th promoter variant corresponds to base identity $b$ and $\xi_{b,i,l} = 0$ otherwise. Note that in comparison to Eq~\ref{eq:exp_shift}, here we are calculating the relative change in expression, which is easier to interpret than the absolute change in expression. An example of an expression shift matrix is shown in panel (5) of Fig~\ref{fig1}.

Using the same procedure as described above, we can also produce synthetic datasets for other classes of regulatory architectures. In Fig~\ref{fig3}, we demonstrate that we can recover the expected binding sites based on synthetic datasets for six common types of regulatory architectures~\cite{Ireland2020-bp}. The states-and-weights diagrams and $p_\mathrm{bound}$ expressions used to produce these synthetic datasets are shown in~\ref{S3_Appendix}.

\subsubsection{Changing mutation rates and adding mutational biases} \label{sec:mutation}

One key parameter in the MPRA pipeline is the level of mutation for each sequence variant in the library. Here, we again consider a gene with the simple repression regulatory architecture as a case study and we examine how varying mutation rates and mutational biases changes the signals in the information footprints. We quantify the level of signal, $S$, by calculating the average mutual information at each of the binding sites. This is given by
\begin{align} \label{eq:signal}
    S = \langle I \rangle_B = \frac{1}{l} \sum \limits_{i \in B} I_i,
\end{align}
where $B$ represents the set of bases within a given binding site, $I_i$ represents the mutual information at base position $i$, and $l$ is the length of $B$, i.e. the number of bases in the binding site.

\begin{figure}[!h]
\centering
\includegraphics[width=0.6\textwidth]{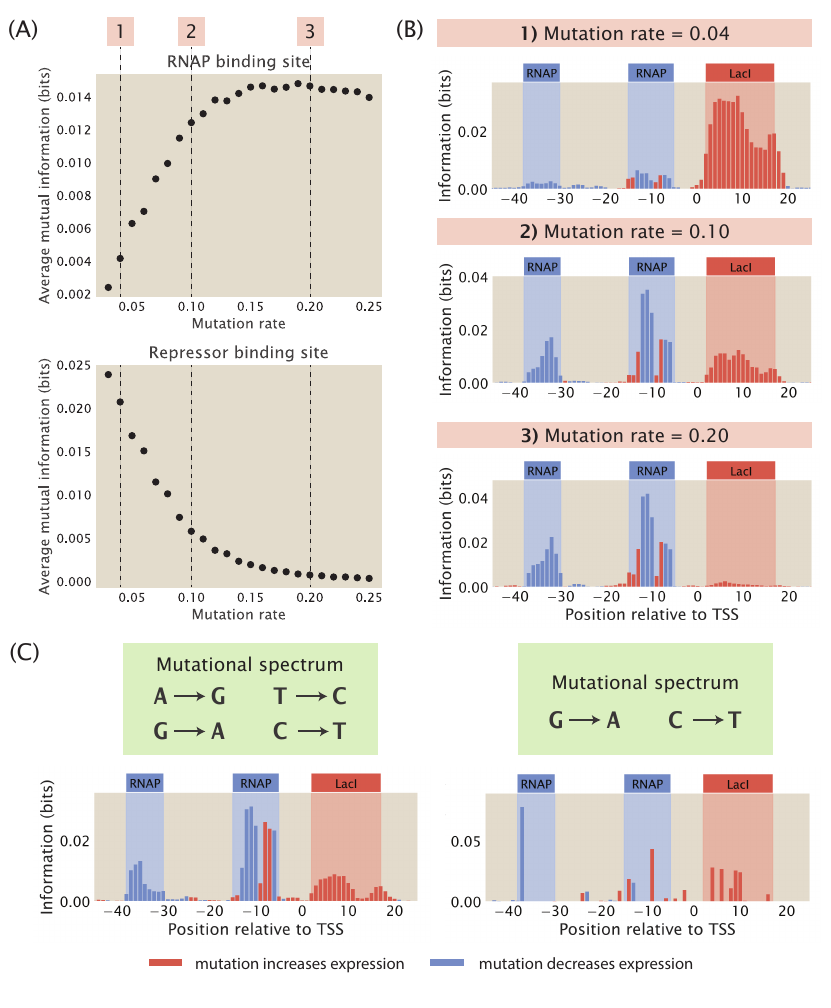}
\caption{{\bf Changing mutation rate and adding mutational biases.}
(A)~Changes in the average mutual information at the RNAP and at the repressor binding sites when the mutation rate of the mutant library is increased. Average mutual information is calculated according to Eq~\ref{eq:signal}. Each data point is the mean of average mutual information across 20 synthetic datasets with the corresponding mutation rate. The numbered labels correspond to information footprints shown in (B). (B)~Representative information footprints built from synthetic datasets with mutation rates of 0.04, 0.1, and 0.2. (C)~Information footprints built from synthetic datasets where the mutant library has a limited mutational spectrum. The left panel shows a footprint where mutations from A to G, G to A, T to C, and C to T are allowed. The right panel shows a footprint where only mutations from G to A and from C to T are allowed.}
\label{fig4}
\end{figure}

As shown in Fig~\ref{fig4}(A) and~\ref{fig4}(B), in general, when there is a higher rate of mutation, the average mutual information at the RNAP binding site increases relative to the average mutual information at the repressor binding site. To explain this effect, we consider $\kappa$, the ratio between the Boltzmann weights of the repressor and RNAP
\begin{align} \label{eq:boltzmann-ratio}
    \kappa = \frac{R \cdot e^{-\beta(\Delta \varepsilon_{\mathrm{rd}} + m_r \Delta \Delta \varepsilon_{\mathrm{rd}})}}{P \cdot e^{-\beta(\Delta \varepsilon_{\mathrm{pd}} + m_p \Delta \Delta \varepsilon_{\mathrm{pd}})}} = \frac{R}{P} \cdot e^{-\beta(\Delta \varepsilon_{\mathrm{rd}} - \Delta \varepsilon_{\mathrm{pd}} + m_r \Delta \Delta \varepsilon_{\mathrm{rd}} - m_p \Delta\Delta \varepsilon_{\mathrm{pd}})},
\end{align}
where $m_r$ and $m_p$ are the number of mutations at the repressor and RNAP binding sites, and $\Delta \Delta \varepsilon_{\mathrm{rd}}$ and $\Delta \Delta \varepsilon_{\mathrm{pd}}$ are the change in binding energies due to each mutation at the repressor and RNAP binding sites. To express $\kappa$ as a function of the mutation rate $\theta$, we can rewrite $m_r$ and $m_p$ as a product of $\theta$ and the lengths of repressor and RNAP binding sites $l_r$ and $l_p$,
\begin{align}
    \kappa = \frac{R}{P} \cdot e^{-\beta E},\ \mathrm{where}\ E = \Delta \varepsilon_{\mathrm{rd}} - \Delta \varepsilon_{\mathrm{pd}} + \theta (l_r \Delta \Delta \varepsilon_{\mathrm{rd}} - l_p \Delta\Delta \varepsilon_{\mathrm{pd}})
\end{align}

We assume that $\Delta \Delta \varepsilon_{\mathrm{rd}}$ and $\Delta \Delta \varepsilon_{\mathrm{pd}}$ are equal to the average effect of mutations per base pair within each binding site, which can be calculated using the formula
\begin{align}
    \Delta \Delta \varepsilon = \frac{1}{3l} \sum \limits_{i=1}^{l} \sum \limits_{b \neq b_i}^{\Lambda} \varepsilon_{i, b},\ \mathrm{where}\ \Lambda = {\mathrm{\{A, T, C, G\}}},
\end{align}
where $\varepsilon_{i,b}$ is the energy contribution from position $i$ when the base identity is $b$. As we are using energy matrices where the energies corresponding to the wild-type base identities, $b_i$, are set to 0, we only need to compute the sum of the energy terms for the mutant bases $b \neq b_i \in \Lambda$ at each position. Since there are three possible mutant bases at each site, it follows that to find the average effect of mutations, we divide the sum of the energy matrix by 3 times the length of the binding site $l$.

By applying this formula to the energy matrices in Fig~\ref{fig2}(B), we see that $\Delta \Delta \varepsilon_{\mathrm{rd}} \approx 2.24\ k_BT$ and $\Delta \Delta \varepsilon_{\mathrm{pd}} \approx 0.36\ k_BT$, where $\Delta \Delta \varepsilon_{\mathrm{pd}}$ is averaged over the 20 bases surrounding the -35 and -10 binding sites. Moreover, $l_r = l_p \approx 20$ base pairs. Therefore,
\begin{align}
    l_r \Delta \Delta \varepsilon_{\mathrm{rd}} - l_p \Delta\Delta \varepsilon_{\mathrm{pd}} = 20 \times 2.24\ k_BT - 20 \times 0.36\ k_BT = 37.6\ k_BT.
\end{align}
Since the above value is positive, $\kappa$ decreases with increasing mutation rate, making the repressor bound state less likely compared to the RNAP bound state. With the repressor bound state becoming less likely, the signal in the repressor binding site goes down, since mutations changing the binding energy of the repressor change the transcription rate less significantly. As shown in~\ref{S6_Appendix}, when we reduce the effect of mutations on the binding energy of the repressor, we recover the signal at the repressor binding site. Conversely, the average mutual information at the repressor binding site increases when the rate of mutation is decreased. This is because when there are very few mutations, the energy $E$ will be less than 0 and therefore $\kappa$ will be greater than 1. As a result, the repressor will be preferentially bound, which blocks RNAP binding and leads to a low signal at the RNAP binding site. We can recover the signal at the RNAP binding site by increasing the binding energy between RNAP and the wild type promoter, as shown in~\ref{S6_Appendix}. 

Importantly, the use of $\kappa$ lays the groundwork for finding the optimal rate of mutation in a mutant library. Specifically, we would like to determine the choice of mutation rate that will give us high and balanced signals for both the RNAP and transcription factor binding sites in the information footprints. To find the optimal rate of mutation, we need to satisfy the condition
\begin{align}
    \begin{split}
        \kappa = \frac{R}{P} e^{-\beta(\Delta \varepsilon_{\mathrm{rd}} - \Delta \varepsilon_{\mathrm{pd}} + \theta (l_r \Delta \Delta \varepsilon_{\mathrm{rd}} - l_p \Delta\Delta \varepsilon_{\mathrm{pd}}))} &= 1,
    \end{split}
    \label{equ:opt_mut_rate}
\end{align}
which puts repressor and RNAP binding on an equal footing. Plugging in the values of $R$, $P$, and the energy terms and solving for $\theta$, we get that $\theta \approx 0.10$. This shows that an intermediate mutation rate is optimal for maintaining high signals at all binding sites. It should be acknowledged that the optimal mutation rate is sensitive to the underlying parameters such as transcription factor copy numbers and binding energies. In~\ref{S7_Appendix}, we numerically compute the optimal mutation rate for parameters within physiological ranges, and we found that the optimal mutation rate can vary from 0.01 to 0.50. Nevertheless, for the remaining analysis shown in this work, we fix the mutation rate at 10\%, which is similar to the mutation rate typically used in MPRAs such as Sort-Seq~\cite{Kinney2010-ue} and Reg-Seq~\cite{Ireland2020-bp}.

In addition to mutation rate, another important variation in the design of the mutant library is the presence of mutational biases. For example, some mutagenesis techniques, including CRISPR-Cas9, often carry mutational biases whereby mutations within the family of purines and the family of pyrimidines have a higher efficiency compared to mutations between purines and pyrimidines~\cite{Anzalone2020-si}. We build mutant libraries that incorporate two different mutational spectrums. In the first case, we allow only swaps between A and G and between C and T. For this library, we observe that the signals at both the RNAP binding site and the repressor binding site are well preserved, as shown in the left panel of Fig~\ref{fig4}(C). In the second case, we only allow mutations from G to A and from C to T without allowing the reverse mutations. As shown in the right panel of Fig~\ref{fig4}(C), due to only two bases being allowed to mutate, only a few, possibly low-effect mutations are observed, making small regions such as the -10 and -35 sites hard to detect. These results show that information footprints are robust to mutational biases provided that most sites are allowed to mutate.

\subsubsection{Noise as a function of library size} \label{sec:library-size}

\begin{figure}[!h]
\centering
\includegraphics[width=0.8\textwidth]{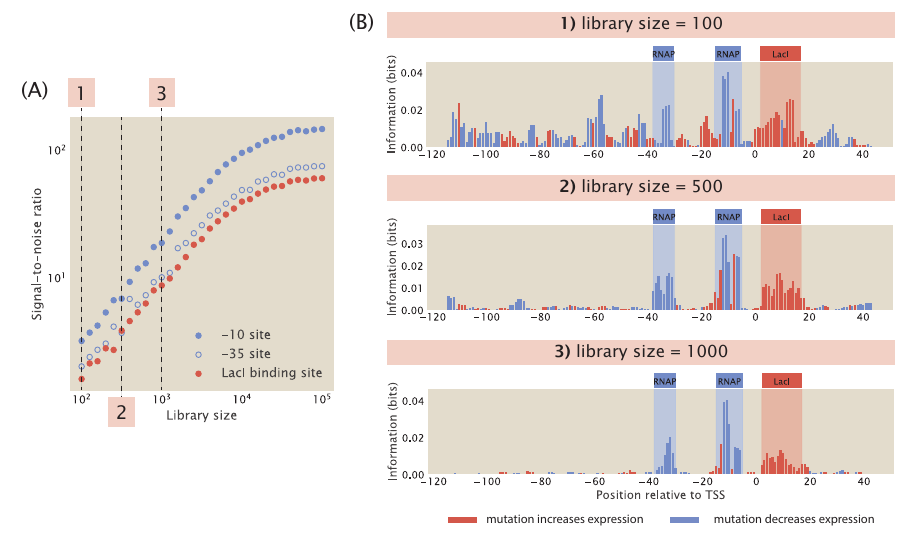}
\caption{{\bf Noise as a function of library size.}
(A)~Signal-to-noise ratio increases as library size increases. Signal-to-noise ratio is calculated according to Eq~\ref{eq:signal-to-noise}. Each data point is the mean of average mutual information across 20 synthetic datasets with the corresponding library size. The numbered labels correspond to footprints in (B). (B)~Representative information footprints with a library size of 100, 500, and 1000.}
\label{fig5}
\end{figure}

Another parameter that is important for library design is the total number of sequence variants in the mutant library. We build synthetic datasets with varying library sizes and computed the information footprints. To quantify the quality of signal in information footprints, we calculate signal-to-noise ratio, $\sigma$, according to the formula
\begin{align} \label{eq:signal-to-noise}
    \sigma = \frac{\langle I \rangle_B}{\langle I \rangle_{NB}},\ \mathrm{where}\ \langle I \rangle_B = \frac{1}{l_B} \sum \limits_{i \in B} I_i\ \mathrm{and}\ \langle I \rangle_{NB} = \frac{1}{l_{NB}} \sum \limits_{i \in NB} I_i.
\end{align}
Here, $I_i$ represents the mutual information at position $i$, $B$ is the set of bases within each binding site, $NB$ is the set of bases outside the binding sites, $l_B$ is the length of the specific binding site, and $l_{NB}$ is the total length of the non-binding sites. As shown in Fig~\ref{fig5}(A) and~\ref{fig5}(B), we observe that signal-to-noise ratio increases as the library size increases. This may be explained by the ``hitch-hiking'' effect: since mutations are random, mutations outside of specific binding sites can co-occur with mutations in specific binding sites. As a result, when the library is small, there is an increased likelihood that a mutation outside of specific binding sites and a mutation at a specific binding site become correlated by chance, leading to artificial signal at the non-binding sites.

\begin{figure}[!h]
\centering
\includegraphics[width=0.8\textwidth]{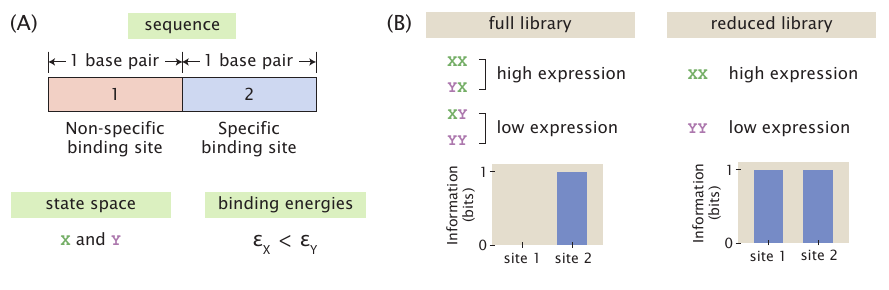}
\caption{{\bf Hitch-hiking effect in the hypothetical minimal promoter.}
(A)~Set-up of the hypothetical minimal promoter. The specific binding sites and the non-binding sites of the minimal promoter are each 1 base-pair long. There are two possible bases at each binding site, X and Y. Strong binding occurs when the base is X, whereas weak binding occurs when the base is Y. (B)~Effect of library size on the information footprint of the minimal promoter. A full mutant library consists of all four possible sequences and leads to a footprint with no signal outside of the specific binding sites. On the other hand, a reduced mutant library with only two sequences creates noise outside of the specific binding sites. In this case, the noise at the non-binding site has the same magnitude as the signal at the specific binding site.}
\label{fig6}
\end{figure}

To demonstrate the hitch-hiking effect analytically, we consider a hypothetical promoter that is constitutively transcribed and only two base pairs long, as illustrated in Fig~\ref{fig6}(A). Without loss of generality, we assume that there are only two letters in the nucleotide alphabet, X and Y. Therefore, a complete and unbiased library contains four sequences: XX, YX, XY, and YY. We designate that $\varepsilon_X < \varepsilon_Y$, i.e. the RNAP is strongly bound at the binding site when the base identity is X and weakly bound when the base identity is Y. We also assume that there is active transcription only when RNAP is bound to the second site. Under these assumptions, there are high expression levels when the promoter sequence is XX or YX and low expression levels when the promoter sequence is XY and YY.

We first consider a mutant library with full diversity and no bias, i.e. the four possible sequences, XX, YX, XY, and YY, are each present in the library exactly once. The marginal probability distribution for expression levels is
\begin{align}
    \Pr(\mu) = \begin{cases}
        0.5,\ &\mathrm{if}\ \mu=0 \\
        0.5,\ &\mathrm{if}\ \mu=1.
    \end{cases}
\end{align}
The marginal probability distributions of base identity at the two sites are
\begin{align}
    \Pr{_1(b)} = \Pr{_2(b)} = \begin{cases}
        0.5,\ &\mathrm{if}\ b=\mathrm{X} \\
        0.5,\ &\mathrm{if}\ b=\mathrm{Y}.
    \end{cases}
\end{align}
The joint probability distribution at the first site is
\begin{align}
    \Pr{_1(\mu, b)} = \begin{cases}
        0.25,\ &\mathrm{if}\ \mu=0\ \mathrm{and}\ b=\mathrm{X} \\
        0.25,\ &\mathrm{if}\ \mu=0\ \mathrm{and}\ b=\mathrm{Y} \\
        0.25,\ &\mathrm{if}\ \mu=1\ \mathrm{and}\ b=\mathrm{X} \\
        0.25,\ &\mathrm{if}\ \mu=1\ \mathrm{and}\ b=\mathrm{Y}.
    \end{cases}
\end{align}
On the other hand, the joint probability distribution at the second site is
\begin{align}
    \Pr{_2(\mu, b)} = \begin{cases}
        0,\ &\mathrm{if}\ \mu=0\ \mathrm{and}\ b=\mathrm{X} \\
        0.5,\ &\mathrm{if}\ \mu=0\ \mathrm{and}\ b=\mathrm{Y} \\
        0.5,\ &\mathrm{if}\ \mu=1\ \mathrm{and}\ b=\mathrm{X} \\
        0,\ &\mathrm{if}\ \mu=1\ \mathrm{and}\ b=\mathrm{Y}.
    \end{cases}
\end{align}
We can calculate the mutual information at each site according to Eq~\ref{eq:MI},
\begin{align}
    I_1 = 4 \left( \frac{1}{4} \log_2 \left(\frac{1/4}{1/2 \cdot 1/2}\right)\right) = 0\\
    I_2 = 2 \left( \frac{1}{2} \log_2 \left(\frac{1/2}{1/2 \cdot 1/2}\right)\right) = 1.
\end{align}
Therefore, when the library has the maximum size, there is perfect signal at the specific binding site and no signal outside of the specific binding site, as shown in Fig~\ref{fig6}(B).

On the other hand, consider a reduced library that only consists of XX and YY. According to the assumptions stated above, XX has high expression and YY has low expression. In this case, there is an apparent correlation between the base identity at the non-binding site and expression levels, where a base identity of X at the non-binding site appears to lead to high expression levels and a base identity of Y at the non-binding site appears to lead to low expression levels. To demonstrate this analytically, we again write down the relevant probability distributions required for calculating mutual information. The marginal probability distributions for expression levels and base identity are the same as the case where we have a full library. However, the joint probability distributions at both of the two sites become
\begin{align}
    \Pr{_1(\mu, b)} = \Pr{_2(\mu, b)} = \begin{cases}
        0,\ &\mathrm{if}\ \mu=0\ \mathrm{and}\ b=\mathrm{X} \\
        0.5,\ &\mathrm{if}\ \mu=0\ \mathrm{and}\ b=\mathrm{Y} \\
        0.5,\ &\mathrm{if}\ \mu=1\ \mathrm{and}\ b=\mathrm{X} \\
        0,\ &\mathrm{if}\ \mu=1\ \mathrm{and}\ b=\mathrm{Y}.
    \end{cases}
\end{align}
This means that for both the non-binding site and the specific binding site, the mutual information is
\begin{align}
    I_1 = I_2 = 2 \left( \frac{1}{2} \log_2 \left(\frac{1/2}{1/2 \cdot 1/2}\right)\right) = 1.
\end{align}
As shown in Fig~\ref{fig6}(B), this creates an artificial signal, or noise, outside of the specific binding sites that cannot be distinguished from the signal at the specific binding site.

For the remaining analyzes shown in this work, we use a library size of 5,000 in order to minimize noise from hitch-hiking effects. We choose not to use a larger library because it would significantly increase the computational cost during parameter searches. Moreover, we would like to use a library size that is experimentally feasible. In Reg-Seq, the average library size is 1,500~\cite{Ireland2020-bp}. A larger library would make MPRAs cost prohibitive as a high-throughput method.

\subsection{Perturbing biological parameters in the computational pipeline}

\subsubsection{Tuning the free energy of transcription factor binding} \label{sec:free-energy}

So far, we have demonstrated that we can build synthetic datasets for the most common regulatory architectures and we have chosen the appropriate mutation rate and library size to construct mutant libraries. Next, we proceed to perturb parameters that affect the probability of RNAP being bound and observe the effects of these perturbations. These analyzes will elucidate the physiological conditions required for obtaining clear signals from transcription factor binding events and delineate the limits of MPRA procedures in identifying unannotated transcription factor binding sites.

We again begin by considering the promoter with the simple repression motif, for which the probability of RNAP being bound is given by Eq~\ref{eq:pbound}. It is known that in \textit{E. coli} grown in minimal media, the copy number of RNAP is $P \approx 10^3$ \cite{Belliveau2021-lu, Bakshi2012-ha} and the copy number of the repressor is $R \approx 10$ \cite{Kalisky2007-kq}. The binding energy of RNAP is $\Delta \varepsilon_\mathrm{pd} \approx -5\ k_BT$ \cite{Brewster2012-ds} and the binding energy of the repressor is $\Delta \varepsilon_\mathrm{rd} \approx -15\ k_BT$ \cite{Garcia2011-np}. Moreover, assuming that the number of non-binding sites is equal to the size of the \textit{E. coli} genome, we have that $N_{\mathrm{NS}} \approx 4 \times 10^6$. Given these values, we can estimate that
\begin{align}
    \frac{P}{N_\mathrm{NS}}e^{-\beta \Delta \varepsilon_\mathrm{pd}} \approx \frac{10^3}{4 \times 10^6} \times e^5 \approx 0.04
\end{align}
and
\begin{align} \label{eq:eFR}
    \frac{R}{N_\mathrm{NS}}e^{-\beta \Delta \varepsilon_\mathrm{rd}} \approx \frac{10}{4 \times 10^6} \times e^{15} \approx 8.
\end{align}
Since $\frac{P}{N_\mathrm{NS}}e^{-\beta \Delta \varepsilon_\mathrm{pd}} \ll 1$, we can neglect this term from the denominator in Eq~\ref{eq:pbound} and simplify $p_{\mathrm{bound}}$ for the simple repression motif to
\begin{align} \label{eq:pbound2}
    p_{\mathrm{bound}} = \frac{\frac{P}{N_\mathrm{NS}}e^{-\beta \Delta \varepsilon_\mathrm{pd}}}{1 + \frac{R}{N_\mathrm{NS}}e^{-\beta \Delta \varepsilon_\mathrm{rd}}}.
\end{align}
Furthermore, we define the free energy of RNAP binding as
\begin{align}
    F_P = \Delta \varepsilon_\mathrm{pd} - k_BT \ln \frac{P}{N_\mathrm{NS}}
\end{align}
and the free energy of repressor binding as
\begin{align}
    F_R = \Delta \varepsilon_\mathrm{rd} - k_BT \ln \frac{R}{N_\mathrm{NS}}.
\end{align}
Both expressions are written according to the definition of Gibbs free energy, where the first terms correspond to enthalpy and the second terms correspond to entropy. Using these definitions, we can rewrite $p_{\mathrm{bound}}$ as
\begin{align}
    p_{\mathrm{bound}} = \frac{e^{-\beta F_P}}{1 + e^{-\beta F_R}}.
\end{align}

\begin{figure}[!h]
\centering
\includegraphics[width=0.8\textwidth]{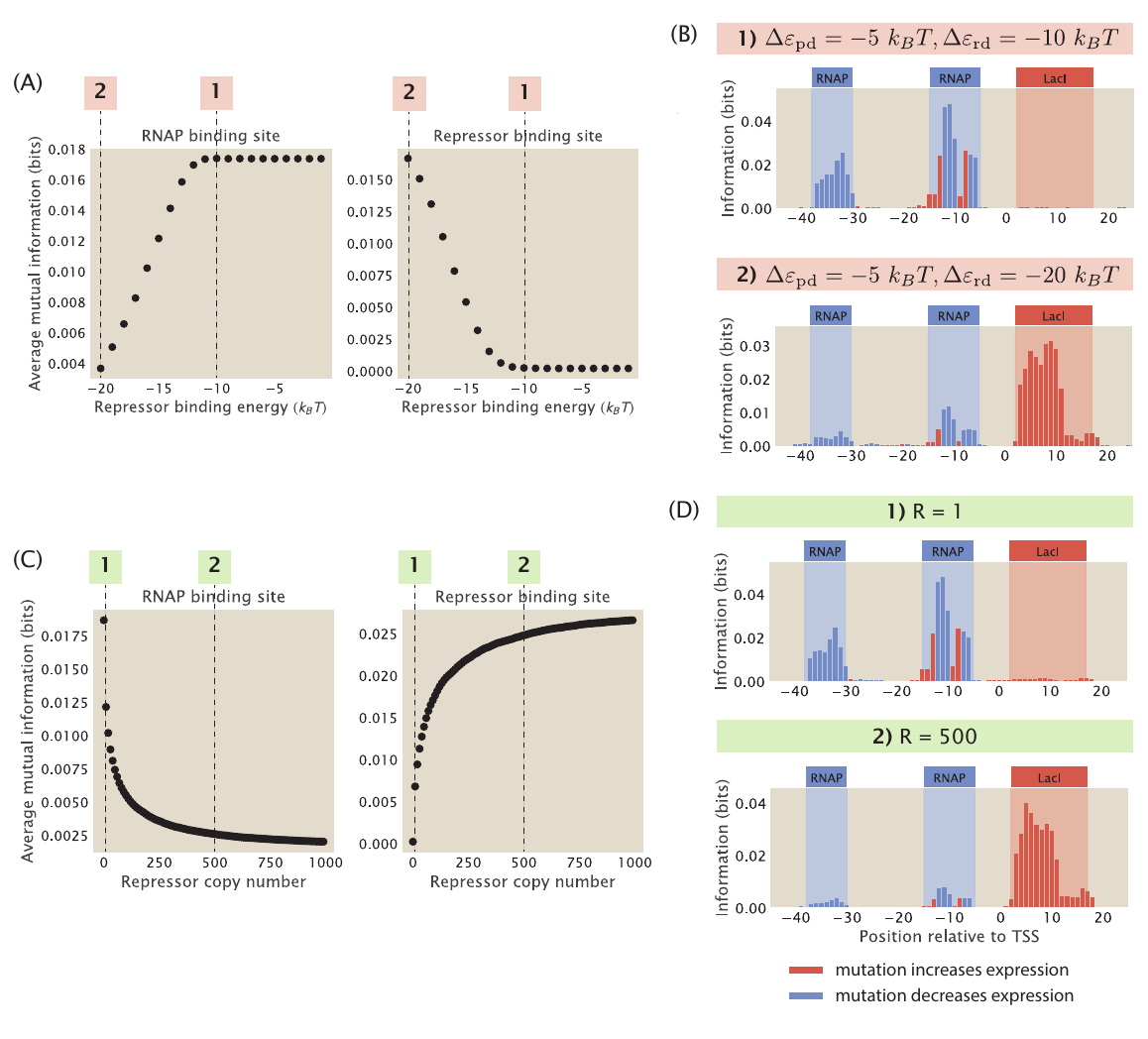}
\caption{{\bf The strength of the signal at binding sites depends on the free energy of repressor binding.}
(A)~Increasing the binding energy of the repressor leads to an increase in average mutual information at the RNAP binding site and a decrease in average mutual information at the repressor binding site. $\Delta \varepsilon_{\mathrm{pd}}$ is fixed at -5 $k_BT$, RNAP copy number is fixed at 1000, and repressor copy number is fixed at 10. Each data point is the mean of average mutual information across 20 synthetic datasets with the corresponding repressor binding energy. Numbered labels correspond to footprints in (B). (B)~Representative information footprints where $\Delta \varepsilon_\mathrm{rd}$ is set to $-20\ k_BT$ and $-10\ k_BT$. (C)~Increasing the copy number of the repressor leads to a decrease in average mutual information at the RNAP binding site and an increase in average mutual information at the repressor binding site. $\Delta \varepsilon_{\mathrm{pd}}$ is fixed at -5 $k_BT$ and $\Delta \varepsilon_{\mathrm{rd}}$ is fixed at -15 $k_BT$. RNAP copy number is fixed at 1000. Each data point is the mean of average mutual information across 20 synthetic datasets with the corresponding repressor copy number. Numbered labels correspond to footprints in (D). (D)~Representative information footprints where repressor copy numbers are set to 1 and 500.}
\label{fig7}
\end{figure}

In this section, we specifically examine the changes in the information footprints when we tune $F_R$. As shown in Fig~\ref{fig7}(A) and \ref{fig7}(C), if we increase $F_R$ by reducing the magnitude of $\Delta \varepsilon_\mathrm{rd}$ or reducing the copy number of the repressor, we lose the signal at the repressor binding site. For example, compared to the O1 operator, the LacI repressor has weak binding energy at the O3 operator, where $\Delta \varepsilon_\mathrm{rd} \approx -10\ k_BT$~\cite{Garcia2011-np}. Therefore
\begin{align}
    F_R = \Delta \varepsilon_\mathrm{rd} - k_BT \ln \frac{R}{N_\mathrm{NS}} \approx (-10 - \ln \frac{10}{4 \times 10^6})\ k_BT \approx 3\ k_BT,
\end{align}
and
\begin{align}
    e^{-\beta F_R} \approx 0.05.
\end{align}
In these cases, $e^{-\beta F_R} \ll 1$ and therefore $e^{-\beta F_R}$ can be neglected from the denominator and the probability of RNAP being bound can be simplified to
\begin{align}
    p_{\mathrm{bound}}=e^{-\beta F_P}.
\end{align}
This implies that mutations at the repressor binding sites will not have a large effect on $p_{\mathrm{bound}}$ and the mutual information at the repressor binding site will be minimal. 

On the other hand, if we decrease the free energy of binding either by increasing the magnitude of $\Delta \varepsilon_\mathrm{rd}$ or increasing the copy number of the repressor, it leads to a stronger signal at the repressor binding site while significantly reducing the signal at the RNAP binding site, as we can see in Fig~\ref{fig7}(A) and \ref{fig7}(C). For example, when $\Delta \varepsilon_\mathrm{rd} = -20 k_BT$, we have that
\begin{align}
    F_R = \Delta \varepsilon_\mathrm{rd} - k_BT \ln \frac{R}{N_\mathrm{NS}} \approx (-20 - \ln \frac{10}{4 \times 10^6})\ k_BT \approx -7\ k_BT,
\end{align}
and therefore
\begin{align}
    e^{-\beta F_R} \approx 10^3.
\end{align}
Here, the Boltzmann weight of the repressor has been increased a hundred fold compared to Eq~\ref{eq:eFR}. Due to the strong binding of the repressor, mutations at the RNAP binding site do not change expression on measurable levels and therefore the signal is low at the RNAP binding site.

In particular, we see in Fig~\ref{fig7}(A) that when the repressor energy is increased beyond $-11\ k_BT$, the average mutual information at the RNAP binding site saturates and the average mutual information at the repressor binding site remains close to 0. To explain this effect, we again take a look at
the ratio between the Boltzmann weights of the repressor and RNAP, the expression for which is stated in Eq~\ref{eq:boltzmann-ratio}. Here, we fix the copy number of the repressors and RNAP, the wild-type binding energy of RNAP, the number of mutations, and the effect of mutations. Therefore,
\begin{align}
    \kappa &= \frac{R}{P} \cdot e^{-\beta(\Delta \varepsilon_{\mathrm{rd}} - \Delta \varepsilon_{\mathrm{pd}} + m_R \Delta \Delta \varepsilon_{\mathrm{rd}} - m_P \Delta\Delta \varepsilon_{\mathrm{pd}})} \\
    &= \frac{10}{1000} \cdot e^{-\beta \Delta \varepsilon_\mathrm{rd} - 5 - 2 \times 2.24 + 2 \times 0.36} \\
    &= 1.5 \times 10^{-6} \cdot e^{-\beta \Delta \varepsilon_\mathrm{rd}}.
\end{align}
We assume that $\kappa$ needs to be at least 0.1 for there to be an observable signal at the repressor binding site. Solving for $\varepsilon_\mathrm{rd}$ using the above equation, we have that $\Delta \varepsilon_\mathrm{rd} \approx -11\ k_BT$. This matches with our observation that the signal stabilizes when $\Delta \varepsilon_\mathrm{rd} > -11\ k_BT$. Taken together, these result invite us to rethink our interpretation of MPRA data, as the lack of signal may not necessarily indicate the absence of binding site, but it may also be a result of weak binding or low transcription factor copy number.

\subsubsection{Changing the regulatory logic of the promoter} \label{sec:logic}

\begin{figure}[!h]
\centering
\includegraphics[width=\textwidth]{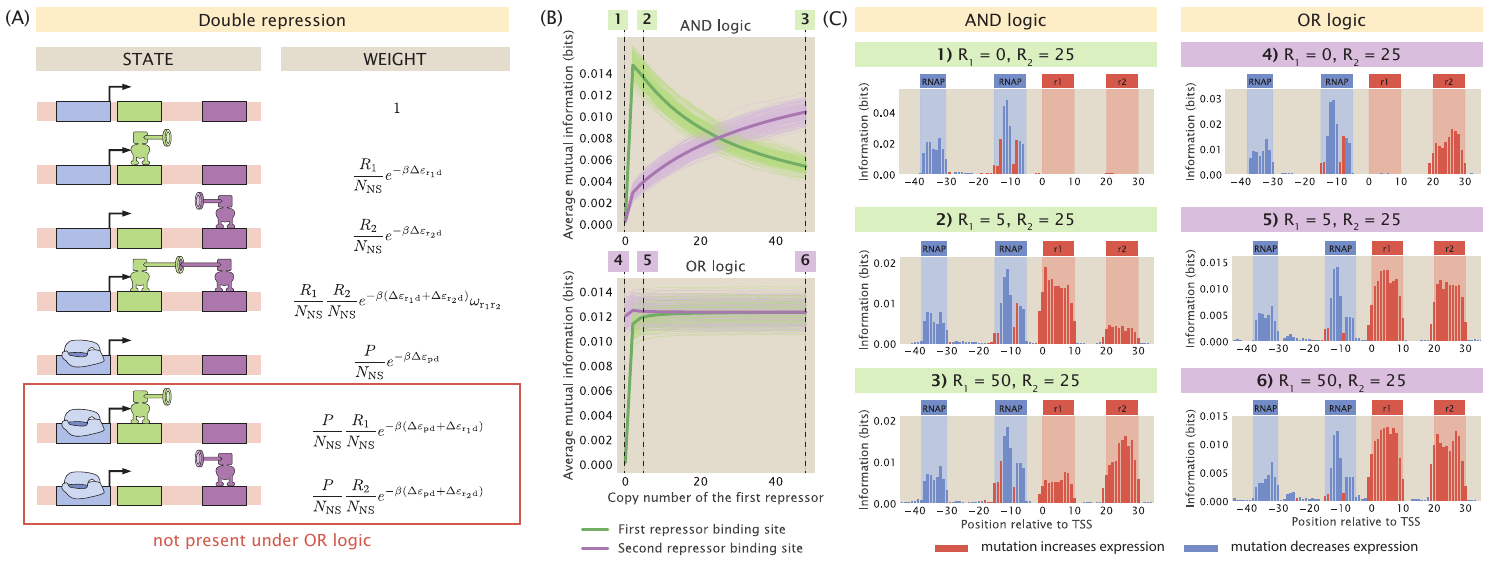}
\caption{{\bf Changing repressor copy number for a double-repression promoter.}
(A)~States-and-weights diagram of a promoter with the double repression regulatory architecture. The bottom two states are only present under AND logic and not present under OR logic. The states-and-weights diagram of a double repression promoter with OR logic is also shown in Fig~\ref{figS2}(E). (B)~Changing the copy number of the first repressor under AND logic and OR logic affects the signal at both repressor binding sites. For the energy matrices of the repressors, the interaction energy between the repressor and a site is set to $0\ k_BT$ if the site has the wild-type base identity and set to $1\ k_BT$ if the site has the mutant base identity. The interaction energy between the repressors is set to $-5\ k_BT$. 200 synthetic datasets are simulated for each copy number. We observe that the average mutual information at binding sites has high variability across synthetic datasets, especially under OR logic. To show variability, the trajectory for each of the synthetic dataset is shown as an individual light green or light purple curve. The average trajectories across all 200 synthetic datasets are shown as the bolded green curves and the bolded purple curves. The numbered labels correspond to footprints in (C).(C)~Representative information footprints of a double repression promoter under AND and OR logic.}
\label{fig8}
\end{figure}

In the previous section, we examined the changes in information footprints when we tune the copy number of the repressors under the simple repression regulatory architecture. The effect of transcription factor copy numbers on the information footprints is more complex when a promoter is regulated by multiple transcription factors. In particular, the changes in the information footprints depend on the regulatory logic of the promoter. To see this, we consider a promoter that is regulated by two repressors. For a double-repression promoter, there are many possible regulatory logics; two of the most common ones are an AND logic gate and an OR logic gate. As shown in Fig~\ref{fig8}(A), if the two repressors operate under AND logic, both repressors are required to be bound for repression to occur. This may happen if each of the two repressors bind weakly at their respective binding sites but bind cooperatively with each other. On the other hand, if the two repressors operate under OR logic, then only one of the repressors is needed for repression. 

We generate synthetic datasets for an AND-logic and an OR-logic double-repression promoter that are regulated by repressors $R_1$ and $R_2$. As shown in Fig~\ref{fig8}(B) and~\ref{fig8}(C), under AND logic, there is no signal at either of the repressor binding sites when $R_1$ is set to 0. This matches our expectation because AND logic dictates that when one of the two repressors is absent, the second repressor is not able to reduce the level of transcription by itself. We recover the signal at both of the repressor binding sites even when there is only one copy of $R_1$. Interestingly, when $R_1 > R_2$, the signal at the $R_1$ binding site is lower than the signal at the $R_2$ binding site. This may be because the higher copy number of $R_2$ compensates for the effects of mutations and therefore expression levels are affected to a greater extent by mutations at the $R_1$ binding site than mutations at the $R_2$ binding site. In comparison, since the two repressors act independently when they are under OR logic, the signal at the $R_2$ binding site is preserved even when $R_1=0$. Moreover, the state where $R_1$ represses transcription competes with the state where $R_2$ represses transcription. As a result, when $R_1$ is increased, the signal at the $R_1$ binding site is increased whereas the signal at the $R_2$ binding site decreases.

As illustrated in~\ref{S8_Appendix} and~\ref{S9_Appendix}, the procedure described above can be extended to other regulatory logic such as the XOR gate and other promoter architectures such as a promoter that is regulated by two activators. These results are informative in the context of transcription factor deletion, which is a key approach for identifying and verifying which transcription factor binds to the putative binding sites discovered in MPRA pipelines \cite{Ireland2020-bp}. The final copy number of the transcription factor depends on which experimental method is chosen to perform the deletion. If the gene coding for the transcription factor is knocked out, no transcription factor will be expressed and the transcription factor copy number will be 0. Therefore, by comparing the footprints from the wild-type strain and the transcription factor deletion strain, we can locate the site where the signal disappears and deduce which transcription factor is bound at that site. On the other hand, if knock-down methods such as RNA interference are used, some leaky expression may take place and the transcription factor copy number may be low but non-zero. In this case, there may not be appreciable differences in the footprints from the wild-type strain and the deletion strain. This would be an important point of consideration in MPRAs where knock-down methods are used to match transcription factors to binding sites.

\subsubsection{Competition between transcription factor binding sites} \label{sec:plasmid}

Thus far, we have assumed that each transcription factor only has one specific binding site in the genome. However, many transcription factors bind to multiple promoters to regulate the transcription of different genes. For example, cyclic-AMP receptor protein (CRP), one of the most important activators in \textit{E. coli}, regulates 330 transcription units~\cite{Keseler2021-ey}. Therefore, it is important to understand how the relationship between sequence and binding energy changes when the copy number of the transcription factor binding site is changed.

\begin{figure}[!h]
\centering
\includegraphics[width=0.7\textwidth]{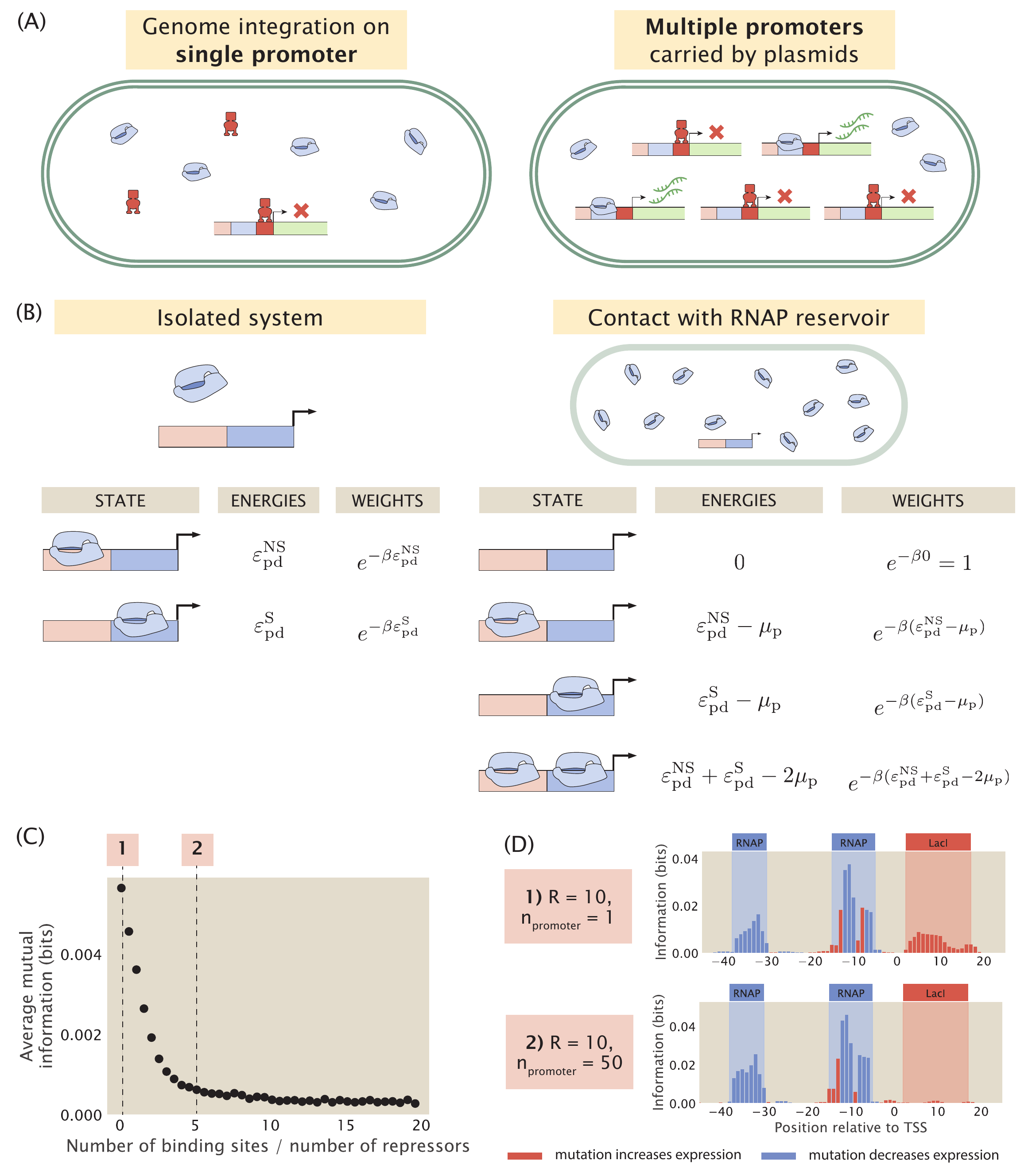}
\caption{{\bf Changing the copy number of transcription factor binding sites.}
(A)~There are two ways of delivering sequence variants into a cell. If the promoter variant is integrated into the genome, the original copy number of the promoter is preserved. On the other hand, if the cells are transformed with plasmids containing the promoter variant, binding site copy number will increase. When the copy number of the binding site is high, the additional binding sites titrate away the repressors and the gene will be expressed at high levels despite the presence of repressors. (B)~States-and-weights models for a constitutive promoter in an isolated system and in contact with a cellular reservoir of RNAPs. To build a thermodynamic model for an isolated system, we assume that there are no free-floating RNAPs and we require that the number of bound RNAPs is equal to the total number of RNAPs in the system. On the other hand, for a system that is in contact with a reservoir, we only need to ensure that the average number of RNAPs bound matches the total number of RNAPs. (C)~Average mutual information at the repressor binding site decreases when the number of the repressor binding site is increased. Repressor copy number is set to 10 for all data points. Each data point is the mean of average mutual information across 20 synthetic datasets with the corresponding number of repressor binding sites. Numbered labels correspond to footprints in (D). (D)~Representative information footprints for cases where there is only 1 repressor binding site and when there are 50 repressor binding sites. }
\label{fig9}
\end{figure}

Binding site copy number is also highly relevant in the context of the experimental MPRA pipeline. When \textit{E. coli} is the target organism, there are two main ways of delivering mutant sequences into the cell. As illustrated in Fig~\ref{fig9}(A), the first is by directly replacing the wild type promoter with the mutant promoter using genome integration methods such as ORBIT~\cite{Murphy2018-rf, Saunders2023-yr}. In this case, we preserve the original copy number of the binding sites. The second method is to transform the bacterial cells with plasmids carrying the promoter variant. If this approach is used, the number of transcription factor binding sites will increase by the copy number of the plasmids. We would like to understand precisely how the signal in the resulting information footprint differs between a genome integrated system and a plasmid system.

To build a synthetic dataset that involves more than one transcription factor binding site, we once again begin by building a thermodynamic model to describe the different binding events. However, in the canonical thermodynamic model that we utilized earlier, introducing multiple transcription factor binding sites would lead to a combinatorial explosion in the number of possible states. To circumvent this issue, we introduce an alternative approach based on the concept of chemical potential. Here, chemical potential corresponds to the free energy required to take an RNAP or a transcription factor out of the cellular reservoir. As shown in Fig~\ref{fig9}(B), it is convenient to use chemical potential because in contrast to an isolated system, the resulting model no longer imposes a constraint on the exact number of RNAP or transcription factor bound to the promoter. Instead, we can tune the chemical potential such that we constrain the average number of bound RNAPs and transcription factors. This decouples the individual binding sites and allows us to write the total partition function as a product of the partition functions at each site.

Using the method of chemical potential, we construct synthetic datasets with different repressor binding site copy numbers. As shown in Fig~\ref{fig9}(C), as the copy number of the repressor binding site is increased, the signal at the repressor binding site decreases rapidly and eventually stabilizes at a near-zero value. In particular, as shown in Fig~\ref{fig9}(D), in a genome integrated system where there is only one copy of the repressor binding site, there is clear signal at the repressor binding site. On the other hand, in a plasmid system where the copy number of the binding site is greater than the copy number of the repressor, the signal for the repressor disappears. Intuitively, this is because the additional binding sites titrate away the repressors, which reduces the effective number of repressors in the system. As a result, the expression of the reporter gene no longer reflects transcriptional regulation by the repressor. This reduces the effect of mutations at the repressor binding site on expression levels, which leads to low mutual information between mutations and base identities at the repressor binding site. 

In wild type \textit{E. coli}, the median ratio of transcription factor copy number and binding site copy number is around 10~\cite{Schmidt2016-wj}, and therefore the titation effects are unlikely to diminish the signals in information footprints when the sequence variant is integrated into the genome. On the other hand, if a plasmid system is used, it is beneficial to make use of a low copy number plasmid. Although we have no knowledge of which transcription factor is potentially regulating the gene of interest and therefore we do not know a priori the copy number of the transcription factor, using a low copy number plasmid has a greater chance of ensuring that the copy number of the transcription factor binding sites is no greater than the copy number of the putative transcription factor.

\subsubsection{Changing the concentration of the inducer} \label{sec:inducer}

So far, in the regulatory architectures that involve repressor binding, we have only considered repressors in the active state, whereas in reality the activity of the repressors can be regulated through inducer binding. Specifically, according to the Monod–Wyman–Changeux (MWC) model~\cite{Monod1965-ij}, the active and inactive states of a repressor exist in thermal equilibrium and inducer binding may shift the equilibrium in either direction. If inducer binding shifts the allosteric equilibrium of the repressor from the active state towards the inactive state, the repressor will bind more weakly to the promoter. This will increase the probability of RNAP being bound and therefore lead to higher expression. In other words, increasing inducer concentration has similar effects to knocking out the repressor from the genome. For example, when lactose is present in the absence of glucose, lactose is converted to allolactose, which acts as an inducer for the Lac repressor and leads to increased expression of genes in the LacZYA operon. Conversely, some inducer binding events may also shift the equilibrium of a repressor from the inactive state towards the active state. One example is the Trp repressor, which is activated upon tryptophan binding and represses gene expression. Here, we use the example of the lacZYA operon and demonstrate how signals in the information footprint depend on the concentration of the allolactose inducer in the system.

\begin{figure}[!h]
\centering
\includegraphics[width=0.9\textwidth]{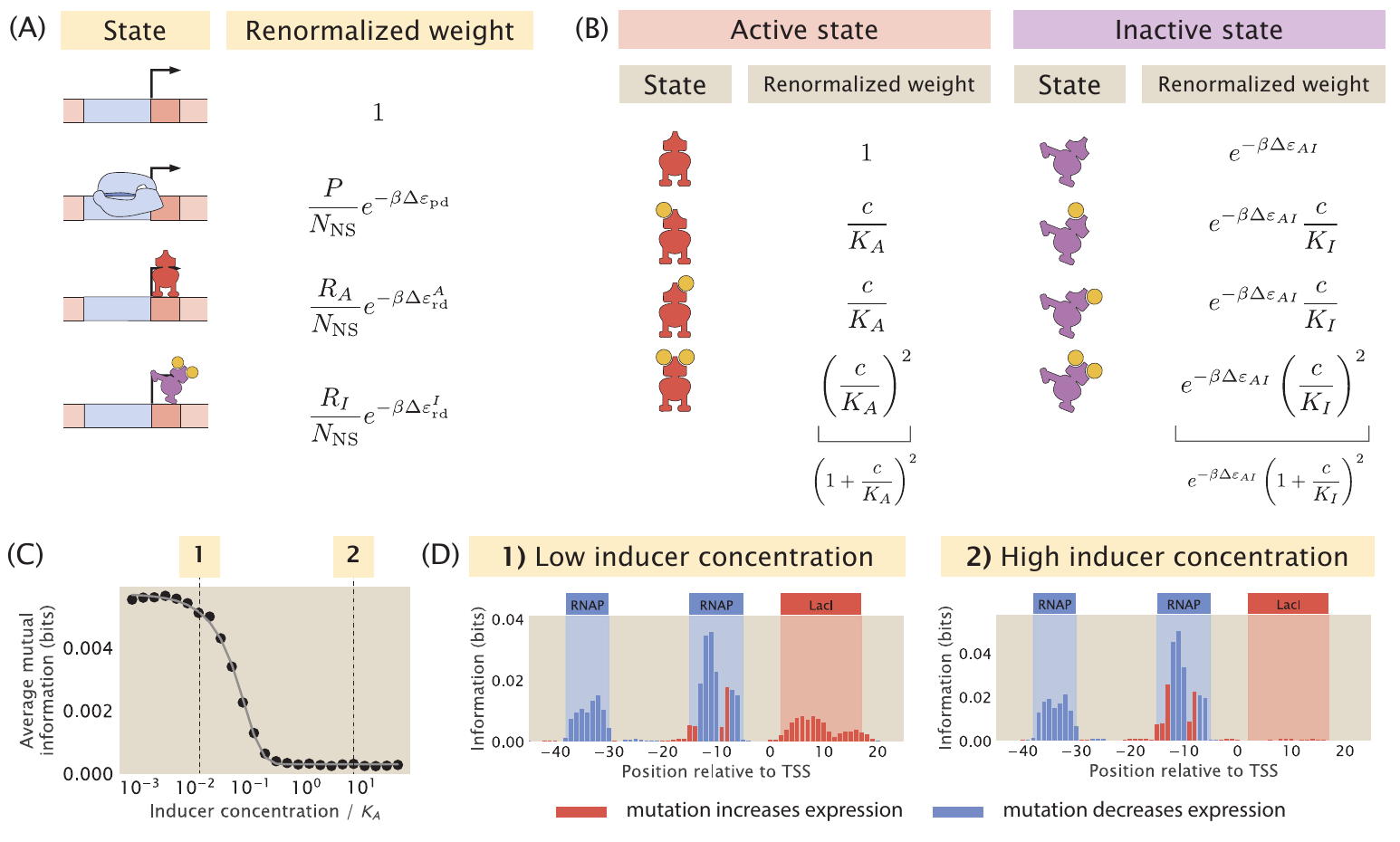}
\caption{{\bf Changing the concentration of the inducer.}
(A) States-and-weights diagram for an inducible repressor. (B) States-and-weights diagram to calculate the probability that the repressor is in the active state. (C) Average mutual information at the repressor binding site decreases as the inducer concentration increases. Here, we let $K_A = 139 \times 10^{-6}$ M, $K_I = 0.53 \times 10^{-6}$ M, and $\Delta \varepsilon_{AI} = 4.5\ k_BT$. The thermodynamic parameters were inferred by Razo-Mejia et al. from predicted IPTG induction curves.~\cite{Razo-Mejia2018-cc} The inducer concentration on the x-axis is normalized with respect to the value of $K_A$. Each data point is the mean of average mutual information across 20 synthetic datasets with the corresponding inducer concentration. The numbered labels correspond to footprints in (D). (D) Representative information footprints with low inducer concentration ($10^{-6}$ M) and high inducer concentration ($10^{-3}$ M).}
\label{fig10}
\end{figure}

As shown in Fig~\ref{fig10}(A), to include an inducible repressor in our thermodynamic model, we add an additional state in the states-and-weights diagram that accounts for binding between the inactivated repressor and the promoter. This additional state is a weak binding state where the repressor is more likely to dissociate from the binding site. In many cases, transcription factors have multiple inducer binding sites. Here, we choose a typical model where the repressor has two inducer binding sites. Based on the new states-and-weights diagram, the probability of RNAP being bound can be rewritten according to the following expression~\cite{Razo-Mejia2018-cc}
\begin{align}
    p_{\mathrm{bound}} = \frac{\frac{P}{N_\mathrm{NS}}e^{- \beta \Delta \varepsilon_\mathrm{pd}}}{1 + \frac{R_A}{N_\mathrm{NS}}e^{-\beta \Delta \varepsilon_\mathrm{rd}^A} + \frac{R_I}{N_\mathrm{NS}}e^{-\beta \Delta \varepsilon_\mathrm{rd}^I} + \frac{P}{N_\mathrm{NS}}e^{-\beta \Delta \varepsilon_\mathrm{pd}}},
\end{align}
where $P$ is the number of RNAPs, $R_A$ is the number of active repressors, $R_I$ is the number of inactive repressors, and $\Delta \varepsilon_\mathrm{pd}$, $\Delta \varepsilon_\mathrm{rd}^A$, $\Delta \varepsilon_\mathrm{rd}^I$ correspond to the energy differences between specific and non-specific binding of the RNAP, the active repressor, and the inactive repressor respectively.

In order to calculate the probability of RNAP being bound, we need to determine the proportion of $R_A$ and $R_I$ with respect to the total number of repressors. To do this, we calculate $p_{\mathrm{active}}(c)$, the probability that the repressor exists in the active conformation as a function of the concentration of the inducer, $c$. To calculate $p_{\mathrm{active}}(c)$, we model the different states of the repressor using another states-and-weights diagram, as illustrated in Fig~\ref{fig10}(B). The probability that the repressor is in the active state is
\begin{align}
    p_{\mathrm{active}}(c) = \frac{\left( 1 + \frac{c}{K_A} \right)^2}{\left(1 + \frac{c}{K_A} \right)^2 + e^{-\Delta \varepsilon_{AI} / k_BT} \left(1 + \frac{c}{K_I}\right)^2},
\end{align}
where $K_A$ is the dissociation constant between the inducer and the active repressor, $K_I$ is the dissociation constant between the inducer and the inactive repressor, and $\Delta \varepsilon_{AI}$ is the structural energy difference between the active repressor and the inactive repressor. This allows us to represent the number of active and inactive repressors as $R_A = p_{\mathrm{active}} R$ and $R_I = (1 - p_{\mathrm{active}}) R$. Therefore, our expression for $p_{\mathrm{bound}}$ can be modified to
\begin{align}
    p_{\mathrm{bound}} = \frac{\frac{P}{N_\mathrm{NS}}e^{-\beta \Delta \varepsilon_\mathrm{pd}}}{1 + p_{\mathrm{active}} \frac{R}{N_\mathrm{NS}}e^{-\beta \Delta \varepsilon_\mathrm{rd}^A} + (1 - p_{\mathrm{active}})\frac{R}{N_\mathrm{NS}}e^{-\beta \Delta \varepsilon_\mathrm{rd}^I} + \frac{P}{N_\mathrm{NS}}e^{-\beta \Delta \varepsilon_\mathrm{pd}}},
\end{align}

We built synthetic datasets for a promoter with the simple repression regulatory architecture with an inducible repressor. As shown in Fig~\ref{fig10}(C) and~\ref{fig10}(D), when the concentration of the inducer is increased from $10^{-2}\ K_A$ to $1\ K_A$, the average signal at the repressor binding site decreases. Interestingly, the average signal is not reduced further when the concentration is increased beyond the value of $K_A$. As shown in \ref{S10_Appendix}, similar results are observed in the case of a simple activation promoter with an inducible activator.

These results show that the presence or absence of inducers can determine whether we will obtain a signal at the transcription factor binding site. This underlies the importance of performing experimental MPRAs under different growth conditions to ensure that we can identify binding sites that are bound by transcription factors that are induced by specific cellular conditions. These efforts may fill in the gap in knowledge on the role of allostery in transcription, which so far has been lacking attention from studies in the field of gene regulation~\cite{Lindsley2006-je}.

\subsection{Identifying transcription factor binding sites from information footprints}

\subsubsection{Noise due to stochastic fluctuations of RNAP and transcription factor copy number} \label{sec:extrinsic}

\begin{figure}[!h]
\centering
\includegraphics[width=0.8\textwidth]{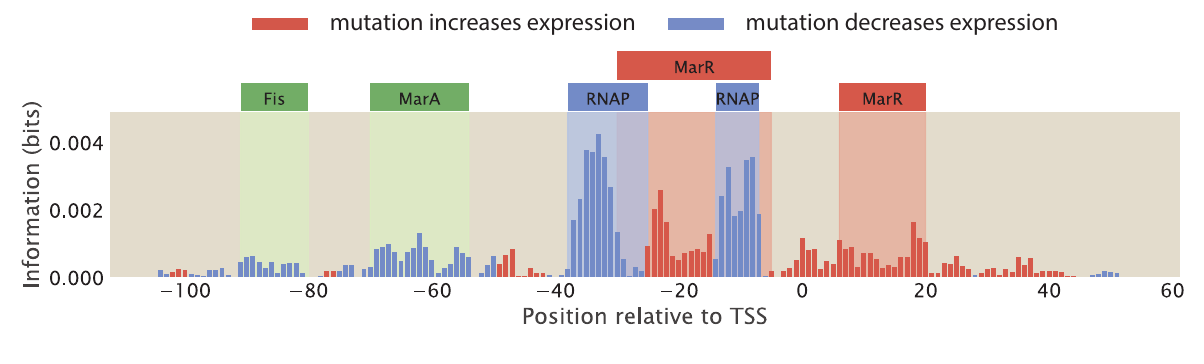}
\caption{{\bf Annotating transcription factor binding sites by identifying sites with high signal.}
The footprint of the \textit{mar} operon, produced by Ireland et al.~\cite{Ireland2020-bp} The binding sites are annotated based on known RNAP and transcription factor binding sites; the signal at some of the binding sites, such as the Fis and MarA binding sites, are not distinguishable from background noise.}
\label{fig11}
\end{figure}

When data from MPRAs is presented in the form of information footprints, one way to annotate transcription factor binding sites is to identify regions where the signal is significantly higher than background noise. Therefore, to be able to precisely and confidently identify RNAP and transcription factor binding sites from information footprints, the footprint is required to have a sufficiently high signal-to-noise ratio. However, this may not always be the case. For example, the footprint shown in Fig~\ref{fig11} is obtained for the \textit{mar} operon by Ireland et al.~\cite{Ireland2020-bp}; while the signal at the RNAP binding sites and the -20 MarR binding sites are clearly above the background noise, the signals at the Fis, MarA, and +10 MarR binding sites may easily be mistaken for noise. 

In Sec \ref{sec:library-size}, we examined how the size of the mutant library may affect the level of noise in information footprints. Here, we continue to examine other factors that may affect signal-to-noise ratio. We first simulated possible sources of experimental noise, including PCR amplification bias and random sampling effects during RNA-Seq library preparation and RNA-Seq itself. However, as shown in~\ref{S11_Appendix}, these experimental processes do not lead to significant levels of noise outside of the specific binding sites. Another potential source of noise is from the biological noise that contributes to stochastic fluctuations in expression levels. These sources of biological noise can be broadly categorized into intrinsic noise and extrinsic noise~\cite{Elowitz2002-kl, Fu2016-iq, Raser2005-fw}. Intrinsic noise arises from the inherent stochasticity in the process of transcription, such as changes in the rate of production or degradation of mRNA. On the other hand, extrinsic noise arises from cell-to-cell variation in the copy number of transcription machineries such as the RNAPs and transcription factors. It has been shown that extrinsic noise occurs on a longer timescale and has a greater effect on phenotypes than intrinsic noise~\cite{Raser2005-fw}. Here, we investigate whether extrinsic noise has an effect on the information footprints.

\begin{figure}[!h]
    \centering
    \includegraphics[width=0.8\textwidth]{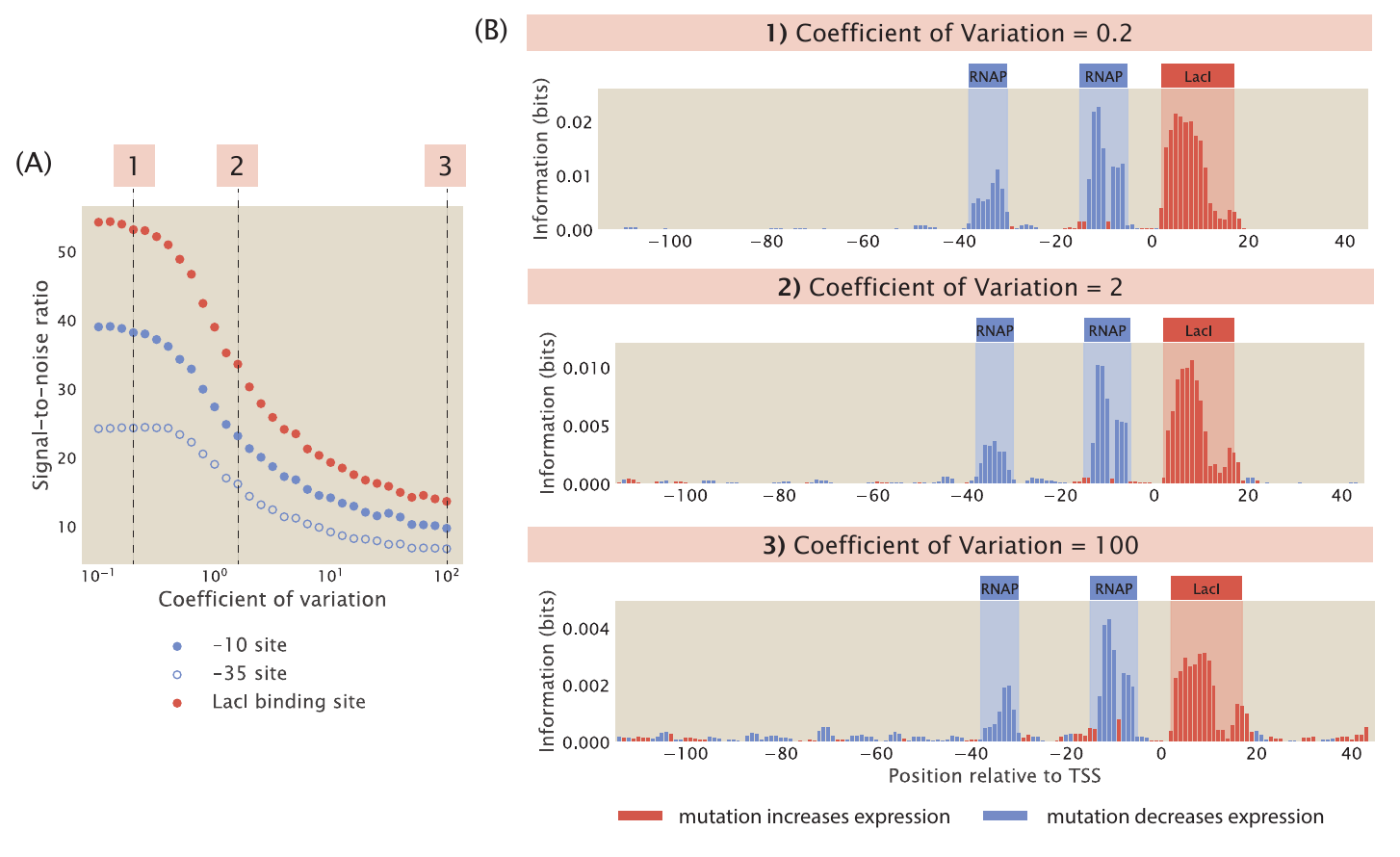}
    \caption{{\bf Adding extrinsic noise to synthetic datasets.}
    (A) Increasing extrinsic noise lowers the signal-to-noise ratio in information footprints. For all synthetic datasets used to generate the data points, the copy numbers of RNAP and repressors are drawn using the Log-Normal distribution described in~\ref{S12_Appendix}. In the Log-Normal distribution, $\mu$ is set to $5000$ for RNAPs and $100$ for repressors. Each data point is the mean of average mutual information across 100 synthetic datasets with the corresponding coefficient of variation. The numbered labels correspond to footprints in (B). (B) Representative information footprints with three levels of extrinsic noise.}
    \label{fig12}
\end{figure}

We build synthetic datasets for a promoter with the simple repression architecture using the same procedure as before, except we no longer specify the copy number of RNAPs and repressors as a constant integer. Instead, as described in \ref{S12_Appendix}, we randomly draw the copy numbers of RNAPs and repressors from a Log-Normal distribution, which is the distribution that the abundance of biomolecules often follow~\cite{Furusawa2005-qs}. Brewster et al.~\cite{Brewster2014-qf} used the dilution method to measure the stochastic fluctuations in transcription factor copy numbers due to asymmetrical partitioning during cell division and they found that transcription factor copy numbers typically vary by less than the 20\% of the mean copy numbers. Moreover, proteomic measurements~\cite{Schmidt2016-wj, Balakrishnan2022-bt} suggest that the coefficient of variation for copy numbers is less than 2 even across diverse growth conditions. Here, we build Log-Normal distributions with a range of coefficients of variations, which cover both the reported levels of extrinsic noise as well as coefficients of variation as high as 100, which is physiologically extreme. As shown in Fig~\ref{fig12}(A) and~\ref{fig12}(B), we observe that the signal-to-noise ratio does decrease when the extrinsic noise is higher. However, we can still distinguish between signal and noise even when we specify a large coefficient of variation. Moreover, as seen in~\ref{S13_Appendix}, even when the signal at the repressor binding site is low due to other factors such as low repressor binding energy, the signal from both binding sites is still distinguishable from the noise caused by copy number fluctuations. In addition, as shown in~\ref{S14_Appendix}, signal-to-noise ratio remains high in different regulatory architectures in the presence of extrinsic noise. These results suggest that information footprints as a read-out are robust to extrinsic noise.

This phenomenon may be explained by the fact that the changes in binding energies due to mutations in the promoter sequence have a much greater contribution to the probability of RNAP being bound than changes in the copy number of transcription factors. Assuming that RNAP binds weakly to the promoter, the expression for $p_{\mathrm{bound}}$ in Eq~\ref{eq:pbound} can be simplified to
\begin{align}
    p_{\mathrm{bound}} = \frac{\frac{P}{N_\mathrm{NS}}e^{- \beta \Delta \varepsilon_\mathrm{pd}}}{1 + \frac{R}{N_\mathrm{NS}}e^{-\beta \Delta \varepsilon_\mathrm{rd}}}.
\end{align}
Based on the experimentally measured energy matrix for LacI \cite{Brewster2012-ds}, the average increase in $\Delta \varepsilon_{\mathrm{rd}}$ due to one mutation is approximately $2\ k_BT$. The LacI binding site is around 20 base pairs long. Therefore, with a 10\% mutation rate, there are on average 2 mutations within the LacI binding site, and the total change in $\Delta \varepsilon_{\mathrm{rd}}$ would be approximately $4\ k_BT$. This leads to a $e^{-\beta \Delta \Delta \varepsilon_\mathrm{rd}} = 0.01$ fold change in the magnitude of $\frac{R}{N_\mathrm{NS}}e^{-\beta \Delta \varepsilon_\mathrm{rd}}$. This means that the copy number of the repressor would have to change by a factor of 100 to overcome the effect of mutations, and this is not possible through fluctuations due to extrinsic noise. Therefore, extrinsic noise by itself will not lead to a sufficiently large change in expression levels to affect the signals in the information footprint.

\subsubsection{Non-specific binding along the promoter}
\label{sec:spurious}

\begin{figure}[!h]
\centering
\includegraphics[width=0.9\textwidth]{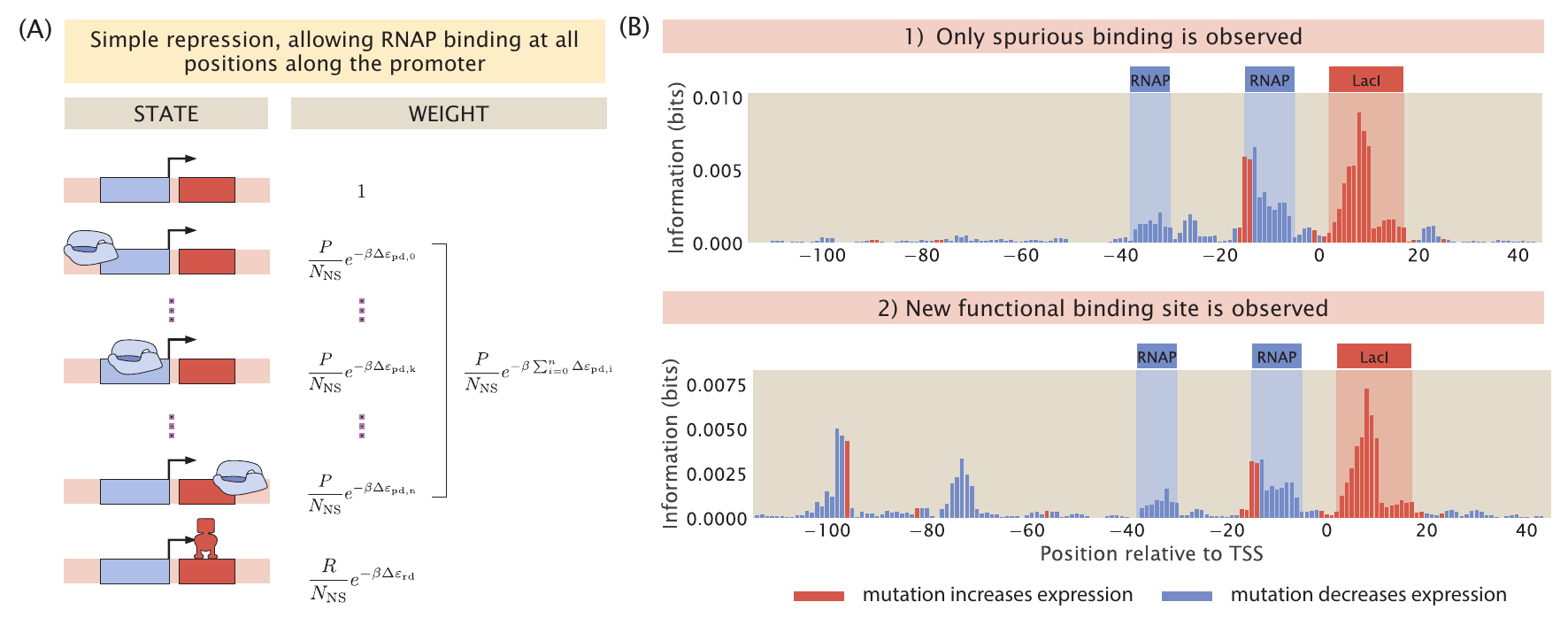}
\caption{{\bf Non-specific RNAP binding can create low levels of noise and lead to non-canonical functional binding sites.}
(A) States-and-weights diagram of a simple repression promoter where spurious RNAP binding is allowed. For each of the RNAP spurious binding events, the binding energy, $\Delta \varepsilon_{\mathrm{pd, i}}$, is computed by mapping the RNAP energy matrix to the spurious binding site sequence. The index $i$ correspond to the position of the first base pair to which RNAP binds along the promoter. $0$ is at the start of the promoter sequence; $k$ is at the canonical RNAP binding site; $n=160-l_p$ is index of the most downstream binding site where the promoter is assumed to be 160 base-pair long and $l_p$ is the length of the RNAP binding site. (B) Information footprints of a promoter under the simple repression regulatory architecture with non-specific binding (top) and with a new functional binding site (bottom). The bottom plot is created by inserting the sequence ``TAGAAT'', which is one letter away from the TATA-box sequence, at the -80 position.}
\label{fig13}
\end{figure}

In the earlier sections of the paper, our thermodynamic models only allow RNAPs to bind at binding sites at specific positions along the promoter. However, in reality, non-specific binding events along the rest of the promoter also occur, albeit at low frequencies. To investigate the effect of non-specific binding on information footprints, we build a thermodynamic model that allow for RNAP binding at every possible position along the promoter as a unique state. The states-and-weights diagram of this expanded thermodynamic model is illustrated in Fig~\ref{fig13}(A). The weight of each state is calculated by mapping the energy matrix to the corresponding non-specific binding site sequence at each position along the promoter. As shown in Fig~\ref{fig13}(B) top panel, in general, non-specific binding only leads to a small amount of noise. Similar to the case of extrinsic noise, this source of noise is not at a sufficiently high level to interfere with our ability to delineate binding site positions. This implies that reducing the hitch-hiking effects described in Sec~\ref{sec:library-size} should be the primary focus when a high signal-to-noise ratio is desired.

On the other hand, a more interesting phenomenon when RNAP is allowed to bind along the entire promoter is the presence of strong signal at non-canonical binding sites. In particular, signal may arise at these sites due to the presence of TATA-like motifs. RNAPs with $\sigma_{70}$ typically bind to the TATA box, which is a motif with the consensus sequence TATAAT located at the -10 site along the promoter. However, since the sequence motif is short, it is likely that TATA-like motifs with a short mutational distance from the TATA-box sequence may exist away from the -10 site. A mutation may easily convert these motifs into a functional TATA-box, allowing RNAP to initiate transcription from a different transcription start site. In the bottom information footprint of Fig~\ref{fig13}(B), the promoter is engineered to contain a TATA-like motif upstream of the canonical binding sites. As shown in the footprint, this leads to a strong signal at the -100 and -75 positions. This analysis unveils a feature of information footprints that may complicate interpretation of signal and noise in real-world MPRA datasets.

\subsubsection{Overlapping binding sites} \label{sec:overlapping}

Other than a low signal-to-noise ratio, another factor that may contribute to the challenge of deciphering regulatory architectures is the presence of overlapping binding events. This is especially common with RNAP and repressor binding sites, since a common mechanism by which repressors act to reduce expression is by binding to the RNAP binding site and thereby sterically blocking RNAP binding. For example, in Fig~\ref{fig11}, we can see that the MarR binding site overlaps with the RNAP binding site. Assuming perfect binding sites, a mutation in the RNAP binding site will decrease expression and a mutation in the repressor binding site will increase expression. However, if the binding sites overlap, we expect either the signals from the two binding events will cancel out or one signal will dominate the other. Here, we build synthetic datasets with different degrees of overlap between RNAP and the repressor and we examine how much overlap can be tolerated before the two binding sites are no longer distinguishable from each other.

\begin{figure}[!h]
    \centering
    \includegraphics[width=0.6\textwidth]{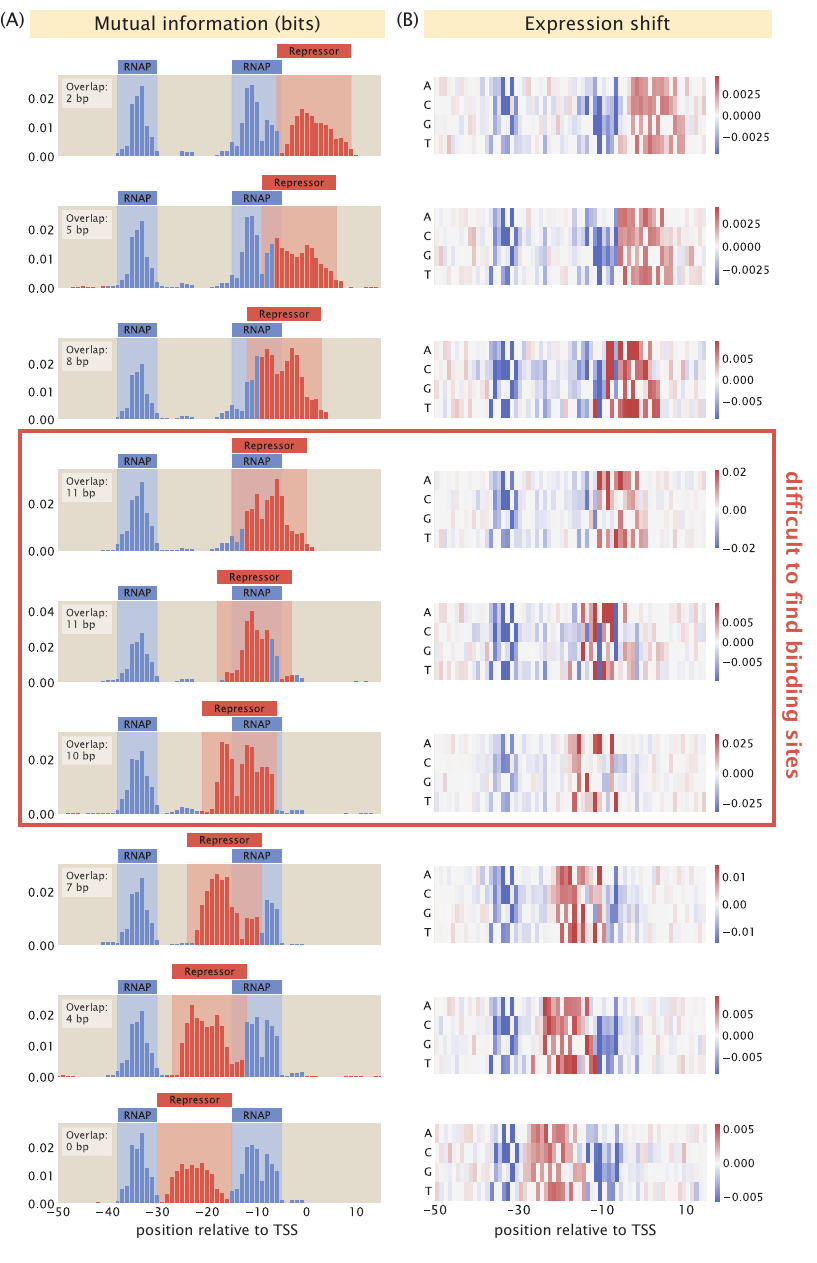}
    \caption{{\bf Changing the degree of overlap between the RNAP and repressor binding sites.}
    (A - B) Information footprints and expression shift matrices of a simple repression promoter with overlapping binding sites. The promoters are designed to maximize binding strength given the known energy matrices of the RNAP~\cite{Brewster2012-ds} and the LacI repressor~\cite{Barnes2019-dx}. The degree of overlap in the information footprints and expression shift matrices in each row is noted at the upper left hand corner of the footprints.}
    \label{fig14}
\end{figure}

In Fig~\ref{fig15}, we show a series of information footprints and expression shift matrices where the repressor is slid along the full range of the RNAP binding site. The promoter sequences are engineered to maximize binding strengths based on the energy matrices of the RNAP and the repressor shown in Fig~\ref{fig2}(B). At positions along the promoter that are only bound by RNAP, the base that minimizes the binding energy of the RNAP is chosen. The same applies for positions that are only bound by the repressor. On the other hand, at overlapping binding sites, base identities that minimize the total binding energy by the RNAP and the repressor are chosen. The base identities of the rest of the promoter sequence are selected at random. In the information footprints shown in Fig~\ref{fig14}(A), the signals from the two binding events are clearly segmented and distinguishable from each other when less than 50\% of the binding sites are overlapping. However, when the vast majority of the base positions are overlapping, signal from repressor binding dominates the signal from RNAP binding. In the expression shift matrices shown in Fig~\ref{fig14}(B), when the repressor binds directly on top of the -10 RNAP binding site, some signal from RNAP binding is still preserved. However, these signals are not strong enough to highlight the presence of an RNAP binding site. Without prior knowledge that RNAP binds at this position, such a footprint could lead to the erroneous conclusion that only repressors bind to this site. These analyzes demonstrate the challenge of deciphering regulatory architectures in the presence of overlapping binding sites. This may be overcome by tuning growth conditions to reduce binding by some of the overlapping binding partners, such that we can obtain cleaner footprints with signal indicating individual binding events.

\subsection{Building synthetic datasets under non-equilibrium conditions} \label{sec:nonequilibrium}

So far in this paper, one important assumption underlying our thermodynamic models is that the processes involved in transcription initiation are in quasi-equilibrium. The success of thermodynamic models in predicting experimental outcomes in previous works lends credibility to the use of equilibrium models \cite{Garcia2011-np, Brewster2012-ds, Brewster2014-qf, Razo-Mejia2018-cc, Phillips2019-ld} in prokaryotic systems such as \textit{E. coli}. However, in transcriptional regulation more generally, there are known to be energy consuming processes such as phosphorylation and nucleosome remodelling. Therefore, it is important to consider how our pipeline may be extended to systems where detailed balance is broken.

To account for non-equilibrium processes in the construction of synthetic MPRA datasets, we invoke the graph-theoretic approach proposed by Gunawardena~\cite{Gunawardena2012-gf} and used by Mahdavi, Salmon et al.~\cite{Mahdavi2023-vr} to calculate the probability of transcriptionally active states, which allows us to consider the full kinetic picture of trancription initiation. With the probability of transcriptionally active states calculated using this approach, we can then predict the expression levels of the promoter variants without imposing equilibrium constraints. This makes it possible for us to build a synthetic dataset that does not rely on the quasi-equilibrium assumptions, and will provide further clues about how to interpret MPRA datasets.

\begin{figure}[!h]
\centering
\includegraphics[width=0.8\textwidth]{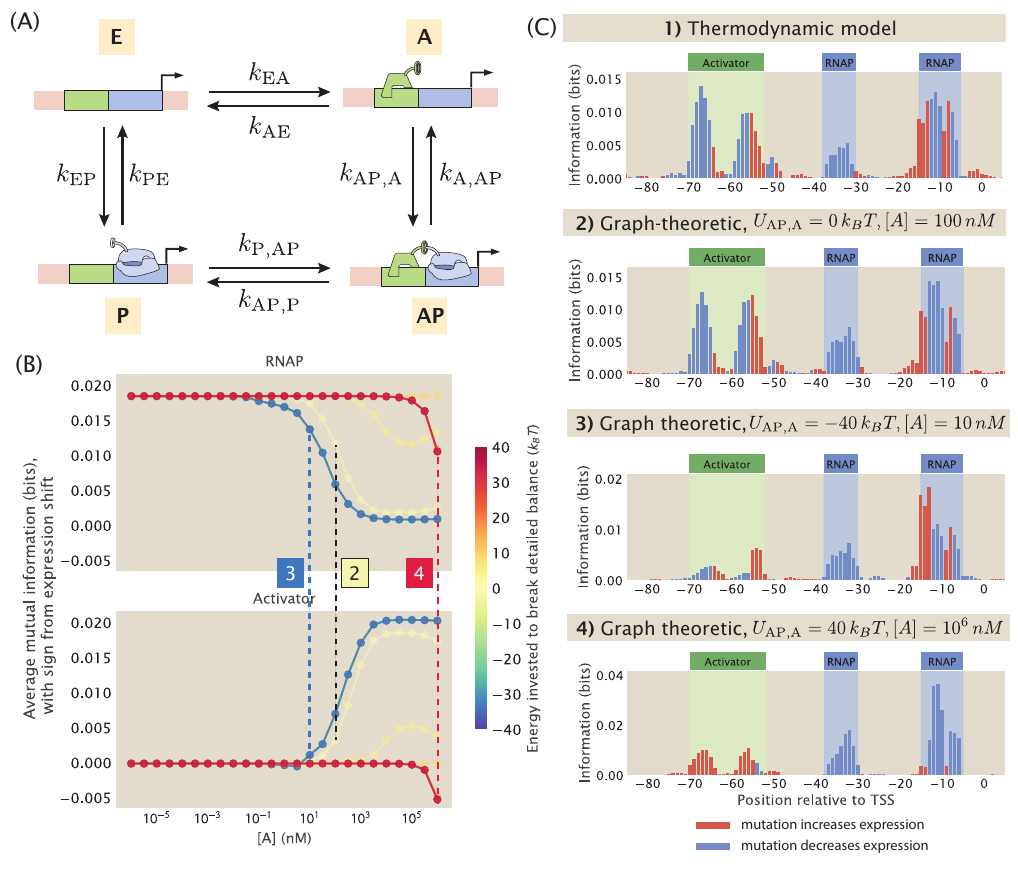}
\caption{{\bf Building synthetic datasets with broken detailed balance.}
(A) Directed square graph that describes the kinetic processes of a simple activation promoter. (B) Changes in average mutual information at the RNAP and activator binding sites when the concentration of the activator ($[A]$) and the energy invested to break the detailed balance at the $\mathrm{AP} \rightarrow \mathrm{A}$ edge ($U_\mathrm{AP,A}$) are changed. The sign of the value on the y-axis is based on the expression shift values calculated using Eq~\ref{eq:exp_shift}. Each data point is the mean of average mutual information across 20 synthetic datasets with the corresponding $U_\mathrm{AP,A}$ and $[A]$. The numbered labels indicate datapoints for which the corresponding information footprints are shown in (C). (C) Information footprints built using the thermodynamic model and the graph-theoretic model. The second footprint is a graph-theoretic treatment of the equilibrium case.}
\label{fig15}
\end{figure}

To see how this can be done, consider a promoter with the simple activation regulatory architecture. Recall that such a promoter can be in one of four possible states: empty (E), bound by RNAP (P), bound by the activator (A), or bound by both RNAP and the activator (AP). As shown in Fig~\ref{fig15}(A), we can describe this architecture using a directed square graph with four vertices and eight edges. Each vertex corresponds to one of the four states; each edge describes the transition between two connected states and is associated with a rate constant. Having written down the architecture using a graph, the probability of each possible state can be derived using the Matrix Tree Theorem, which states that the probability of a given state at steady state is proportional to the sum of products of rate constants across all spanning trees that are rooted in that state. The expression for the probability of each state is given in~\ref{S15_Appendix}. With this, we can write down the total probability of the transcriptionally active states
\begin{align}
    p_\mathrm{active} = p_\mathrm{A} + p_\mathrm{AP},
\end{align}
where $p_\mathrm{A}$ is the probability that the activator is bound to the promoter and $p_\mathrm{AP}$ is the probability that both the activator and RNAP are bound to the promoter.

Importantly, to construct a synthetic MPRA dataset with broken detailed balance, the sequence-dependence of $p_\mathrm{active}$ needs to be preserved. Here, we make the simplifying assumption that all the on-rates are diffusion limited and therefore independent of promoter sequence. This means that the sequence-dependence of $p_\mathrm{active}$ comes from the mapping between the off rates and the sequences of the promoter variants. One way to create this mapping is to leverage the fact that the dissociation constant $K_d$ is sequence-dependent. Specifically, we have that
\begin{align}
    K_d = c_0\,e^{\beta \Delta \varepsilon},
\end{align}
where $c_0 = 1\ M$ is the reference concentration of the standard state. As we have demonstrated in Fig~\ref{fig2}(A), $\Delta \varepsilon_D$ can be calculated using energy matrices in a sequence-dependent manner, which confers sequence-specificity to $K_d$. Since $k_\mathrm{off}^\mathrm{eq} = k_\mathrm{on}^\mathrm{eq} \times K_d$, we can also calculate $k_\mathrm{off}$ in a sequence-dependent manner. To demonstrate that the graph-theoretic approach allows us to build synthetic datasets, we first use the method outlined above to calculate $k_\mathrm{on}$ and $k_\mathrm{off}$ for each promoter variant at equilibrium and used these values to predict the expression levels of the promoter variants with the simple activation regulatory architecture. As shown in Fig~\ref{fig15}(C), the information footprint built from the graph-theoretic synthetic dataset under equilibrium is comparable to the information footprint built from the synthetic dataset built from our thermodynamic models.

Finally, to build a synthetic dataset under non-equilibrium, we can estimate $k_\mathrm{off}$ by incorporating the energy $U$ invested to break detailed balance, where
\begin{align}
    k_\mathrm{off} = k_\mathrm{off}^\mathrm{eq} \times e^{\beta U}.
\end{align}
The diffusion-limited $k_\mathrm{on}$ and the sequence-dependent $k_\mathrm{off}$ for each edge can then be used to calculate $p_\mathrm{active}$ for each promoter variant and predict the expression levels of mutant promoters under non-equilibrium conditions. For example, we can break detailed balance at the edge where RNAP unbinds from the state where both the activator and RNAP are bound. That is to say, we invest an energy $U_\mathrm{AP,A}$ such that
\begin{align}
    k_\mathrm{AP,A} = k_\mathrm{AP,A}^\mathrm{eq} \times e^{\beta U_\mathrm{AP,A}}.
\end{align}
As shown in Fig~\ref{fig15}(B), when detailed balance is broken at this edge, varying $U$ and the concentration of the activator leads to interesting behaviour in the information footprints. Under some conditions, the footprints are similar to what is obtained under equilibrium conditions. However, as shown in panel 4 of Fig~\ref{fig15}(C), in the cases where a positive energy is invested to break detailed balance at the $\mathrm{AP} \rightarrow \mathrm{A}$ edge and when activator concentration is high, the binding of activator displaces the binding of RNAP. As a result, the activator effectively behaves as a repressor. This generalized approach presents opportunities to explore MPRA datasets without being constrained by equilibrium assumptions.

\bigskip

\section*{Discussion}

In this paper, we explore the landscape of sequence-energy-phenotype mapping by utilizing a computational pipeline that simulates MPRA pipelines. More generally, our computational pipeline makes it possible to use statistical mechanical models of gene expression to systematically explore the connection between mutations and level of gene expression.   Using this pipeline, we have examined the effects of perturbing various experimental and biological parameters. These perturbations occur at multiple stages of the pipeline. Some parameters pertain to the initial library design, such as library size, mutation rate, and presence of mutational bias. Other parameters are built into the model itself, such as the copy number of the promoter and the transcription factors, which are parameters that may vary biologically or be affected by the design of experimental procedures.

We have demonstrated that our computational pipeline has high flexibility and can easily be adapted to examine the effects of other perturbations not included in this paper. Furthermore, the computational nature of the pipeline allows full parameter searches to be done precisely and efficiently. For example, it would be both time-consuming and cost-prohibitive to experimentally determine the optimal library size and mutation rate as it would involve performing a large array of experimental tests. On the other hand, using our computational pipeline, we can efficiently build a series of synthetic datasets with different mutant libraries and determine the strategy for library design that is optimal for deciphering regulatory architectures.

Apart from informing the choice of experimental parameters, the pipeline also helps to anticipate challenges involved in parsing information footprints. For example, in Sec~\ref{sec:overlapping}, we predict how the signal in information footprints would be affected when there are overlapping binding sites. One potential usage of this pipeline is for building synthetic datasets that involve features that could lead to information footprints that are hard to parse. Since the synthetic datasets are built with prior knowledge of the underlying regulatory architectures, these datasets can be used to develop and improve algorithms for deciphering these architectures. This will increase confidence in the results when the same algorithms are used to analyze experimental datasets and determine the location of binding sites. Moreover, this will pave the way for automatically annotating binding sites for any given information footprint given MPRA data. To enable others who perform MPRAs in the context of transcriptional regulation to use our computational platform, we have made our code publicly available and we are developing an \href{https://www.rpgroup.caltech.edu/tregs-web/}{interactive website} where the users may generate footprints given their own parameters of interest.

One limitation of our computational MPRA pipeline is that since we rely on writing down states and weights models in order to predict the probability of transcriptionally active states, the combinatorial explosion can make it challenging for us to consider promoters that are regulated by three or more transcription factors. However, data from RegulonDB would suggest that over 80\% of the promoters in \textit{E. coli} fall under the six regulatory architectures that we discussed in Fig~\ref{fig3}~\cite{Mahdavi2023-vr}. Therefore, at least in prokaryotic genomes, we are confident that our computational pipeline can be used to simulate MPRA datasets for the vast majority of promoters.

Furthermore, our current computational pipeline only considers transcription initiation factors, while other types of transcription factors, such as elongation factors and termination factors, are also important for determining gene expression. It is challenging to include these factors into our model as it would involve additional kinetic terms that would have to be worked out, though we are extremely interested in developing these approaches as well in our future work. While this work does not directly address the challenge in understanding the role of transcription elongation factors and termination factors, we believe that by achieving a full understanding of transcription initiation factor binding through the efforts of both the computational and experimental MPRA pipelines, it will help streamline strategies needed to decipher the roles of other types of transcription factors.

In addition, our computational pipeline neglects the interaction between different genes in regulatory networks, which affects expression levels and may alter the expected signal in MPRA summary statistics such as information footprints. A future direction, therefore, involves building synthetic datasets of genetic networks. This would require an additional step where we modify the expression levels of each gene based on its dependency on other genes. This would not only improve the reliability of our prediction of expression levels, but these multi-gene synthetic datasets may also be used to test approaches for deciphering the architecture of regulatory networks.

Finally, while the vast amount of literature discussed in~\ref{S1_Appendix} gives us confidence on the validity of thermodynamic models, we acknowledge that there are many cases of transcriptional regulation in which detailed balance may be broken and thermodynamic models may no longer be appropriate. Our final results section (Sec~\ref{sec:nonequilibrium}) is a preliminary effort where we combine graph-theoretic models of transcriptional regulation and our computational pipeline to produce synthetic datasets and summary statistics without enforcing equilibrium constraints. We expect that many more interesting and informative results can come out of the angle of non-equilibrium synthetic MPRA datasets. We are excited to further pursue this direction in our future work.

In summary, we have developed a theoretical framework for a widely used category of experiments in the field of transcriptional regulation. Our simulation platform establishes a systematic way of testing how well high-throughput methods such as the MPRAs can be used to recover the ground truth of how the expression of a gene is transcriptionally regulated. This demonstrates the importance of developing theories of experiments in general, and we believe there is much untapped potential in extending similar types of theories to other areas of experimental work as well.  Finally, we anticipate that this approach will also be very useful in performing systematic studies on the relation between mutations in regulatory binding sites and the corresponding level of gene expression in a way that will shed light on both physiological and evolutionary adaptation.

\section*{Data availability}
All code used in this work and the presented figures are available open source at \href{https://github.com/RPGroup-PBoC/theoretical_regseq}{https://github.com/RPGroup-PBoC/theoretical\_regseq}.

\section*{Acknowledgments}
We would like to thank Justin Kinney, Sara Mahdavi, and Gabriel Salmon for helpful discussions and feedback on this manuscript. This work in the Rob Phillips group is supported by the NIH Maximizing Investigators' Research Award (MIRA) 1R35 GM118043. Tom Röschinger was supported by Boehringer Ingelheim Fonds.

\section*{Author contributions}

\textbf{Conceptualization}: Rosalind Wenshan Pan, Tom Röschinger, Rob Phillips

\noindent \textbf{Formal analysis}: Rosalind Wenshan Pan, Tom Röschinger

\noindent \textbf{Investigation}: Rosalind Wenshan Pan, Tom Röschinger, Kian Faizi

\noindent \textbf{Methodology}: Rosalind Wenshan Pan, Tom Röschinger, Kian Faizi

\noindent \textbf{Software}: Rosalind Wenshan Pan, Tom Röschinger, Kian Faizi

\noindent \textbf{Visualization}: Rosalind Wenshan Pan

\noindent \textbf{Writing – original draft}: Rosalind Wenshan Pan

\noindent \textbf{Writing – review and editing}: Rosalind Wenshan Pan, Tom Röschinger, Kian Faizi, Hernan Garcia, Rob Phillips

\noindent \textbf{Supervision}: Rob Phillips

\noindent \textbf{Funding acquisition}: Rob Phillips

%
%
%

\printbibliography
\end{refsection}

\include{appendix.tex}

\end{document}

%% file: appendix.tex





\renewcommand{\thesection}{S\arabic{section} Appendix}
\renewcommand{\thepage}{\arabic{page}}
\renewcommand{\thefigure}{S\arabic{figure}}
\renewcommand{\theequation}{S\arabic{equation}}
\renewcommand\refname{SI references}
\setcounter{page}{1}
\setcounter{figure}{0}
\setcounter{equation}{0}

\begin{flushleft}
{\Large
\textbf\newline{Supporting Information for Deciphering regulatory architectures from synthetic single-cell expression patterns} 
}
\newline
\\
Rosalind Wenshan Pan\textsuperscript{1*},
Tom Röschinger\textsuperscript{1},
Kian Faizi\textsuperscript{1},
Hernan Garcia\textsuperscript{2,3,4,5},
Rob Phillips\textsuperscript{1,6*}
\\
\bigskip
\textsuperscript{1}Division of Biology and Biological Engineering, California Institute of Technology, Pasadena, CA; 
\textsuperscript{2}Biophysics Graduate Group, University of California, Berkeley, CA;
\textsuperscript{3}Department of Physics, University of California, Berkeley, CA;
\textsuperscript{4}Department of Molecular and Cell Biology, University of California, Berkeley, CA;
\textsuperscript{5}Institute for Quantitative Biosciences-QB3, University of California, Berkeley, CA;
\textsuperscript{6}Department of Physics, California Institute of Technology, Pasadena, CA
\bigskip

* Correspondence: rosalind@caltech.edu and phillips@pboc.caltech.edu

\end{flushleft}


\linenumbers

\begin{refsection}

\section{Models of the probabilities of microscopic states during transcription} \label{S1_Appendix}

For the purposes of this paper, the first step in the modeling of transcription is the identification of some set of microscopic states that represent the entire set of states available to the promoter dictating our gene of interest. Already, such models represent a coarse-graining of the complex microscopic reality of DNA and its attendant proteins within a cell. For example, DNA conformation at all scales (chromatin structure, supercoiling, etc.) is not included in the most naive application of these models. This kind of coarse-graining is consistent with the depiction of regulatory architectures as a region of DNA with some collection of binding sites and their allied transcription factors such as seen in RegulonDB or EcoCyc and represented in Fig~\ref{fig3} in the paper. Though the assumption of discrete states itself deserves further scrutiny, in this paper we accept these assumptions about a set of discrete states with no further discussion. Once we accept this constellation of microscopic states, many models focus on the steady-state probabilities of those states. Of course, the full dynamical trajectories of these different microstates are of great interest and one approach to computing those dynamics is the use of coupled chemical master equations.  With the advent of experimental approaches to measuring the {\it dynamics} of transcription, the analysis of gene expression dynamics requires something more than is offered by the steady-state probabilities considered here.

In general, the steady-state probabilities of the different microstates can be written in the form
\begin{equation} \label{pRationalFunction}
    p_i(x_1,x_2, \cdots x_n) = {P_i(x_1,x_2, \cdots x_n) \over Q(x_1,x_2, \cdots x_n)},
\end{equation}
where here we examine the probability of the $i^{th}$ state as a rational function (i.e. a ratio of two polynomials)  where $x_j$ is the concentration of the $j^{th}$ transcription factor (or RNA polymerase) and with $Q(x_1,x_2, \cdots x_n)$ serving as a kind of generalized partition function gotten by summing over all states (or paths). In some instances, it is justified to specialize Eq~\ref{pRationalFunction} to the form
\begin{equation}
p_{i}(x)= {P_i(x) \over Q(x)},
\end{equation}
where for example our regulatory architecture of interest features only a single repressor or activator whose concentration is measured by $x$. If we think of the huge topic of input-output functions in biology, then $P_i(x)/Q(x)$  includes a representation of leakiness (the amount of output $p_{i}(x)$ even in the absence of input, $x=0$), dynamic range,  $EC_{50}$ (the concentration at which the output reaches half its maximum) and the sensitivity as measured by the slope of the input-output curve (usually in logarithmic variables) at the midpoint~\cite{Razo-Mejia2018-cc}.

In the present paper, we exploit two different ways of thinking about these rational functions, one of which is a subset of the other.  In particular, the most general models we consider provide the steady-state probabilities using the tools of graph theory and make no assumption of a quasi-equilibrium~\cite{Gunawardena2012-gf, Ahsendorf2014-my}.  Indeed, such models are a convenient platform for exploring the precise consequences of breaking detailed balance associated with some kinetic step connecting two different microstates~\cite{Mahdavi2023-vr}. Within the study of transcriptional regulation, the more common class of useful models is sometimes referred to as ``thermodynamic models,'' which have a deep and interesting history in the context of both test-tube biochemistry and the study of signaling, regulation and physiology within living organisms. Such models have played an important role as a conceptual framework for more than a century and a convenient and inspiring point of departure is the work of Archibald V Hill after whom the famed Hill function
\begin{equation}
    p_\mathrm{bound}(x)={\left({x \over K}\right)^n \over 1+ \left({x \over K}\right)^n }
\end{equation}
is named. In this case, $x$ is the concentration of some ligand and $K$ is an effective dissociation constant. As Hill himself tells us, this functional form was hypothesized to describe the occupancy of hemoglobin by oxygen (the example he used, though it applies and has been applied much more broadly). Already more than a century ago, Hill argued of  the function that now bears his name: ``My object was rather to see whether an equation {\it of this type} can satisfy all the observations, than to base any direct physical meaning on $n$ and $K$.''~\cite{Hill1910-hu} He goes further in his 1913 paper noting ``In point of fact $n$ does not turn out to be a whole number, but this is due simply to the fact that aggregation is not into one particular type of molecule, but rather into a whole series of different molecules: so that equation (1) is a rough mathematical expression for the sum of several similar quantities with $n$ equal to 1, 2, 3, 4 and possibly higher integers.''~\cite{Hill1913-ml}

In the subsequent decades, the equilibrium analysis of biochemical interactions became increasingly sophisticated with Pauling formulating a model that goes far beyond the Hill function by computing the average number of oxygen molecules bound to hemoglobin in the form
\begin{equation}
    \langle N_\mathrm{bound} \rangle = \frac{4x+12 x^2 j+12 x^3j^3+4x^4 j^6}{1+4x+6x^2j+4x^3j^3+x^4j^6},
\end{equation}
where we adopt the simplifying notation $x=[O_2]$~\cite{phillips2012, Pauling1935-pt}. The parameter $j$ is an interaction energy that imposes cooperativity in the sense that once one $O_2$ molecule is bound, the binding of the next one is easier. The equilibrium model of Adair went even further~\cite{Adair1925-tx}, positing that the average number of oxygen molecules bound to hemoglobin is given by
\begin{equation}
\langle N_\mathrm{bound} \rangle =\frac{4x+12 x^2 j+12 x^3j^3 k+4 x^4j^6k^4 l}{1+4x+6x^2j+4x^3j^3k+x^4j^6k^4l},
\end{equation}
where now the parameter $k$ captures 3-body interactions between $O_2$ molecules and the parameter $l$ captures 4-body interactions~\cite{phillips2012}.  For more than a century, equilibrium thinking has suffused the study of biochemical reactions, even when promoted from the sterile setting of test-tube biochemistry to the messy world of hemoglobin molecules enclosed within red blood cells that are themselves cycling rapidly through the circulatory systems of animals ranging from high-flying birds such as the bar-headed geese to elite divers such as blue whales.

It was not a huge leap to go from the idea of examples such as oxygen binding to hemoglobin (a special case of receptor-ligand binding) to the idea of DNA itself as the receptor and various proteins such as transcription factors and RNA polymerase as the ``ligand.''  Classic work from Ackers and Shea~\cite{Ackers1982-ax, Shea1985-js} formalized the kind of ``regulated recruitment'' thinking that had been diligently pursued by Ptashne and coworkers~\cite{ptashne2004}, turning it into a formal mathematical structure for evaluating the state probabilities for various occupancies of a promoter of interest.  More recently, these ideas have been developed deeply by a battery of researchers several examples of which are given here~\cite{Vilar2003-rj, Buchler2003-yt, Sherman2012-et}. Similarly, right from the get-go, the study of gene regulation made it clear that the phenomenon of induction (i.e. the use of effector molecules to tune the state of expression) would require a quantitative description,  and beautiful work in the 1960s articulated a wide range of different equilibrium allosteric models such as the MWC model~\cite{Monod1965-ij}, the KNF model~\cite{Koshland1966-zf} and the generalization of these ideas in the model of Manfred Eigen~\cite{Eigen1968-pz}.  All of these models, even if no one explicitly says so, are ``thermodynamic models'' in that they provide a systematic protocol for using statistical mechanics to find the probabilities of {\it all} of the allowed states.  Note that these different models differ not in whether they are quasi-equilibrium or not, but rather in which states they permit.

The vast majority of approaches to generating a specific functional form for state probabilities in transcription like those given in Eq~\ref{pRationalFunction} are either: (a) some version of a thermodynamic model which appeals in one way or another to states and Boltzmann weights, (b) phenomenological guesses in which some convenient functional form is adopted (usually a Hill function) and more rarely, (c) using the tools of non-equilibrium physics and graph theory, a version of $p_{i}$ is adopted that reflects broken detailed balance.  Approaches (a) and (c) both require some sort of mechanistic commitment about the classes of states that the system can adopt, and we note that (a) is a special case of (c).  In the case of thermodynamic models, these polynomials have a very special form dictated by the Boltzmann weights of the different states of binding between transcription factors and their target DNA.

Part of the reason for the importance of this appendix is because there are so many different opinions in play on the subject of thermodynamic models in the biological setting in the literature.  Part of the passion associated with the subject is that some authors are explicit in naming their work as thermodynamic or equilibrium models and others are not.  Some fragment of the scientific population has an intrinsic belief that because ``living organisms are out of equilibrium,'' thermodynamic or equilibrium models have no place.  We think this extreme view can be replaced by a more nuanced perspective given that even within the fiery interior of a star, equilibrium ideas are used routinely and successfully (see the Saha equation).  Rather, we are going to highlight several alternative and more nuanced pictures including: (i)  justification based upon separation of time scales, (ii) a null model which serves as a first and simplest regulatory hypothesis and (iii) phenomenology.

First, we consider the status of thermodynamic models as a null model for the steady-state probabilities of biochemical phenomena with special emphasis on the application to transcription.  Then, we subject such models to the harshest scrutiny: what is their track record as a conceptual framework for thinking about experimental data, and what is the nature of their shortcomings? Here we use the title ``thermodynamic models'' as a shorthand  to refer to {\it all} models of occupancy of transcription factors, nucleosomes and polymerases in which the state probabilities are obtained from the Boltzmann distribution, or some phenomenological approximation (i.e. Hill function) to the equilibrium state probabilities.  These models are ubiquitous not only in the theory literature of transcription, but also as an interpretive null model for huge classes of experimental data.  To give a flavor for the use of these models, we provide several key case studies followed by a smorgasbord of citations which the reader is urged to consult. Aside from providing references, we decided to forego our own extensive efforts at using and scrutinizing thermodynamic models because we wanted to highlight the ubiquitous nature of such thinking beyond our own work.  The original theory work of Ackers and Shea~\cite{Ackers1982-ax, Shea1985-js} focused primarily on the example of phage lambda, itself already introduced non-mathematically in thermodynamic model format by Ptashne and collaborators~\cite{ptashne2004}. Perhaps no example is more famous than the bacterial example of the {\it lac} operon and its synthetic variants.  M\"{u}ller-Hill and Oehler and coworkers made an impressive series of quantitative and rigorous studies of synthetic variants of the {\it lac} operon which in modern parlance we would view as having ``tuned the knobs'' of transcription such as the strength of DNA binding sites for Lac repressor, the copy number of the Lac repressor and even the length of the DNA loop formed by binding two sites simultaneously, done with exquisite single-base pair precision~\cite{Oehler1994-gr, Oehler2006-cz}. In a large number of papers, Vilar and Leibler~\cite{Vilar2003-rj} and subsequently, Saiz and Vilar~\cite{Vilar2005-hq, Vilar2013-gf, Vilar2023-ux} have provided a corresponding theoretical study of this data (and much more).  Kuhlman et al. used the tools of molecular biology to construct strains of {\it E. coli} such that they could explicitly test thermodynamic models~\cite{Kuhlman2007-qj}(which they expertly modeled using thermodynamic models following their own earlier theory work showing how thermodynamic models could be used to dissect logic gates~\cite{Buchler2003-yt}). Similar studies in the context of the {\it ara} operon by Schleif and co-workers provided a picture of how DNA looping can be treated quantitatively within the confines of thermodynamic models~\cite{Dunn1984-tw, Ogden1980-tb, Schleif1975-xb}. We are strong advocates for those cases in which ultimately models of transcription are confronted with well-designed experiments that tune the same knobs that were controlled in the theoretical models.  Another example of this kind of regulatory dissection for bacterial promoters is offered by work on MarA which activates transcription~\cite{Martin2008-tv, Wall2009-ny}. One reason for skepticism concerning these apparent successes is the possibility that in some cases non-equilibrium and equilibrium models will ``agree'' on some particular set of data.  To distinguish them may involve tuning some knob that has not yet been tuned.  This paragraph only scratches the surface of the vast array of work based upon these models.  A more detailed sense of the reach of this work can be gleaned by looking at the hundreds of citations of papers such as those of Buchler, Gerland and Hwa~\cite{Buchler2003-yt}, Bintu et al.~\cite{Bintu2005-ib, Bintu2005-oz}, and Sherman and Cohen~\cite{Sherman2012-et}.

One of the most important measures of the ``success'' of thermodynamics is in their power to unify apparently quite distinct data in the form of data collapse.  Two extremely impressive examples of such data collapse were explored in the context of chemotaxis (see Fig 5 of Keymer et al.~\cite{Keymer2006-ck}) and quorum sensing (see Fig 6 of Swem et al.~\cite{Swem2008-rv}), where in both cases the activity of a signaling pathway was modeled using the equilibrium MWC model of allosteric receptors.  The key finding is that the receptor activity for an entire suite of mutants could be collapsed onto one single master curve in the same way that for simple ligand-receptor binding (Hill function with Hill coefficient $n=1$), if we plot $p_\mathrm{bound}$ vs $c/K_d$ rather than $c$, we find that all ligand-receptor curves fall on one universal curve.  For the chemotaxis and quorum sensing examples, the data collapse is much more subtle.  Though we cannot consider this a bulletproof demonstration that the thermodynamic models are ``right,'' they certainly provide a powerful, unifying and parameter-free predictive framework for thinking about experiments.  In the context of transcription, similar data collapse was achieved featuring a very demanding parameter-free collapse of a broad array of experimental data in which binding site strength, transcription factor copy number and gene copy number were systematically varied (see Fig 4 of Weinert et al.~\cite{Weinert2014-zw}) and for these same constructs as a function of inducer concentration (see Fig 7(B) of Razo et al.~\cite{Razo-Mejia2018-cc}).

Before briefly turning to a survey of some of the shortcomings of the thermodynamic models, we present examples in which differential equations are used to model the dynamics of either mRNA or protein production and that implicitly feature thermodynamic models to describe gene regulation.  Our key point here is to note that often these equations take the form (for example, Eq 5 in Cherry and Adler~\cite{Cherry2000-nv})
\begin{equation}
{{\rm d}A \over {\rm d}t}= -\gamma A + f_\mathrm{production}(A),
\end{equation}
where almost always the production term can be written in the form
\begin{equation}
f_\mathrm{production}(A) = {P(A) \over Q(A)},
\end{equation}
where $P(A)$ and $Q(A)$ are polynomials. Further, in most instances, these rational functions are either of the phenomenological (or quasi-equilibrium) Hill form or appeal directly to Boltzmann states and weights.  To be concrete, in two of the classic examples of synthetic biology, the genetic switch and the repressilator, the dynamical models were of the form described above~\cite{Gardner2000-yq, Elowitz2000-fv}. For example, for the
genetic switch the production of the two species of repressor are written in
dimensionless form (see Eq 1a and Eq 1b of Gardner et al.~\cite{Gardner2000-yq})  as
\begin{equation}\label{eq:genswitchratedim}
    \begin{split}
    {{\rm d}r_1\over {\rm d}\tau} & = - r_1 + {\alpha\over 1 + r_2^n}, \\
    {{\rm d}r_2\over {\rm d}\tau} & = - r_2 + {\alpha\over 1 + r_1^n}.
    \end{split}
\end{equation}
Here $r_1$ and $r_2$ are the dimensionless concentrations of the two mutually repressing repressors, time is measured in units of $\tau = \gamma t$ where $\gamma$ is the degradation rate and $\alpha$ is a dimensionless protein production rate.  Our main point here is to note that although the words ``thermodynamic model'' or ``equilibrium'' are never used, the right-hand side of these equations is explicitly computing the probability of binding site occupancy by repressors.

We are hopeful that the subject of transcription will be held to the highest quantitative standards.  We are excited by the many examples highlighted here (which only scratches the surface) and hope for a time when we have a deep and predictive understanding of all the genes in key model organisms and that these insights will serve as the basis for the study of non-model organisms such as redwood trees and blue whales. One route to such predictive understanding is the critical scrutiny offered by a dialogue between theory and experiment. In this paragraph, we note in passing a number of examples where it appears that the thermodynamic null models do not pass muster.   One subject of intense effort over the last few decades is the study of DNA packing in eukaryotes and its implications for gene expression.  As usual, the literature of this topic is immense.  Intense debates have unfolded on the position of nucleosomes on genomic DNA with the conclusion likely that equilibrium models by themselves will {\it not} explain all the extant data~\cite{Segal2006-yg, Struhl2013-bu, Kaplan2009-bd}. The connection between expression and nucleosomal occupancy has been taken farther recently using single-cell methods  with the result that the thermodynamic null model must be superseded by a more detailed model~\cite{Doughty2024-gy}. Similarly, in the context of the {\it Pho5} promoter in yeast, a quite amazing set of experiments was done to measure the occupancies of nucleosomes on this promoter.  This data was analyzed using tens of thousands of models and the only models consistent with all of the data appear to require broken detailed balance~\cite{Brown2013-wx, Wolff2021-pk}. In a beautiful use of the MWC model highlighted above, Mirny worked out the probability that nucleosomes will be present essentially regulating some gene of interest and even went so far as to make an analogy with the Bohr effect in hemoglobin in which the post-translational modification of nucleosomes could be thought of as a kind of Bohr effect~\cite{Mirny2010-na}. Though it took nearly a decade, it appears that this model is not sufficient to explain chromatin accessibility and gene expression~\cite{Eck2020-ma}.  We note that this example is common: to really make the comparison between theory and experiment often means that the data that exists is just not quite right to make the acid test (see the example of hemoglobin where, in our view, misuses of the MWC model led to the conclusion that the ``model doesn't work'' whereas in reality, it was rather that the set of states needed to be expanded to include other effectors~\cite{Yonetani2002-ck, Rapp2017-ot, Eck2020-ma}). This is also carefully explained in Chapter 7 of Ref.~\cite{phillips2020}. However, failure of the thermodynamic framework reaches well beyond the example of chromatin where it is clear that energy-consuming processes such as nucleosome remodeling can break detailed balance.  Beautiful single-molecule experiments in {\it E. coli} revealed that even in the process of transcription factor binding to its DNA target, the results were not consistent with the thermodynamic framework~\cite{Hammar2014-fp}.  Similarly, application of the thermodynamic modeling framework in the context of the {\it Pseudomonas aeruginosa} genes associated with the transcription factor BqsR have thus far been unsuccessful~\cite{Kreamer2015-gy}.  One particularly interesting example of the shortcomings of the thermodynamic approach was in the context of the glucocorticoid receptor where it was found that the rank ordering of gene expression did not scale with the rank ordering of binding site strength~\cite{Meijsing2009-mc}.  As with our list of ``successes'' this list of failures of the thermodynamic framework is superficial and incomplete. Further, we argue that to actually make claims of ``right'' or ``wrong'' requires diligent and often frustrating dialogue between theory and experiment where dedicated efforts need to be made to make sure that the same knobs are being tuned in both the theory and the corresponding experiments.

In summary, we would put the place of thermodynamic models in thinking about in vivo biochemistry (including the case of transcription) on par with other seemingly naive but incredibly productive null models such as the apparently crazy idea of a noninteracting electron gas to describe metals or the Ising model as a way to describe magnetic phenomena. In the context of transcription, though it is  routine to critique thermodynamic models, we would argue that in fact, the thermodynamic models are just as plausible as the equally popular ``two-state promoter'' used so often to explain noise in transcription.  Several beautiful examples of the use of the two-state framework are given in So at al. and Zenklusen et al.~\cite{So2011-cd, Zenklusen2008-ue}. For the purposes of the present paper, the thermodynamic model framework is a convenient null model which allows us to self-consistently generate synthetic datasets that can then be analyzed by the tools used to study real data with the added benefit that we ``know the answer'' from the outset. This appendix is a very superficial rendering of a vast subject. We hope at the least that it provides an entry into the literature which has used the so-called thermodynamic models and that attempts a balanced view of their successes and shortcomings.  In our view, ultimately, the status of all models of transcription will only be really clarified by a painstaking dialogue between theory and experiment.

\section{Predicting the probability of RNA polymerase being bound using thermodynamic models} \label{S2_Appendix}

To estimate the expression levels of a gene with a given promoter, one key step is to calculate the probability of RNA polymerase (RNAP) being bound, which is needed in computing the rate of mRNA production using Eq~\ref{eq:dmdt}. This can be done by building the so-called thermodynamic models of transcriptional regulation. In this appendix, we outline the protocol for building thermodynamic models, using a promoter that is constitutively expressed and a promoter with the simple repression architecture as examples.

\begin{figure}[!h]
\centering
\includegraphics[width=1\textwidth]{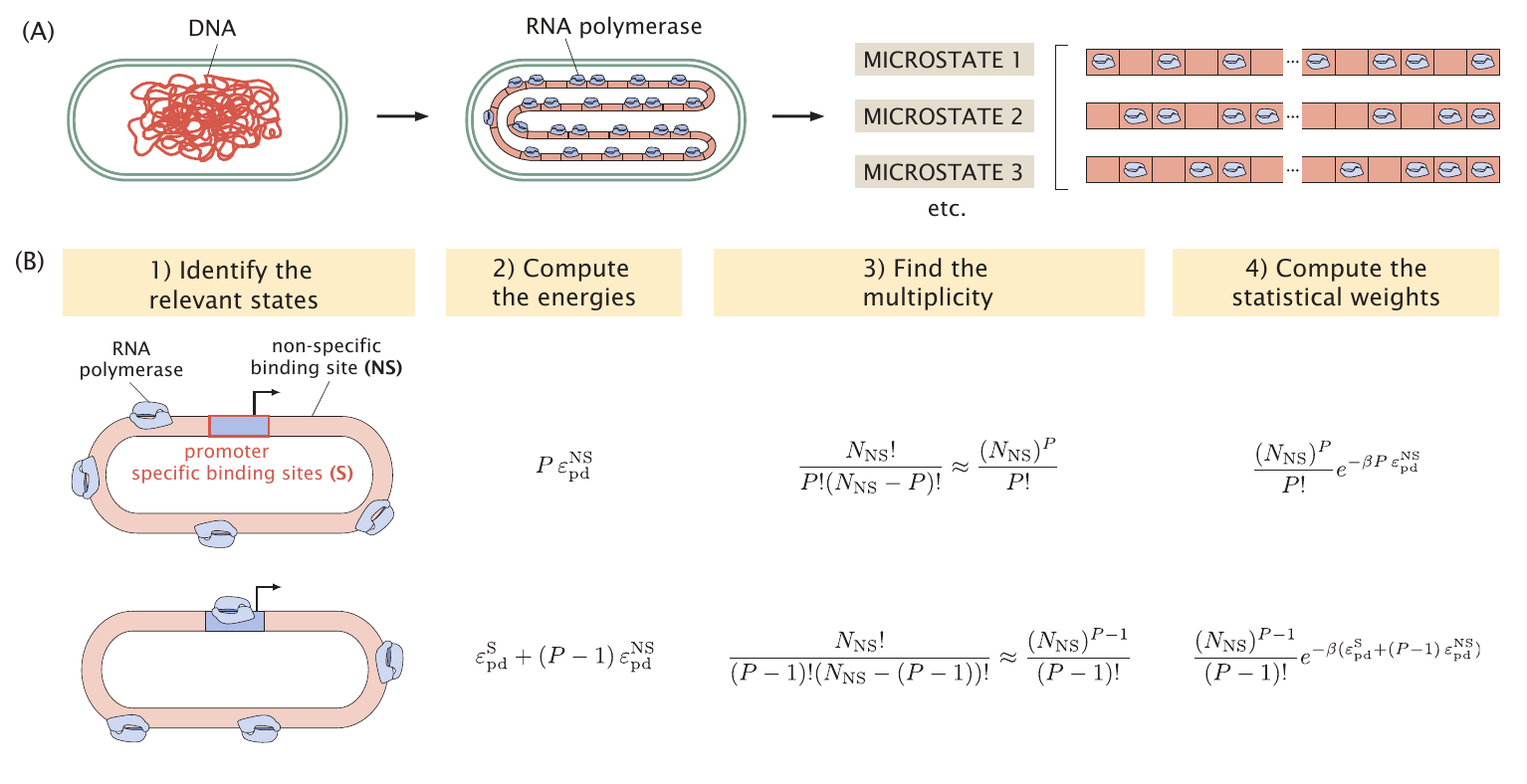}
\caption{{\bf Writing down thermodynamic models for transcriptional regulation.}\\
(A) Conceptualizing binding events along the genome as microstates. Each of the red boxes is a site along the genome to which RNAP and transcription factors can bind. Each of the possible configurations of binding is considered a microstate. (B) Protocol for writing down a thermodynamic model. The protocol involves four important steps. Firstly, all the relevant promoter states need to be identified. Secondly, the energies of each state is computed. Next, we compute the multiplicity, which tells us the number of possible configurations associated with each state. Finally, statistical weights are written down based on the energies and the multiplicity terms. For the energy terms $\varepsilon$, the superscript $\mathrm{NS}$ refers to non-specific binding, the superscript $\mathrm{S}$ refers to specific binding, the subscript $\mathrm{pd}$ refers to RNAP (p) binding to DNA (d).}
\label{figS2}
\end{figure}

On a broader level, the first step to build a thermodynamic model involves abstracting the genome into discrete microstates. As shown in Fig~\ref{figS2}(A), our approach entails disregarding the three-dimensional topology of the genome and conceptualizing it as a linear sequence of discrete binding sites. RNAPs and transcription factors can bind to these sites in a number of configurations, each of which can be defined as a microstate. With this way of defining the microstates, we can then write down a general protocol for building thermodynamic models, which is shown in Fig~\ref{figS2}(B). First, we need to identify all the relative promoter states. Second, we compute the energies for each of the states. Next, we need to calculate the multiplicity for each of the states. This is needed because each of the states encompass a range of possible configurations, and we need to count the total number of configurations associated with each state. Finally, we can compute the statistical weight of each state, which is the product of multiplicity and the Boltzmann weight calculated using the Boltzmann law of statistical mechanics. The statistical weight is in the form of
\begin{align}
    \omega_i = W\, e^{- \beta \varepsilon_i},
\end{align}
where $W$ is the multiplicity, $\beta = 1 / k_BT$ with $k_B$ representing the Boltzmann constant and $T$ representing temperature, and $\varepsilon_i$ is the energy of the $i$-th state. The probability of any given state is then calculated by dividing $\omega_i$ by the partition function, which is the sum of the weights of all states.

To demonstrate how this protocol is used, let us consider a constitutive promoter, which is not regulated by any transcription factor. As shown in Fig~\ref{figS2}(B), there are two possible states of binding: the state where one RNAP is bound to the promoter and the state where the promoter is empty. In the latter state, all RNAPs are bound to a non-specific binding site along the rest of the genome. Let us suppose that the specific binding energy at the promoter is $\varepsilon_\mathrm{pd}^\mathrm{S}$ and the non-specific binding energy along the rest of the genome is $\varepsilon_\mathrm{pd}^\mathrm{NS}$. If there are $P$ molecules of RNAPs in the system, then the energy of the empty promoter is $P \varepsilon_\mathrm{pd}^\mathrm{NS}$ and the energy when one RNAP is bound to the promoter and the remaining $(P-1)$ RNAPs are bound non-specifically is $\varepsilon_\mathrm{pd}^\mathrm{S} + (P - 1) \varepsilon_\mathrm{pd}^\mathrm{NS}$.

Following the protocol, we next find the multiplicity of each promoter state. Let $N_\mathrm{NS}$ be the number of non-specific binding sites, then the multiplicity of the empty promoter state is given by
\begin{align}
    W_\mathrm{NS}(P, N_\mathrm{NS}) = \frac{N_\mathrm{NS}!}{P! (N_\mathrm{NS} - P)!} \approx \frac{(N_\mathrm{NS})^P}{P!}.
\end{align}
The approximation in the second step holds true because $N_\mathrm{NS}$ is typically taken to be the length of the genome, which is on the order of $10^6$ for \textit{E. coli}. Therefore, we have that $N_\mathrm{NS} \gg P$ and $\frac{N_\mathrm{NS}!}{(N_\mathrm{NS} - P)!} \approx (N_\mathrm{NS})^P$. We can use the same procedure to count the number of configurations associated with the state where RNAP is bound to the promoter. Since one RNAP molecule is bound to the promoter, there remain $(P-1)$ RNAP molecules that can bind to the non-specific binding sites in the rest of the genome. Therefore, the multiplicity of the RNAP-bound state is given by
\begin{align}
    W_\mathrm{S}(P-1, N_\mathrm{NS}) = \frac{N_\mathrm{NS}!}{(P-1)! (N_\mathrm{NS} - (P-1))!} \approx \frac{(N_\mathrm{NS})^{P-1}}{(P-1)!}.
\end{align}

Having written down the energies and multiplicity terms of the two promoter states, we can compute the statistical weights of the states, which are given in the fourth column on Fig~\ref{figS2}(B). Now we're ready to write down the probability of RNAP being bound for a constitutive promoter
\begin{align}
    p_{\mathrm{bound}} = \frac{\frac{(N_\mathrm{NS})^{P-1}}{(P-1)!} e^{-\beta \varepsilon_\mathrm{pd}^\mathrm{S}} e^{-\beta (P-1) \varepsilon_\mathrm{pd}^\mathrm{NS}}}{\frac{(N_\mathrm{NS})^P}{P!} e ^{-\beta P \varepsilon_\mathrm{pd}^\mathrm{NS}} + \frac{(N_\mathrm{NS})^{P-1}}{(P-1)!}  e^{-\beta \varepsilon_\mathrm{pd}^\mathrm{S}} e^{-\beta (P-1) \varepsilon_\mathrm{pd}^\mathrm{NS}}},
\end{align}
where $\varepsilon_\mathrm{pd}^\mathrm{S}$ is the binding energy of RNAP at the promoter and $\varepsilon_\mathrm{pd}^\mathrm{NS}$ is the binding energy of RNAP at the non-specific binding sites. In particular, we assume that the binding energy is same across all non-specific binding sites. To simplify the expression, we can multiply all the terms in the numerator and the denominator by $\frac{P!}{(N_\mathrm{NS})^P} e^{\beta P \varepsilon_\mathrm{pd}^\mathrm{NS}}$. This gives us
\begin{align}
    p_{\mathrm{bound}} = \frac{\frac{P}{N_{\mathrm{NS}}}{e^{- \beta \Delta \varepsilon_{\mathrm{pd}}}}}{1 + \frac{P}{N_{\mathrm{NS}}} e^{- \beta \Delta \varepsilon_{\mathrm{pd}}}},
\end{align}
where $\Delta \varepsilon_{\mathrm{pd}} = \varepsilon_\mathrm{pd}^\mathrm{S} - \varepsilon_\mathrm{pd}^\mathrm{NS}$ is the binding energy of the RNAP at the promoter relative to the binding energy at the non-specific binding sites.

\begin{figure}[!h]
\centering
\includegraphics[width=0.95\textwidth]{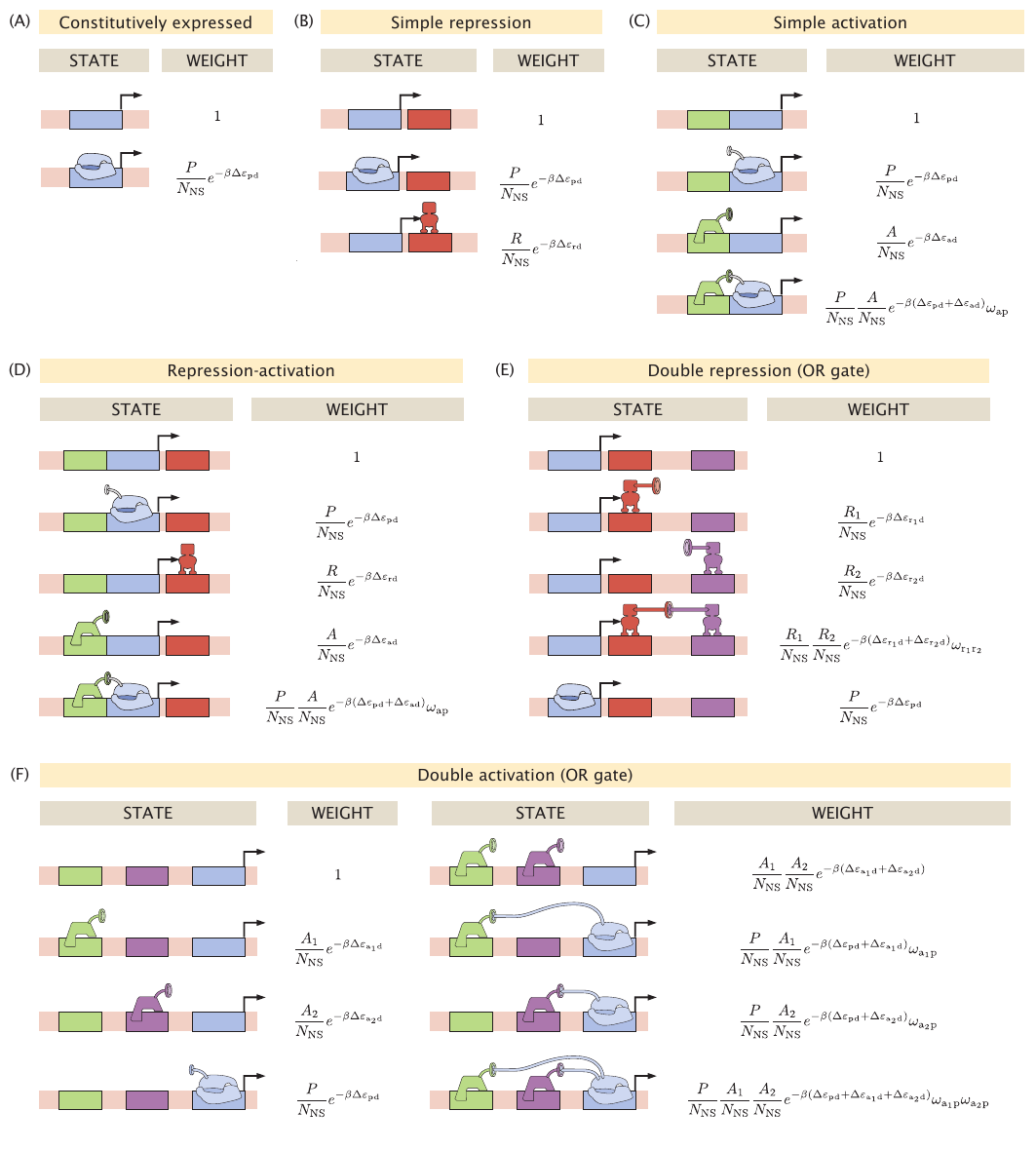}
\caption{{\bf States-and-weights models for common regulatory architectures.}
In all the diagrams, $P$ represents the number of RNAP; $R$ represents the number of repressors; $A$ represents the number of activators; $N_{\mathrm{NS}}$ represents the number of non-specific binding sites; $\Delta \varepsilon_{\mathrm{pd}}$ represents the binding energy of the RNAP; $\Delta \varepsilon_{\mathrm{rd}}$ represents the binding energy of the repressor; $\Delta \varepsilon_{\mathrm{ad}}$ represents the binding energy of the activator; $\omega_{\mathrm{ab}} = e^{-\beta \varepsilon_\mathrm{int}}$ represents the interaction energy between a and b. (A) States-and-weights model for a promoter that is constitutively expressed. (B) States-and-weights model for a promoter under the simple repression regulatory architecture. (C) States-and-weights model for a promoter under the simple activation regulatory architecture. (D) States-and-weights model for a promoter under the repression-activation regulatory architecture. (E) States-and-weights model for a promoter under the double repression regulatory architecture with OR logic. (F) States-and-weights model for a promoter under the double activation regulatory architecture with OR logic.}
\label{figS3}
\end{figure}

This protocol can be easily extended to cases where the promoter is regulated by transcription factors. The simplest architecture that involves a transcription factor is where the promoter is regulated by a single repressor. As shown in Fig~\ref{figS3}(B), for a promoter with the simple repression regulatory architecture, there are three possible state of binding: the state with an empty promoter, the state where RNAP is bound to the promoter, and the state where the repressor is bound to the promoter. Following the protocol above, we can write down the following statistical weights for each of the three states
\begin{align}
    Z_\mathrm{empty\, promoter} &= \frac{N_\mathrm{NS}!}{P!R!(N_\mathrm{NS} - P - R)!} e^{-\beta P \varepsilon_\mathrm{pd}^\mathrm{NS}} e^{-\beta R \varepsilon_\mathrm{rd}^\mathrm{NS}}\\
    &\approx \frac{(N_\mathrm{NS})^P}{P!} \frac{(N_\mathrm{NS})^R}{R!} e^{-\beta P \varepsilon_\mathrm{pd}^\mathrm{NS}} e^{-\beta R \varepsilon_\mathrm{rd}^\mathrm{NS}} \\
    Z_\mathrm{RNAP\, on\, promoter} &= \frac{N_\mathrm{NS}!}{(P-1)!R!(N_\mathrm{NS} - (P-1) - R)!} e^{-\beta (P-1) \varepsilon_\mathrm{pd}^\mathrm{NS}} e^{-\beta R \varepsilon_\mathrm{rd}^\mathrm{NS}} e^{-\beta \varepsilon_\mathrm{pd}^\mathrm{S}} \\
    &\approx \frac{(N_\mathrm{NS})^{P-1}}{(P-1)!} \frac{(N_\mathrm{NS})^R}{R!} e^{-\beta (P-1) \varepsilon_\mathrm{pd}^\mathrm{NS}} e^{-\beta R \varepsilon_\mathrm{rd}^\mathrm{NS}} e^{-\beta \varepsilon_\mathrm{pd}^\mathrm{S}} \\
    Z_\mathrm{repressor\, on\, promoter} &= \frac{N_\mathrm{NS}!}{P!(R-1)!(N_\mathrm{NS} - P - (R-1))!} e^{-\beta P \varepsilon_\mathrm{pd}^\mathrm{NS}} e^{-\beta (R-1) \varepsilon_\mathrm{rd}^\mathrm{NS}} e^{-\beta \varepsilon_\mathrm{rd}^\mathrm{S}} \\
    &\approx \frac{(N_\mathrm{NS})^{P}}{P!} \frac{(N_\mathrm{NS})^{R-1}}{(R-1)!} e^{-\beta P \varepsilon_\mathrm{pd}^\mathrm{NS}} e^{-\beta (R-1) \varepsilon_\mathrm{rd}^\mathrm{NS}} e^{-\beta \varepsilon_\mathrm{rd}^\mathrm{S}}
\end{align}
where $N_{\mathrm{NS}}$ is the number of non-specific binding sites; $P$ is the number of RNAP; $R$ is the number of repressors; $\Delta \varepsilon_{\mathrm{pd}}$ is the binding energy of the RNAP; $\varepsilon_{\mathrm{pd}}^\mathrm{S}$ and $\varepsilon_{\mathrm{pd}}^\mathrm{NS}$ are the specific binding energy of the RNAP at the promoter and the binding energy of the RNAP at the non-specific binding site; $\varepsilon_{\mathrm{rd}}^\mathrm{S}$ and $\varepsilon_{\mathrm{rd}}^\mathrm{NS}$ are the specific binding energy of the repressor at the promoter and the binding energy of the repressor at the non-specific binding site. This allows us to write down the probability of RNAP binding as
\begin{align}
    p_\mathrm{bound} = \frac{Z_\mathrm{RNAP\, on\, promoter}}{Z_\mathrm{empty\, promoter} + Z_\mathrm{RNAP\, on\, promoter} + Z_\mathrm{repressor\, on\, promoter}}.
\end{align}
Again, we simplify the expression by multiplying both the numerator and the denominator by $\frac{P!}{(N_\mathrm{NS})^P} \frac{R!}{(N_\mathrm{NS})^R} e^{\beta \varepsilon_\mathrm{pd}^\mathrm{NS}} e^{\beta \varepsilon_\mathrm{rd}^\mathrm{NS}}$. This gives us the following expression for the probability of RNAP being bound
\begin{align}
    p_{\mathrm{bound}} = \frac{\frac{P}{N_{\mathrm{NS}}}{e^{- \beta \Delta \varepsilon_{\mathrm{pd}}}}}{1 + \frac{P}{N_{\mathrm{NS}}} e^{- \beta \Delta \varepsilon_{\mathrm{pd}}} + \frac{R}{N_{\mathrm{NS}}} e^{- \beta \Delta \varepsilon_{\mathrm{rd}}}},
\end{align}
where $\Delta \varepsilon_{\mathrm{pd}} = \varepsilon_\mathrm{pd}^\mathrm{S} - \varepsilon_\mathrm{pd}^\mathrm{NS}$ and $\Delta \varepsilon_{\mathrm{rd}} = \varepsilon_\mathrm{rd}^\mathrm{S} - \varepsilon_\mathrm{rd}^\mathrm{NS}$ are the binding energies of the RNAP and the repressors at the promoter relative to their non-specific binding energies. Here, the weak promoter approximation is often made, which states that the RNAP binding state has a much lower Boltzmann weight compared to the repressor binding site. Therefore, the expression can often be simplified to
\begin{align}
    p_{\mathrm{bound}} = \frac{\frac{P}{N_{\mathrm{NS}}}{e^{- \beta \Delta \varepsilon_{\mathrm{pd}}}}}{1 + \frac{R}{N_{\mathrm{NS}}} e^{- \beta \Delta \varepsilon_{\mathrm{rd}}}}.
\end{align}

It is important to note that when the expressions for $p_\mathrm{bound}$ are used to predict the expression levels of promoter variants in an MPRA library, the energy terms $\varepsilon_i$ are calculated for each promoter variant by mapping binding site sequences to energy matrices. The procedure for calculating the total energies is explained in Sec~\ref{sec:pipeline} and illustrated in Fig~\ref{fig2}(A).

\section{States-and-weights models for common regulatory architectures} \label{S3_Appendix}

There are six common regulatory architectures for promoters in \textit{E. coli}. In \ref{S2_Appendix}, we have written down $p_{\mathrm{bound}}$, the probability that the RNAP is bound to the promoter, for a constitutively expressed promoter and a promoter with the simple repression regulatory architecture. Based on the states-and-weights diagrams shown in Fig~\ref{figS3} and using the same protocol introduced in \ref{S2_Appendix}, we can write down $p_{\mathrm{bound}}$ the remaining four common regulatory architectures~\cite{Bintu2005-ib}.

For a promoter with the simple activation regulatory architecture, the states-and-weights diagram is shown in Fig~\ref{figS3}(C), and the probability of RNAP being bound is given by
\begin{align}
    p_{\mathrm{bound}} = \frac{\frac{P}{N_{\mathrm{NS}}}{e^{- \beta \Delta \varepsilon_{\mathrm{pd}}}} + \frac{P}{N_{\mathrm{NS}}}\frac{A}{N_{\mathrm{NS}}}{e^{- \beta (\Delta \varepsilon_{\mathrm{pd}} + \Delta \varepsilon_{\mathrm{ad}})} \omega_{\mathrm{ap}}}}{1 + \frac{P}{N_{\mathrm{NS}}} e^{- \beta \Delta \varepsilon_{\mathrm{pd}}} + \frac{A}{N_{\mathrm{NS}}} e^{- \beta \Delta \varepsilon_{\mathrm{rd}}} + \frac{P}{N_{\mathrm{NS}}}\frac{A}{N_{\mathrm{NS}}}{e^{- \beta (\Delta \varepsilon_{\mathrm{pd}} + \Delta \varepsilon_{\mathrm{ad}})} \omega_{\mathrm{ap}}}},
\end{align}
where $N_{\mathrm{NS}}$ is the number of non-specific binding sites; $P$ is the number of RNAP; $A$ is the number of activators; $\Delta \varepsilon_{\mathrm{pd}}$ is the binding energy of the RNAP; $\Delta \varepsilon_{\mathrm{ad}}$ is the binding energy of the activator; $\omega_{\mathrm{a_1a_2}}$ is the interaction energy between the activator and the RNAP.

For a promoter with the repression-activation regulatory architecture, the states-and-weights diagram is shown in Fig~\ref{figS3}(D), and the probability of RNAP being bound is given by
\begin{align}
    p_{\mathrm{bound}} = \frac{\frac{P}{N_{\mathrm{NS}}}{e^{- \beta \Delta \varepsilon_{\mathrm{pd}}}} + \frac{P}{N_{\mathrm{NS}}}\frac{A}{N_{\mathrm{NS}}}{e^{- \beta (\Delta \varepsilon_{\mathrm{pd}} + \Delta \varepsilon_{\mathrm{ad}})} \omega_{\mathrm{ap}}}}{1 + \frac{P}{N_{\mathrm{NS}}} e^{- \beta \Delta \varepsilon_{\mathrm{pd}}} + \frac{R}{N_{\mathrm{NS}}}{e^{- \beta \Delta \varepsilon_{\mathrm{rd}}}} + \frac{A}{N_{\mathrm{NS}}} e^{- \beta \Delta \varepsilon_{\mathrm{ad}}} + \frac{P}{N_{\mathrm{NS}}}\frac{A}{N_{\mathrm{NS}}}{e^{- \beta (\Delta \varepsilon_{\mathrm{pd}} + \Delta \varepsilon_{\mathrm{ad}})} \omega_{\mathrm{ap}}}},
\end{align}
where $N_{\mathrm{NS}}$ is the number of non-specific binding sites; $P$ is the number of RNAP; $R$ is the number of repressors; $A$ is the number of activators; $\Delta \varepsilon_{\mathrm{pd}}$ is the binding energy of the RNAP; $\Delta \varepsilon_{\mathrm{rd}}$ is the binding energy of the repressor; $\Delta \varepsilon_{\mathrm{ad}}$ is the binding energy of the activator; $\omega_{\mathrm{a_1a_2}}$ is the interaction energy between the activator and the RNAP.

Let $r_1 = \frac{R_1}{N_{\mathrm{NS}}} e^{-\beta \Delta \varepsilon_{r_1d}}$, $r_2 = \frac{R_2}{N_{\mathrm{NS}}} e^{-\beta \Delta \varepsilon_{r_2d}}$, and $p = \frac{P}{N_{\mathrm{NS}}} e^{-\beta \Delta \varepsilon_{pd}}$. Then, for a promoter with the double repression regulatory architecture under OR logic, the states-and-weights diagram is shown in Fig~\ref{figS3}(E), and the probability of RNAP being bound is given by
\begin{align}
    p_{\mathrm{bound}} = \frac{p}{1 + r_1 + r_2 + r_1r_2 \omega_{\mathrm{r_1r_2}} + p},
\end{align}
where $N_{\mathrm{NS}}$ is the number of non-specific binding sites; $P$ is the number of RNAP; $R_1$ is the number of the first repressor; $R_2$ is the number of the second repressor; $\Delta \varepsilon_{\mathrm{pd}}$ is the binding energy of the RNAP; $\Delta \varepsilon_{\mathrm{r_1d}}$ is the binding energy of the first repressor; $\Delta \varepsilon_{\mathrm{r_2d}}$ is the binding energy of the second repressor; $\omega_{\mathrm{r_1r_2}}$ is the interaction energy between the two repressors. The states-and-weights diagram of the AND-logic double repression regulatory architecture is shown in Fig~\ref{fig7}(A). In this case, the probability of RNAP being bound is given by
\begin{align}
    p_{\mathrm{bound}} = \frac{p + r_1p + r_2p}{1 + r_1 + r_2 + r_1r_2 \omega_{\mathrm{r_1r_2}} + p + r_1p + r_2p },
\end{align}

Let $a_1 = \frac{A_1}{N_{\mathrm{NS}}} e^{-\beta \Delta \varepsilon_{a_1d}}$, $a_2 = \frac{A_2}{N_{\mathrm{NS}}} e^{-\beta \Delta \varepsilon_{a_2d}}$, and $p = \frac{P}{N_{\mathrm{NS}}} e^{-\beta \Delta \varepsilon_{pd}}$. Then, for a promoter with the double repression regulatory architecture under OR logic, the states-and-weights diagram is shown in Fig~\ref{figS3}(F), and the probability of RNAP being bound is given by
\begin{align}
    p_{\mathrm{bound}} = \frac{p + a_1p\omega_{a_1p} + a_2 p \omega_{a_2p} + a_1a_2p \omega_{a_1p} \omega_{a_2p}}{1 +a_1 + a_2 + p + a_1a_2 \omega_{a_1a_2} p + a_1p\omega_{a_1p} + a_2 p \omega_{a_2p} + a_1a_2p \omega_{a_1p} \omega_{a_2p}},
\end{align}
where $N_{\mathrm{NS}}$ is the number of non-specific binding sites; $P$ is the number of RNAP; $A_1$ is the number of the first activator; $A_2$ is the number of the second activator; $\Delta \varepsilon_{\mathrm{pd}}$ is the binding energy of the RNAP; $\Delta \varepsilon_{\mathrm{a_1d}}$ is the binding energy of the first activator; $\Delta \varepsilon_{\mathrm{a_2d}}$ is the binding energy of the second activator; $\omega_{\mathrm{a_1p}}$ is the interaction energy between the first activator and the RNAP; $\omega_{\mathrm{a_1a_2}}$ is the interaction energy between the second activator and the RNAP. The states-and-weights diagram of the AND-logic double activation regulatory architecture is shown in Fig~\ref{figS9}(A). In this case, the probability of RNAP being bound is given by
\begin{align}
    p_{\mathrm{bound}} = \frac{p + a_1p\omega_{a_1p} + a_2 p \omega_{a_2p} + a_1a_2p \omega_{a_1p} \omega_{a_2p} \omega_{a_1a_2}}{1 +a_1 + a_2 + p + a_1a_2 \omega_{a_1a_2} p + a_1p\omega_{a_1p} + a_2 p \omega_{a_2p} + a_1a_2p \omega_{a_1p} \omega_{a_2p} \omega_{a_1a_2}},
\end{align}
where $\omega_{\mathrm{a_1a_2}}$ is the interaction energy between the two activators.

\section{Definition of probability distributions in the calculation of mutual information} \label{S4_Appendix}

In order to build an information footprint from data, we need to calculate the mutual information between expression levels and the base identity at each position in the sequence, which is defined as
\begin{align}
    I_i = \sum_{b} \sum_{\mu} \Pr{_i}(b, \mu) \log_2{\left(\frac{\Pr{_i}(b, \mu)}{\Pr{_i}(b)\Pr(\mu)}\right)},
\end{align}
where $b$ represents base identity and $\mu$ represents expression levels.

\begin{figure}[!h]
\centering
\includegraphics[width=0.85\textwidth]{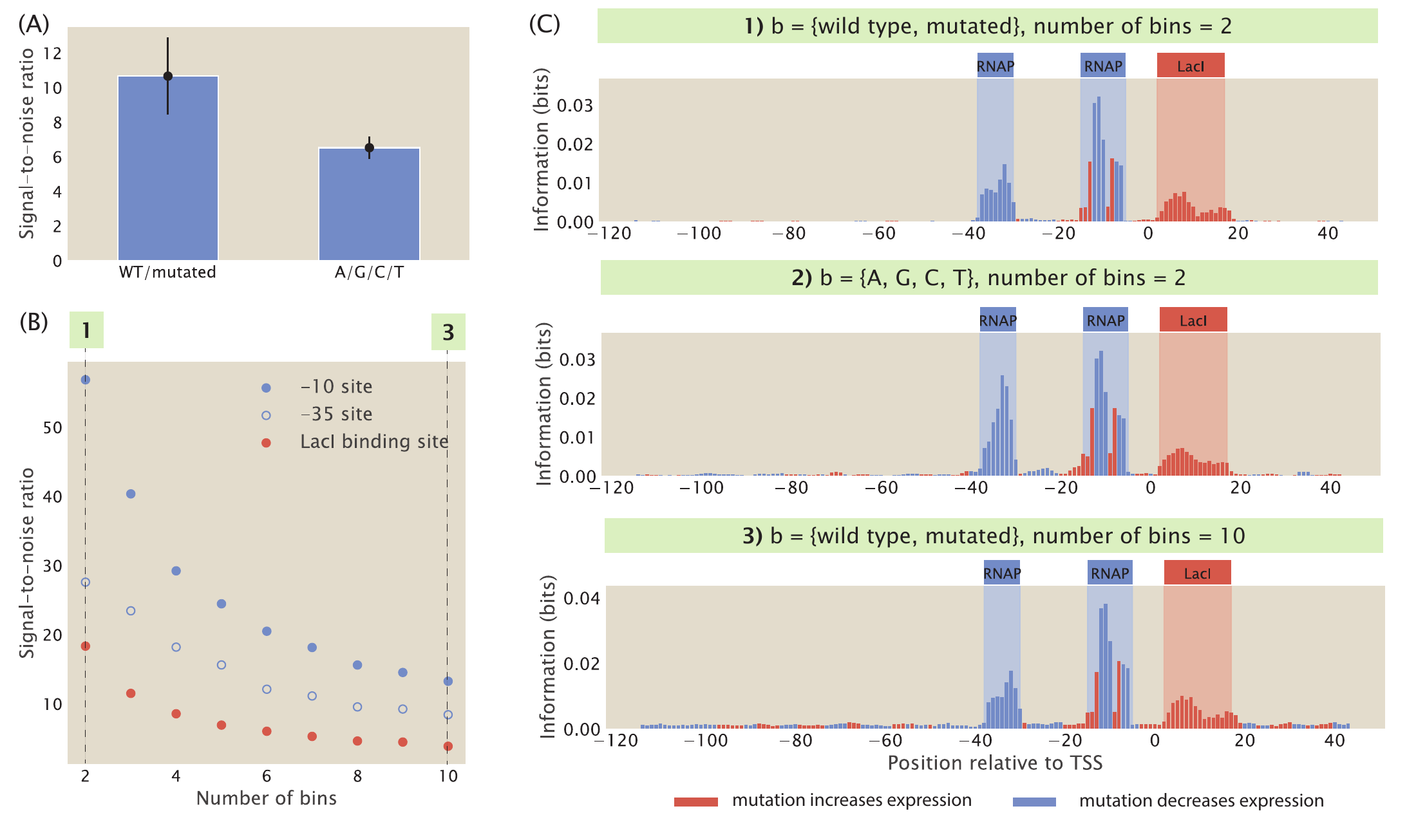}
\caption{{\bf Definition of probability distributions in the calculation of mutual information.}
(A) Using a probability distribution of the four bases leads to a reduced signal-to-noise ratio in the information footprint. The heights of the bars are the average signal-to-noise ratios calculated from the information footprints of 20 synthetic datasets with the repression-activation architecture. (B) Signal-to-noise ratio decreases when the number of bins increases. Each data point is the mean of average mutual information across 20 synthetic datasets with the corresponding number of bins. The numbered labels correspond to footprints in (C). (C) Choosing different probability distributions to calculate the information footprint for a synthetic dataset with the repression-activation genetic architecture. The top footprint uses a probability distribution of wild-type and mutated bases and uses 2 bins to calculate the probability distribution for expression levels. The middle footprint uses a probability distribution of the four bases (A, G, C, T) and uses 2 bins to calculate the probability distribution for expression levels. The bottom footprint uses a probability distribution of wild-type and mutated bases and uses 10 bins to calculate the probability distribution for expression levels.}
\label{figS4}
\end{figure}

As shown in Fig~\ref{figS4}(A) and~\ref{figS4}(C), we find that binding sites have a higher signal-to-noise ratio in information footprints when the coarse grained approach is taken
\begin{align}
    b = 
    \begin{cases}
        0, & \text{if the base is mutated}, \\
        1, & \text{if the base is wild type}.
    \end{cases}
\end{align}

On the other hand, to obtain a distribution of expression levels, sequencing counts are binned into $N$ bins, where the bins are chosen such there is an equal number of sequences in each bin. Again, we observe the highest signal-to-noise ratio for $N=2$, and see a continuous decrease with increasing number of bins. As shown in Fig~\ref{figS4}(B) and~\ref{figS4}(C),  we observe that signal-to-noise ratio is higher when fewer bins are used to partition expression levels. 

These observations may be explained by the fact that increasing the number of states or bins increases the level of noise. When there are more states or more bins, fewer sequences will be present in each bin. This amplifies the hitch-hiking effects discussed in Sec~\ref{sec:library-size}, leading to a higher level of noise. In addition, since the boundaries between the bins are artificially set, a sequence may be randomly grouped into the N-th bin rather than the adjacent (N-1)-th and (N+1)-th bins simply because the boundaries are set a particular level. This randomness occurs in both the marginal probability distribution for $\mu$ and the joint probability distribution, resulting in noise that is increased when more bins are added.

\section{Analytical calculation of information footprint} \label{S5_Appendix}

To better understand how mutations in the binding sites create signals in the information footprint, we derive the information footprint analytically for a constitutively expressed gene, i.e. the promoter of the gene only has a binding site for RNAP and is not bound by any transcription factors.


Consider a promoter region where the RNAP binding site is $l$ base pairs long and the probability of mutation at each site is $\theta$. Furthermore, the binding energy of RNAP to the wild-type sequence is denoted by $\Delta\varepsilon$ and we assume that at each position within the binding site, a mutation comes with cost $\Delta\Delta\varepsilon$ to the binding energy. Therefore, if there are $m$ mutations in the binding site, the total binding energy between RNAP and the mutant binding site is $\Delta\varepsilon + m\Delta\Delta\varepsilon$. 

In a sufficiently large data set, the ratio of sequences with mutation at position $i$ is given by
\begin{align}
    \Pr{_i(b)} = \begin{cases}
        1 - \theta,\ &\mathrm{if}\ b=0 \\
        \theta,\ &\mathrm{if}\ b=1.
    \end{cases}
\end{align}

Next, we determine $\Pr(\mu)$. As before, we define $\Pr(\mu)$ as the probability that a given sequence leads to high expression levels or low expression levels. To predict expression levels, we begin by calculating $p_{\mathrm{bound}}$ for each promoter variant. Since the gene is constitutively expressed, the probability of RNAP binding is given by
\begin{align}
    p_{\mathrm{bound}} &= \frac{\frac{P}{N_{\mathrm{NS}}}e^{- \beta (\Delta\varepsilon + m\Delta\Delta\varepsilon)}}{1 + \frac{P}{N_{\mathrm{NS}}}e^{- \beta (\Delta\varepsilon + m\Delta\Delta\varepsilon)}}.
\end{align}

As derived in Eq~\ref{eq:mrna-level}, the steady state copy number of mRNAs is proportional to the probability of the RNAP bound state. Therefore, expression level is only dependent on the number of mutations in the RNAP binding site. 

\begin{figure}[!h]
\centering
\includegraphics[width=0.7\textwidth]{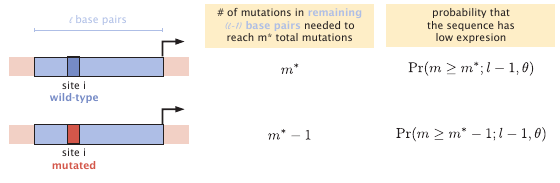}
\caption{{\bf Calculating number of mutations needed to reach the threshold between low expression and high expression bins.}
The joint probability distribution at site $i$ is a product of the probability that site $i$ is mutated or wild type and the probability that the sequence has high or low expression level. Since the expression of the sequence depends on the presence of a mutation at site $i$, we need to consider two different cases in order to calculate the probability of expression. In the case where site $i$ has wild-type base identity, there need to be $m^*$ mutations outside of site $i$ in the RNAP binding site for the sequence to reach the threshold $m^*$. Therefore, the probability that the sequence has low expression is $\Pr(m \geq m^*; l-1, \theta)$. On the other hand, in the case where site $i$ is mutated, since one mutation is known to exist, there only need to be $m^* - 1$ mutations outside of site $i$ in the RNAP binding site for the sequence to reach the threshold. In this case, the probability that the sequence has low expression is $\Pr(m \geq m^*-1; l-1, \theta)$.}
\label{figS5}
\end{figure}

The probability distribution for the number of mutations in the RNAP binding site can be expressed using the binomial distribution, where the probability of $k$ mutations in the binding site is given by
\begin{align}
    \Pr(m=k; l, \theta) = \binom{l}{k} \theta^k (1-\theta)^{l-k}.
\end{align}
As illustrated in Fig~\ref{figS5}, since expression levels are solely determined by the number of mutations in the binding site, and sequences are binned by expression levels to obtain $P(\mu)$, there is a threshold number of mutations, $m^*$, where sequences with $m^*$ or more than $m^*$ mutations fall into the lower expression bin. Hence, $P(\mu)$ is given by
\begin{align}
    \Pr(\mu) =
    \begin{cases}
        \Pr(m \geq m^*;l, \theta) = \sum \limits_{k=m^*}^l \binom{l}{k} \theta^k (1-\theta)^{l-k},\ &\mathrm{if}\ \mu=0 \\
        \Pr(m < m^*;l, \theta) = 1 - \Pr(m \geq m^*; l, \theta),\ &\mathrm{if}\ \mu=1.
    \end{cases}
\end{align}

Finally, we determine the expression for $\Pr{_i}(b, \mu)$. To do this, we consider two cases, one where the position $i$ is outside of the RNAP binding site and one where the position $i$ is within the RNAP binding site. If $i$ is not in the RNAP binding site $\mathcal{B}$, then a mutation would have no effect on the expression levels, therefore 
\begin{align}
    \Pr{_{i \notin \mathcal{B}}}(b, \mu) = 
    \begin{cases}
        (1-\theta) \cdot \Pr(m \geq m^*; l, \theta),\ &\mathrm{if}\ b=0\ \mathrm{and}\ \mu=0 \\
        (1-\theta) \cdot \Pr(m < m^*; l, \theta),\ &\mathrm{if}\ b=0\ \mathrm{and}\ \mu=1 \\
        \theta \cdot \Pr(m \geq m^*; l, \theta),\ &\mathrm{if}\ b=1\ \mathrm{and}\ \mu=0 \\
        \theta \cdot \Pr(m < m^*; l, \theta),\ &\mathrm{if}\ b=1\ \mathrm{and}\ \mu=1.
    \end{cases}
\end{align}

Having derived all the marginal probability distributions and the joint probability distributions, we can then write down mutual information at a non-binding site and at a binding site. If position $i$ is outside the RNAP binding site, then
\begin{align}
    \begin{split}
        I_i = &(1-\theta) \cdot \Pr(m \geq m^*; l, \theta) \log_2 \frac{(1-\theta) \cdot \Pr(m \geq m^*; l, \theta)}{(1-\theta) \cdot \Pr(m \geq m^*; l, \theta)} \\
        &+ (1-\theta) \cdot \Pr(m < m^*; l, \theta) \log_2 \frac{(1-\theta) \cdot \Pr(m < m^*; l, \theta)}{(1-\theta) \cdot \Pr(m < m^*; l, \theta)} \\
        &+ \theta \cdot \Pr(m \geq m^*; l, \theta) \log_2 \frac{\theta \cdot \Pr(m \geq m^*; l, \theta)}{\theta \cdot \Pr(m \geq m^*; l, \theta)} \\
        &+ \theta \cdot \Pr(m < m^*; l, \theta) \log_2 \frac{\theta \cdot \Pr(m < m^*; l, \theta)}{\theta \cdot \Pr(m < m^*; l, \theta)}.
    \end{split}
\end{align}
We can see that $I_i = 0$ since the fractions within the logarithms all cancel out to be 1. This is because the joint probability $\Pr{_i}(b, \mu)$ for bases outside the binding site is simply given by the product of the marginal distributions,
\begin{align}
    \Pr{_{i \notin \mathcal{B}}}(b, \mu) = \Pr{_{i \notin \mathcal{B}}}(\mu) \Pr{_{i \notin \mathcal{B}}}(b).
\end{align}

If the position $i$ is in the RNAP binding site, the calculation for $\Pr{_i}(b, \mu)$ is more complex. As illustrated in Fig~\ref{figS5}, if the position $i$ has wild-type base identity, then the sequence would have low expression levels if there are more than $m^*$ mutations in the remaining $l-1$ bases in the RNAP binding site and the sequence would have high expression levels if there are less than $m^*$ mutations in the remaining $l-1$ bases in the RNAP binding site. On the other hand, if the position $i$ is mutated, then the sequence would have low expression levels if there are more than $m^* - 1$ mutations in the remaining $l-1$ bases and the sequence would have high expression levels if there are less than $m^* - 1$ mutations in the remaining $l-1$ bases. Taken together, we can write down the joint probability distribution as
\begin{align}
    \Pr{_{i \in \mathcal{B}}}(b, \mu) = 
    \begin{cases}
        (1-\theta) \cdot \Pr(m \geq m^*; l-1, \theta),\ &\mathrm{if}\ b=0\ \mathrm{and}\ \mu=0 \\
        (1-\theta) \cdot \Pr(m < m^*; l-1, \theta),\ &\mathrm{if}\ b=0\ \mathrm{and}\ \mu=1 \\
        \theta \cdot \Pr(m \geq m^*-1; l-1, \theta),\ &\mathrm{if}\ b=1\ \mathrm{and}\ \mu=0 \\
        \theta \cdot \Pr(m < m^*-1; l-1, \theta),\ &\mathrm{if}\ b=1\ \mathrm{and}\ \mu=1. \\
    \end{cases}
\end{align}

In this case, the joint distribution does not factor into the marginal distributions, 
\begin{align}
    \Pr{_{i \in \mathcal{B}}}(b, \mu) \neq \Pr{_{i \in \mathcal{B}}}(\mu) \Pr{_{i \in \mathcal{B}}}(b),
\end{align}
and therefore, mutual information has to be larger than zero, $I_i > 0$, clearly distinguishing positions that are within the binding site from positions outside. Specifically,
\begin{align}
    \begin{split}
        I_i = &(1-\theta) \cdot \Pr(m \geq m^*; l-1, \theta) \log_2 \frac{(1-\theta) \cdot \Pr(m \geq m^*; l-1, \theta)}{(1-\theta) \cdot \Pr(m \geq m^*; l, \theta)} \\
        &+ (1-\theta) \cdot \Pr(m < m^*; l-1, \theta) \log_2 \frac{(1-\theta) \cdot \Pr(m < m^*; l-1, \theta)}{(1-\theta) \cdot \Pr(m < m^*; l, \theta)} \\
        &+ \theta \cdot \Pr(m \geq m^* - 1; l-1, \theta) \log_2 \frac{\theta \cdot \Pr(m \geq m^*-1; l-1, \theta)}{\theta \cdot \Pr(m \geq m^*; l, \theta)} \\
        &+ \theta \cdot \Pr(m < m^* -1; l-1, \theta) \log_2 \frac{\theta \cdot \Pr(m < m^*-1; l-1, \theta)}{\theta \cdot \Pr(m < m^*; l, \theta)} \\
        = &(1-\theta) \cdot \Pr(m \geq m^*; l-1, \theta) \log_2 \frac{\Pr(m \geq m^*; l-1, \theta)}{\Pr(m \geq m^*; l, \theta)} \\
        &+ (1-\theta) \cdot \Pr(m < m^*; l-1, \theta) \log_2 \frac{\Pr(m < m^*; l-1, \theta)}{\Pr(m < m^*; l, \theta)} \\
        &+ \theta \cdot \Pr(m \geq m^* - 1; l-1, \theta) \log_2 \frac{\Pr(m \geq m^*-1; l-1, \theta)}{\Pr(m \geq m^*; l, \theta)} \\
        &+ \theta \cdot \Pr(m < m^* -1; l-1, \theta) \log_2 \frac{\Pr(m < m^*-1; l-1, \theta)}{\Pr(m < m^*; l, \theta)}.
    \end{split}
\end{align}

\section{Recovering binding site signal under extreme mutation rates} \label{S6_Appendix}

\begin{figure}[!h]
\centering
\includegraphics[width=0.8\textwidth]{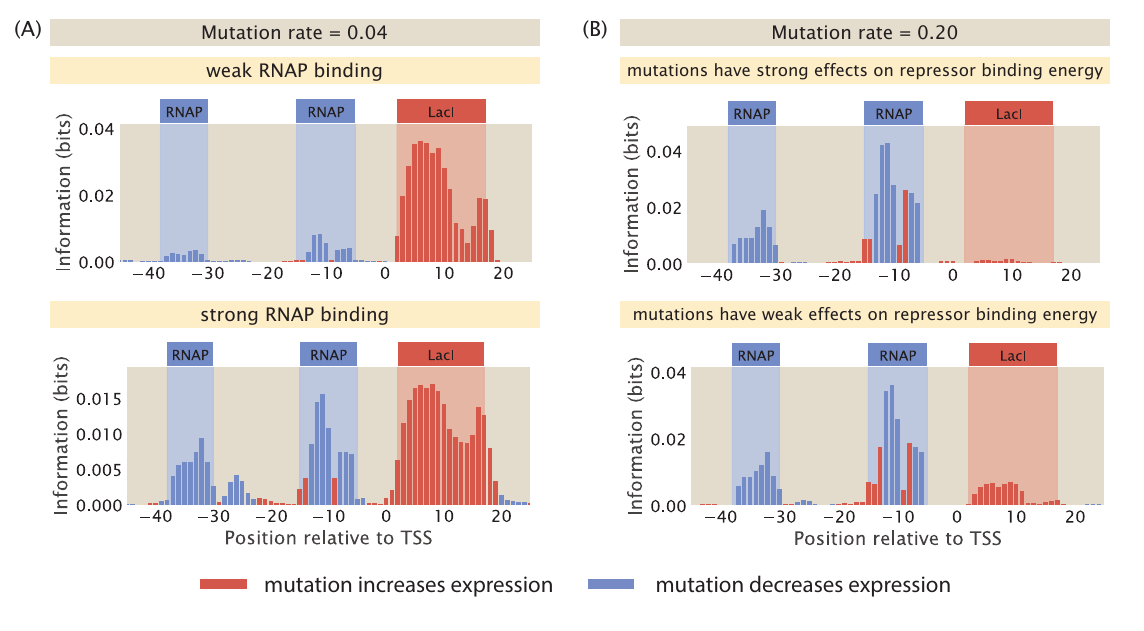}
\caption{{\bf Recovering signals from information footprints under extreme mutation rates.}
(A) We generated two synthetic datasets with a mutation rate of 0.04 in the mutant library. In the first dataset, we set the RNAP binding energy $\Delta \varepsilon_{rd}$ to be $-5\ k_BT$, which is typical of RNAP binding at the wild type -10 and -35 binding sites. In the footprint produced from this dataset, there is low mutual information at the RNAP binding site due to the low mutation rate. On the other hand, in the second dataset, we increased $\Delta \varepsilon_{rd}$ to $-12\ k_BT$. This allows us to recover the signal at the RNAP binding site. (B) We generated two synthetic datasets with a mutation rate of 0.20. In the first dataset, we used the experimentally measured energy matrix for LacI at the O1 operator shown in Fig~\ref{fig2}(B), where the average effect of mutations on binding energy is $2.24\ k_BT$. In the footprint produced from this dataset, there is low mutual information at the repressor binding site due to the high mutation rate. In the second dataset, we reduced the average effect of mutations five-fold and are able to recover the signal at the repressor binding site.}
\label{figS6}
\end{figure}

As we have shown in Sec \ref{sec:mutation}, when the rate of mutation in the mutant library is low, we lose the signal at the RNAP binding site. We hypothesize that this is because RNAP binds weakly at the promoter. We generated a synthetic dataset that has low mutation rate but stronger binding energy at the RNAP binding site. As shown in Fig~\ref{figS6}(A), the information footprint built from this dataset has a much higher level of mutual information at the RNAP binding site compared to the information footprint built from a dataset with the same mutation rate but weak RNAP binding energy, which supports our hypothesis.

We also showed that when the rate of mutation in the mutant library is high, there is low mutual information at the repressor binding site. Our hypothesis is that this is caused by the large effects of mutations on the repressor binding energy. We generated a synthetic dataset with high mutation rate while reducing the effect of mutation on binding energy by five fold. As shown in Fig~\ref{figS6}(B), this allows us to recover the signal at the repressor binding site, which is also in line with our hypothesis.

\section{Optimal mutation rate for various parameters} \label{S7_Appendix}

An essential part of designing libraries for MPRAs is the rate at which bases in the sequences are mutated. In Sort-Seq~\cite{Kinney2010-ue} and Reg-Seq~\cite{Ireland2020-bp}, the mutation rate for promoter sequences is chosen as 0.1 per base. In Sec~\ref{sec:mutation}, we have calculated that this is on par with the optimal mutation rate for a promoter regulated by one transcription factor, given a specific set of parameters for copy numbers and binding energies of the transcription factor and RNAP. Here, we explore how the result for the optimal mutation rate depends on the specific choice of parameters.

In Eq~\ref{equ:opt_mut_rate}, we have defined the optimal mutation rate as the rate where the Boltzmann weights of RNAP binding and transcription factor binding are equal. Here we explore how this mutation rate depends on the binding energy of RNAP $\Delta \varepsilon_\mathrm{pd}$, the binding energy of the transcription factor $\Delta \varepsilon_\mathrm{rd}$, the copy number of RNAP $P$, and the copy number of the transcription factor $R$.

\begin{figure}[!h]
    \centering
    \includegraphics[width=\textwidth]{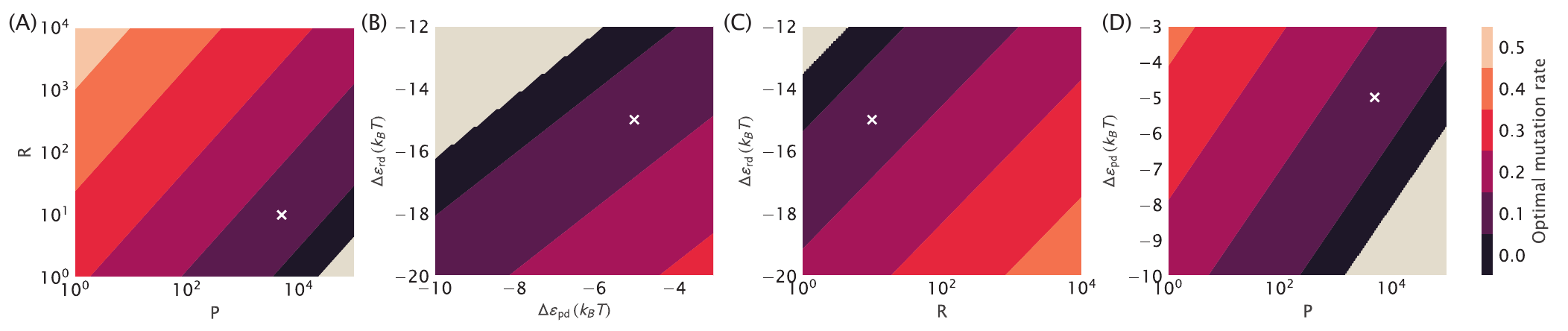}
    \caption{\textbf{Optimal mutation rate for single transcription factor binding.}
    Numerical solutions for the mutation rate calculated by finding the mutation rate that leads to equal Boltzmann weights between the RNAP bound state and the transcription factor bound state using Eq~\ref{equ:opt_mut_rate}. The white crosses mark the standard set of parameters: $R = 10$, $P = 5000$, $\Delta\varepsilon_\mathrm{rd} = -15\, k_\mathrm{B}T$ and $\Delta\varepsilon_\mathrm{pd} = -15\, k_\mathrm{B}T$. Panels (A) to (D) vary 2 of the 4 parameters, while the other two stay constant. Gray color indicates regimes where no mutation rate can be found that fulfills the criteria of $\kappa = 1$.}
    \label{figS7}
\end{figure}

For each set of parameters, we can solve Eq~\ref{equ:opt_mut_rate} numerically for the mutation rate that gives us $\kappa=1$. In Fig~\ref{figS7}, the optimal mutation rate is computed numerically when two of the parameters are varied while the others are kept constant. Increasing the binding energy or copy number of the transcription factor increases the mutation rate, while increasing the binding energy of RNAP and increasing the copy number of RNAP decreases the mutation rate. Changing the binding energy can have drastic effects on the mutation rate, in contrast to changes in copy numbers. This can be explained by the fact that the binding energies contributing exponentially to $\kappa$, while the copy numbers come into play as linear factors. For cases where the transcription factor bound state becomes very unlikely, e.g. in the cases of very weak binding of the transcription factor or very strong binding of the RNAP, there is no optimal mutation rate that can be found given the criteria in Eq~\ref{equ:opt_mut_rate}. These regions can be found in Fig~\ref{figS7}(B)-(D) as grey regions.

Binding sites and transcription factors come with widely different values for the parameters we have tested, e.g., the copy number of the activator CRP can be as high as about 500, while the copy number for an essential transcription factor DicA can be as low as 10 as measured in mass-spectrometry experiments~\cite{Schmidt2016-wj}. The binding energy for the lac-repressor varies on the order of $6\,k_\mathrm{B}T$ ($-15.7\,k_\mathrm{B}T$ for the O1 operator and $-9.3\,k_\mathrm{B}T$ for the O3 operator~\cite{Garcia2011-np}. Hence, it can be very beneficial to create a library of mutant sequences that contains sequences with different mutation rates in order to detect binding sites with these different parameters.

\section{Changing transcription factor copy numbers for a double repression promoter under XOR gate} \label{S8_Appendix}

In Sec~\ref{sec:logic}, we examined how changing transcription factor copy numbers affect the footprint for a double repression promoter under AND and OR logic gates. As discussed by Buchler et al.~\cite{Buchler2003-yt} and de Ronde et al.~\cite{De_Ronde2012-fx}, only a limited number of logic gates are attainable through parameter variations, and a thermodynamic model can be written down for each of the possible logic gates. This means that our thermodynamic-model-based computational pipeline can be easily adapted to consider all possible types of interactions between transcription factors.

\begin{figure}[!h]
    \centering
    \includegraphics[width=\textwidth]{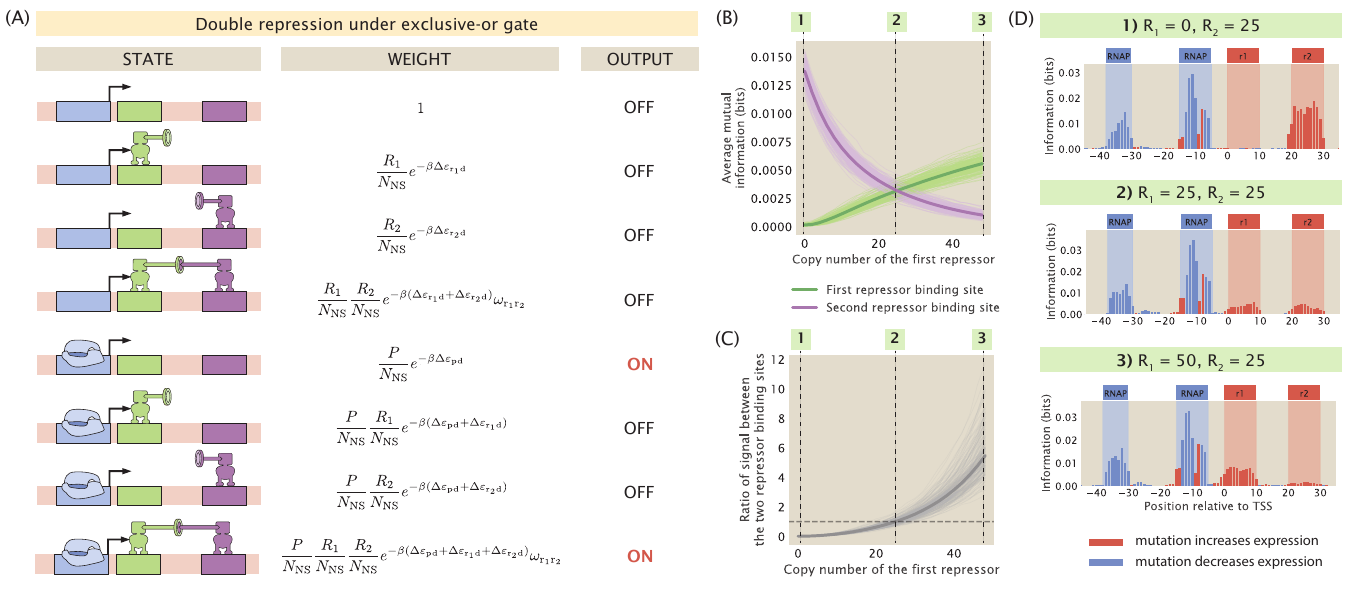}
    \caption{{\bf Changing repressor copy number for a double-repression promoter under the XOR logic gate.}
    (A) States-and-weights diagram of a promoter with the double repression regulatory architecture and under the exclusive-or (XOR) gate. (B) Changes in the average mutual information at the two repressor binding sites when the copy number of the first repressor is increased. For the energy matrices of the repressors, the interaction energy between the repressor and a site is set to $0\, k_BT$ if the site has the wild type base identity and set to $1\, k_BT$ if the site has the mutant base identity. To enforce the XOR logic, the interaction energy between the repressors is set to $5\,k_BT$. 200 synthetic datasets are simulated and the trajectory for each of the synthetic dataset is shown as an individual light green or light purple curve. The average trajectories across all 200 synthetic datasets are shown as the bolded green curves and the bolded purple curves. The three numbered labels correspond to the information footprints shown in (D). (C) Ratio of average mutual information between the two repressor binding sites when the copy number of the first repressor changes. The individual trajectories (plotted in light grey) and mean trajectory (plotted in dark grey) are from the same 200 synthetic datasets used in generate the plot in (B). As expected, the ratio is equal to 1 when the copy number of the first repressor is equal to the copy number of the second repressor. (D) Representative information footprints with three different combination of repressor copy numbers.}
    \label{figS8}
\end{figure}

As an example, let us consider the exclusive-or (XOR) logic gate in a promoter regulated by two repressors, which is another important and interesting logic gate other than the AND and OR gates. As illustrated in Fig~\ref{figS8}(A), under the XOR gate, gene expression is repressed when only one of the repressors is present at high concentrations, but not when both of the repressors are present at high concentrations. One possible mechanism by which this may occur is if the interaction between the two repressors is repulsive. As shown in Fig~\ref{figS8}(B) and~\ref{figS8}(C), when the copy number of the second repressor is kept constant at 25 and the copy number of the first repressor is increased from 0 to 50, the signal at the first repression binding site increases and the signal at the second repressor binding site decreases. This behaviour is consistent with the definition of XOR logic gates and demonstrates that our computational pipeline can handle a diverse range of interaction regimes.

\section{Transcription factor knock-out under double activation} \label{S9_Appendix}

\begin{figure}[!h]
\centering
\includegraphics[width=0.7\textwidth]{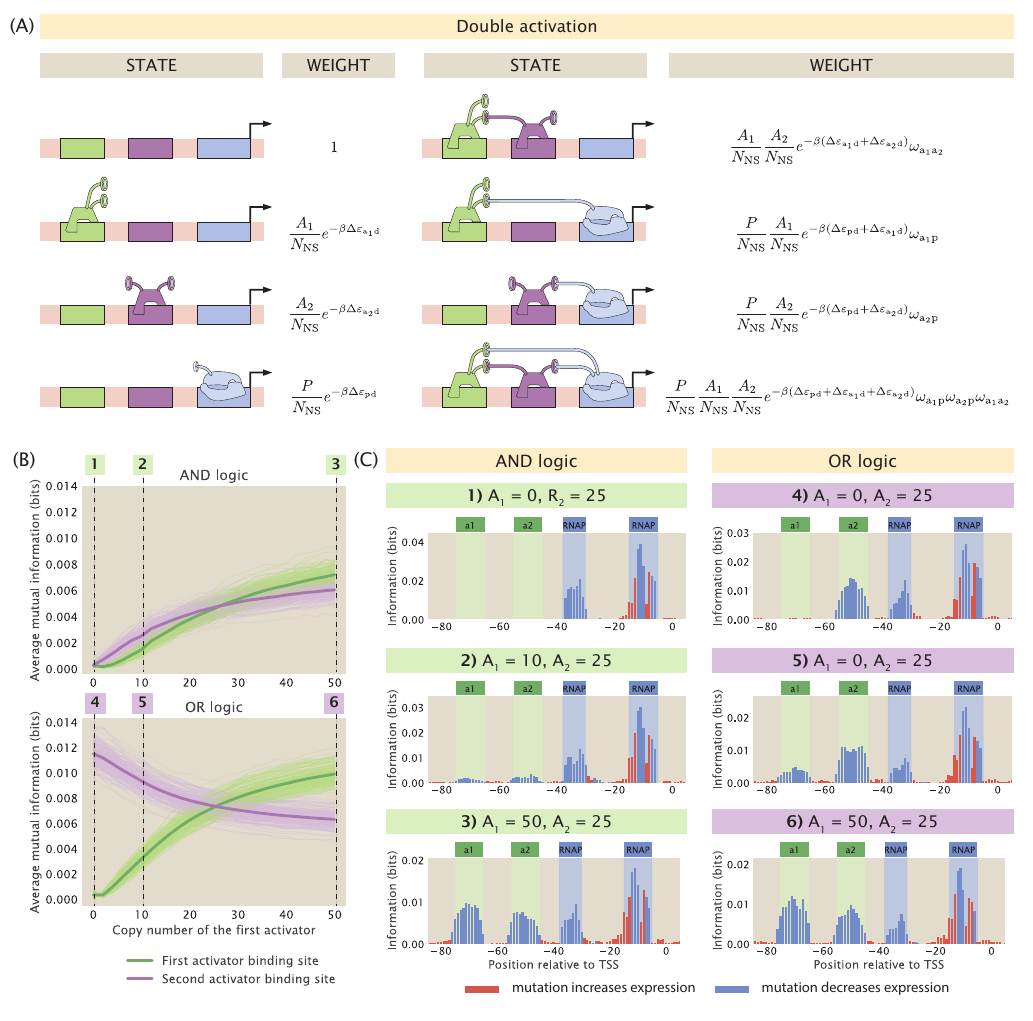}
\caption{{\bf Changing the copy number of activators in a double activation promoter.}
(A)~States-and-weights diagram of a promoter with the double activation regulatory architecture. Under OR logic, the two activators do not exhibit coorperativity and $\omega_{\mathrm{a_1a_2}} = 0\ k_BT$. The states-and-weights diagram of a double activation promoter with OR logic is also shown in Fig~\ref{figS3}(F). (B)~Changing the copy number of the first activator under AND logic and OR logic affects the signal at both activator binding sites. The energy matrices of the activators are randomly generated in the same way as the energy matrices of the repressors in Fig~\ref{fig8}. For the promoter with AND logic, the interaction energies between the activators and between the activator and the RNAP are set to -4 $k_BT$. For the promoter with OR logic, the interaction energies between the activators and between the activator and the RNAP are set to -7 $k_BT$. The higher interaction energy for the OR logic promoter is to ensure that there are similar levels of signal at the activator binding sites compared to the AND logic promoter.  200 synthetic datasets are simulated and the trajectory for each of the synthetic dataset is shown as an individual light green or light purple curve. The average trajectories across all 200 synthetic datasets are shown as the bolded green curves and the bolded purple curves. (C)~Representative information footprints of a double repression promoter under AND and OR logic.}
\label{figS9}
\end{figure}

A double-activation promoter can also operate under an AND or an OR logic gate~\cite{Buchler2003-yt}. The states-and-weights diagram for a double-activation promoter is shown in Fig~\ref{figS9}(A). Under AND logic, the two activators can interact both with the RNAP and with each other. This cooperativity leads to a further increase in expression levels. In contrast, under OR logic, the activators independently interact with RNAP and there is no cooperativity between them. We build synthetic datasets for an AND-logic and an OR-logic double-activation promoter. As shown in Fig~\ref{figS9}(B) and~\ref{figS9}(C), under AND logic, since cooperativity is at play, the signal at both $A_1$ and $A_2$ binding sites increases when $A_1$ is increased. On the other hand, under OR logic, the two activators act independently and there is competition between the signals at the two sites. When $A_1$ is increased, the signal at $A_1$ binding site correspondingly increases but the signal at $A_2$ binding site decreases.

\section{Changing inducer concentration for the inducible activator} \label{S10_Appendix}

\begin{figure}[!h]
\centering
\includegraphics[width=0.9\textwidth]{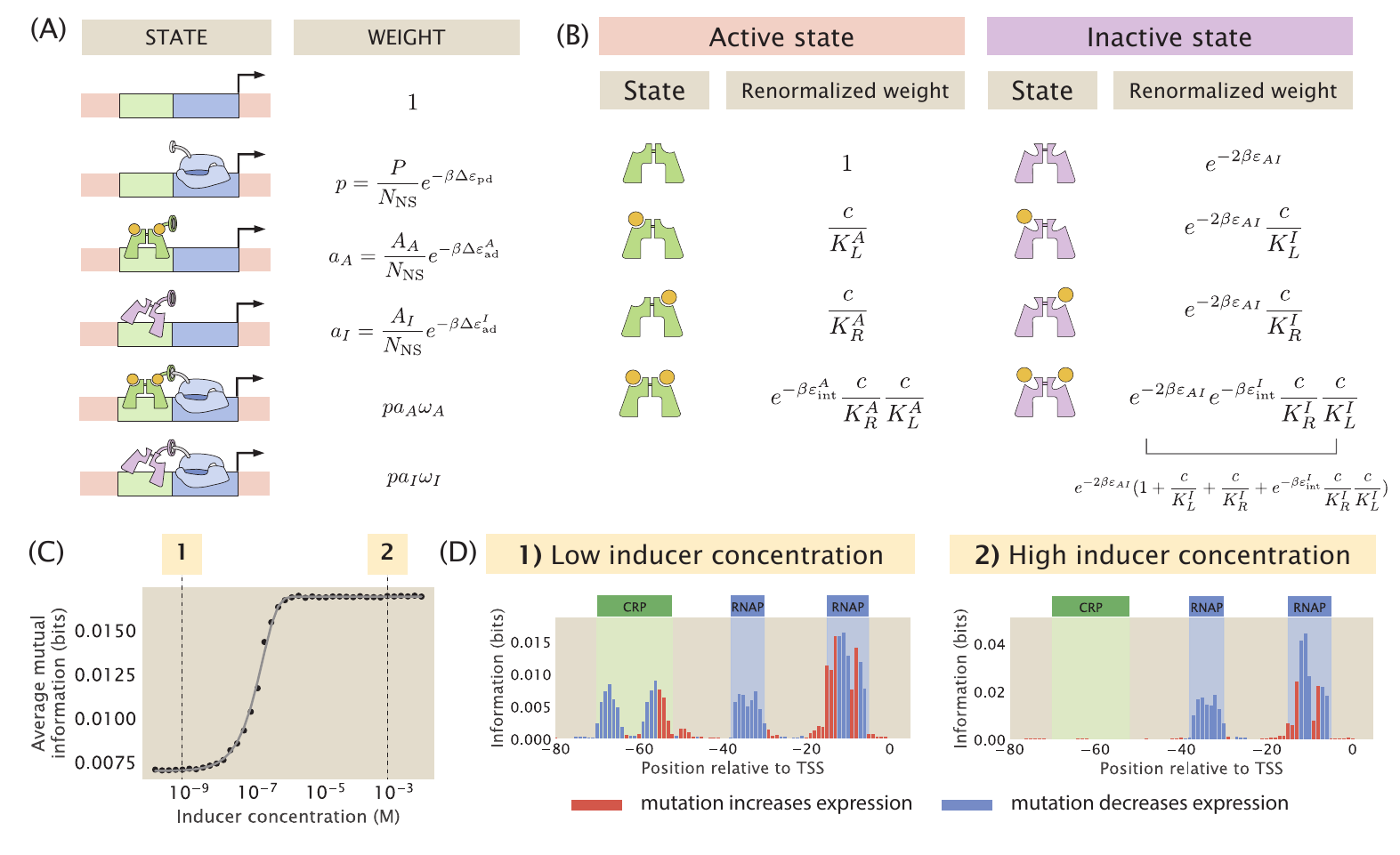}
\caption{{\bf Changing inducer concentration for the inducible activator.}
(A) States-and-weights diagram for an inducible activator. In the diagram, $N_\mathrm{NS}$ is the number of non-binding sites in the genome, $P$ is the copy number of the RNAP, $A_A$ is the copy number of active activators, $A_I$ is the copy number of inactive activators, $\Delta \varepsilon_\mathrm{pd}$ is the binding energy of the RNAP, $\Delta \varepsilon_\mathrm{ad}^A$ is the binding energy of the active activator, $\Delta \varepsilon_\mathrm{ad}^I$ is the binding energy of the inactive activator. $\omega_A = e^{-\beta \varepsilon_{p, a_A}}$ and $\omega_I = e^{-\beta \varepsilon_{p, a_I}}$, where $\varepsilon_{p, a_A}$ is the interaction energy between the RNAP and the active activator and $\varepsilon_{p, a_I}$ is the interaction energy between the RNAP and the inactive activator. (B) States-and-weights diagram to calculate the probability that the activator is in the active state. (C) Average mutual information at the RNAP binding site increases as the inducer concentration increases. Here, we let $K_L^A = K_R^A = 3 \times 10^{-6}$ M, $K_L^I = K_R^A = 10^{-7}$ M, and $\Delta \varepsilon_{AI} = -3\ k_BT$ \cite{Einav2018-ko}. Each data point is the mean of average mutual information across 20 synthetic datasets with the corresponding inducer concentration. The numbered labels correspond to footprints in (D). (D) Representative information footprints with low inducer concentration ($10^{-9}$ M) and high inducer concentration ($10^{-3}$ M).}
\label{figS10}
\end{figure}

In Sec \ref{sec:inducer}, we discussed the effects of inducer concentration on the information footprints of a simple repression promoter with an inducible repressor. Similar effects can also be seen for a simple activation promoter with an inducible activator. One example of an inducible activator is CRP, which changes its conformation when bound to cyclic-AMP and thereby becomes more favorable to DNA binding \cite{Einav2018-ko}. Based on the states-and-weights diagram for such a promoter, which is shown in Fig~\ref{figS10}(A), the probability of RNAP being bound is given by
\begin{align} \label{eq:inducible-activator}
    p_\mathrm{bound} = \frac{p + pa_A\omega_A + pa_I\omega_I}{1 + p + a_A + a_I + pa_A\omega_A + pa_I\omega_I},
\end{align}
where $p$ is the normalized weight of the RNAP bound state, $a_A$ is the normalized weight of the active activator bound state, and $a_I$ is the normalized weight of the inactive activator bound state. $\omega_A$ and $\omega_I$ account for the interaction energy between the RNAP and the active activator and the interaction energy between the RNAP and the inactive activator, respectively. The exact expressions for $p$, $a_A$, $a_I$, $\omega_A$ and $\omega_I$ are given in Fig~\ref{figS10}(A).

To simplify the expression above, we determine the proportion of active and inactive activators with respect to the total number of activators. Similar to the case of simple repression, we calculate $p_\mathrm{active}(c)$, which is the probability that the activator exists in the active conformation as a function of the inducer concentration, $c$. The different states of the activator can be modelled using the states-and-weights diagram shown in Fig~\ref{figS8}(B). Here, we consider two types of cooperativity. The first type of cooperativity is between the two binding sites, where each ligand binding event changes the binding affinity of the unbound site. This is inherent to the classic MWC model and is already encoded in the terms $\omega_A$ and $\omega_I$ in Eq~\ref{eq:inducible-activator}. The second type of cooperativity is between the two ligands, which accounts for the negative cooperativity of CRP in the inactive state. This is accounted for by the cooperative energy terms $\varepsilon_\mathrm{int}^A$ and $\varepsilon_\mathrm{int}^I$, which represent the interaction energies between the two ligands in the active and inactive states, respectively. Given the states-and-weights diagram, $p_\mathrm{active}(c)$ is given by
\begin{align}
    p_\mathrm{active}(c) = \frac{1 + \frac{c}{K_L^A} + \frac{c}{K_R^A} + \frac{c}{K_L^A}\frac{c}{K_R^A}e^{-\beta \varepsilon_\mathrm{int}^A}}{1 + \frac{c}{K_L^A} + \frac{c}{K_R^A} + \frac{c}{K_L^A}\frac{c}{K_R^A}e^{-\beta \varepsilon_\mathrm{int}^A} + e^{-2\beta \varepsilon_{AI}}(1 + \frac{c}{K_L^I} + \frac{c}{K_R^I} + \frac{c}{K_L^I}\frac{c}{K_R^I}e^{-\beta \varepsilon_\mathrm{int}^I})},
\end{align}
where $K^A_L$ is the dissociation constant between the inducer and the left binding pocket of the active activator, $K^A_R$ is the dissociation constant between the inducer and the right binding pocket of the active activator, $K^I_L$ is the dissociation constant between the inducer and the left binding pocket of the inactive activator, and $K^I_R$ is the dissociation constant between the inducer and the right binding pocket of the inactive activator. With this expression, we can represent the number of active and inactive activators as $A_A = p_\mathrm{active} A$ and $A_I = (1 - p_\mathrm{active})A$. Therefore, we have that $a_A = p_\mathrm{active} \frac{A}{N_\mathrm{NS}} e^{-\beta \Delta \varepsilon_{ad}^A}$ and $a_I = (1 - p_\mathrm{inactive}) \frac{A}{N_\mathrm{NS}} e^{-\beta \Delta \varepsilon_{ad}^I}$.

We built synthetic datasets for a simple activation promoter with an inducible activator. As shown in Fig~\ref{figS10}(C) and Fig~\ref{figS10}(D), when the concentration of the inducer is increased, the average signal at the RNAP binding site increases, which corresponds to an increase in expression. At high inducer concentration, the activator is too strongly bound to be affected by mutations, and therefore the signal at the activator binding site is negligible.

\section{Noise from experimental procedures} \label{S11_Appendix}

In MPRAs such as Reg-Seq, the mutant library is grown up in culture. Once the cell culture is prepared, genomic DNA (gDNA) and mRNAs are extracted, the latter of which is used as a template in reverse transcription to make complementary DNA (cDNA). Afterwards, polymerase chain reation (PCR) is performed to amplify the reporter gene from the gDNA and cDNA. Finally, sequencing adapters are attached to the gDNA and cDNA. The gDNA and cDNA are then sequenced to obtain DNA and RNA counts for each sequence variant.

\begin{figure}[!h]
\centering
\includegraphics[width=\textwidth]{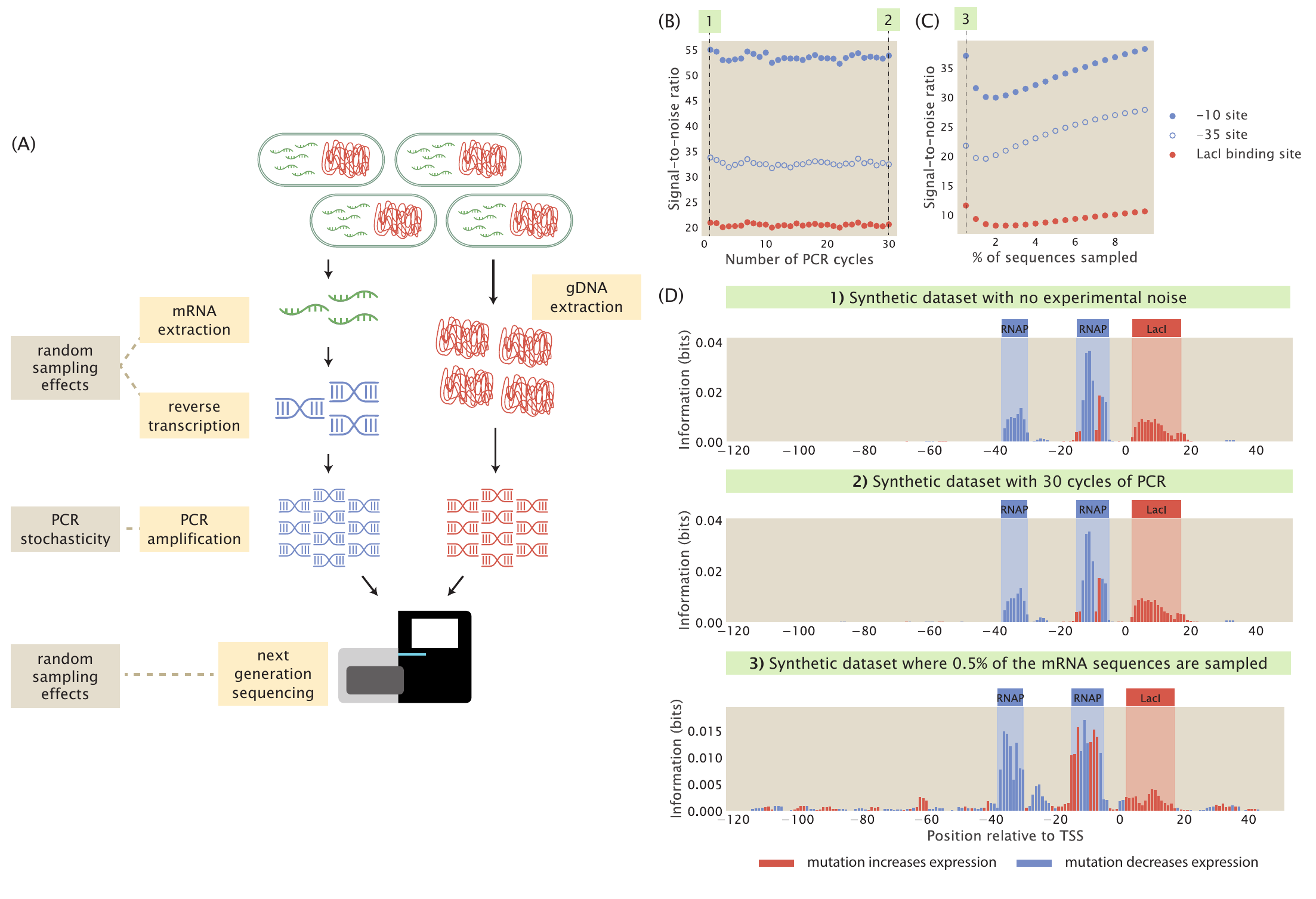}
\caption{{\bf Noise from experimental procedures in the Reg-Seq pipeline.}
(A) The two main sources of noise in the experimental MPRA pipeline are stochasticity from PCR amplification and random sampling effects from RNA extraction, reverse transcription, and RNA-Seq. (B) Signal-to-noise ratio in the information footprints remains high when the number of PCR amplification cycles is increased. Here, $P_\mathrm{amp}=0.5$. Each data point is the mean of average mutual information across 20 synthetic datasets with the corresponding number of PCR cycles. The numbered labels correspond to footprints in (D). (C) Signal-to-noise ratio remains high when only a small percentage of the sequences are randomly sampled. Each data point is the mean of average mutual information across 20 synthetic datasets with the corresponding percentage of sampled sequences. The numbered labels correspond to footprints in (D). (D) Representative information footprints with no experimental noise, PCR stochasticity after 30 cycles, and random sampling effects after 0.5\% of the RNA sequences are sampled.}
\label{figS11}
\end{figure}

As illustrated in Fig~\ref{figS11}(A), there are at least two possible sources of experimental noise. Firstly, PCR amplification is a stochastic process where the probability that a DNA molecule is amplified in a given cycle is less than one. This stochasticity may cause some sequences to have an artificially high RNA count. We note that assuming that the same reporter gene is used for each sample, the only difference in the sequence being amplified would be the barcode. Since barcodes are typically much shorter, it is unlikely to significantly alter the GC-content of the sequence, and therefore we do not discuss the effect of PCR sequence bias. Secondly, during RNA-Seq as well as the prior library preparation procedures such as RNA extraction and reverse transcription, we cannot ensure that every mRNA is extracted, converted to cDNA, and sequenced. Instead, in these steps, only a random subset of the original pool of mRNAs is sampled and included in the final sequencing dataset. As a result, a sequence may have an artificially low RNA count because some copies of the mRNA associated with that sequence are not sampled in one of the experimental steps.

We simulate these two sources of experimental noise in our computational pipeline. To simulate PCR with $n$ cycles of amplification, we start with the original mRNA counts predicted based on the probability of RNAP being bound. Subsequently, we model the number of sequences that are successfully amplified during each cycle using a Binomial distribution~\cite{Kebschull2015-kr}. Hence, for each sequence variant,
\begin{align} \label{eq:PCR}
    n(j+1) = n(j) + B(n(j),P_\mathrm{amp}),
\end{align}
where $n(j)$ is the number of sequences of the promoter variant in cycle $j$, $B(n, P)$ models the Binomial distribution, and $P_\mathrm{amp}$ is the probability that a sequence is successfully amplified in a cycle. We applied Eq~\ref{eq:PCR} to calculate the final count of each sequence variant in a library. As shown in Fig~\ref{figS11}(B) and~\ref{figS11}(D), even when the probability of amplification is set to a low number of $P_\mathrm{amp}=0.5$, increasing the number of PCR cycles does not reduce the signal-to-noise ratio in information footprints. Therefore, we conclude that stochasticity in PCR does not contribute to significant levels of noise in information footprints.

To simulate the random sampling effect during RNA extraction, reverse transcription, and sequencing, we randomly draw a subset of promoter variants in the mutant library and we only consider the expression levels of the selected promoter variants when we calculate mutual information to build the information footprint. As shown in Fig~\ref{figS11}(C) and~\ref{figS11}(D), the levels of noise only becomes significant when less than 1\% of the original pool of sequences is sampled. Therefore, random sampling effects are not a significant source of noise in information footprints either.

\section{Modelling extrinsic noise using a Log-Normal distribution} \label{S12_Appendix}

In order to account for extrinsic noise, we choose to use a Log-Normal distribution to describe the copy number of RNAPs and repressors. Let $X$ be the copy number of the RNAP or the transcription factor, we define the Log-Normal distribution as
\begin{align} \label{eq:lognormal}
    \log {X} \sim \mathrm{Normal}\left(\log{\mu}, (\alpha \log{\mu})^2 \right).
\end{align}

\begin{figure}[h!]
\centering
\includegraphics[width=0.7\textwidth]{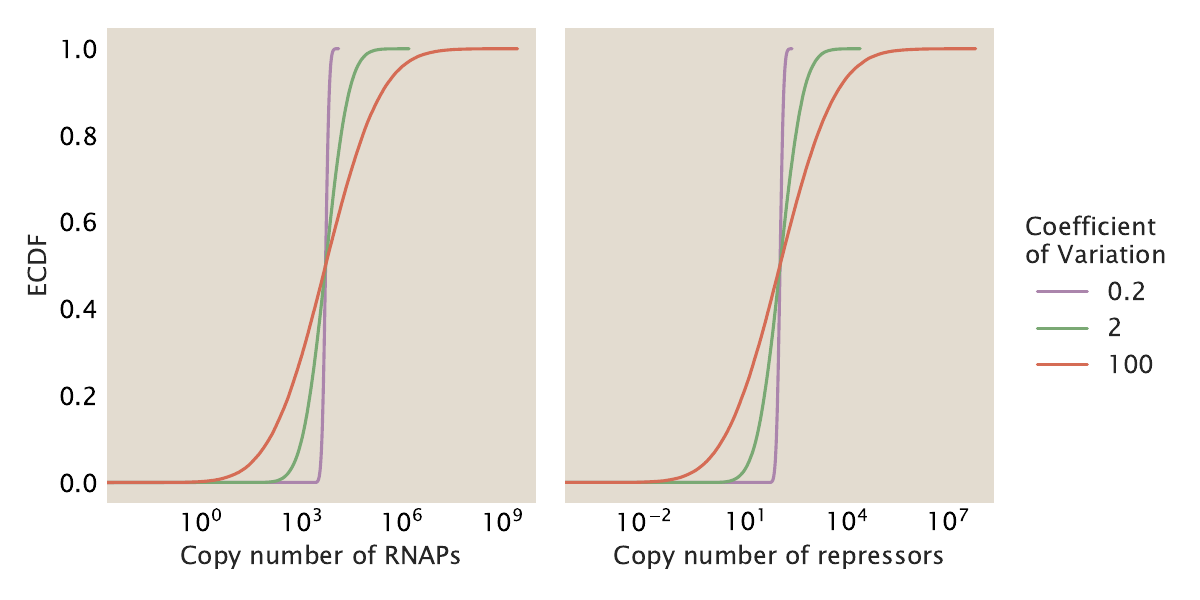}
\caption{{\bf Modelling the copy number of RNAPs and repressors using a Log-Normal distribution.}
CDFs for the copy numbers of RNAP and repressors modelled using a Log-Normal distribution and under three different levels of coefficient of variation.}
\label{figS12}
\end{figure}

Since the goal of our analysis is to better formulate regulatory hypotheses based on real world data, we focus on levels of copy number fluctuations that are physiologically relevant. Assuming that the extrinsic noise in copy numbers primarily comes from asymmetrical partitioning during cell division, one method to measure fluctuations in transcription factor copy numbers between cells is the dilution method developed by Rosenfeld, Young et al.~\cite{Rosenfeld2007-bk}. Based on Brewster et al.~\cite{Brewster2014-qf} who utilized the dilution method, transcription factor copy numbers typically vary by less than $20\%$ of the mean copy number. Consider the proteomic measurements from Schmidt et al.~\cite{Schmidt2016-wj} and Balakrishnan et al.~\cite{Balakrishnan2022-bt}, the coefficient of variation for transcription factor copy numbers is less than 2 even across very different growth conditions. With these empirical data, we can then define our distributions of copy numbers to respect the known levels of fluctuation. Given the known mean and variance of a Log-Normal distribution, we can derive that the coefficient of variation is given by
\begin{align*}
    \mathrm{CoV}(X) = \frac{\mathbb{E}(X)}{\mathrm{Var}{X}} = \sqrt{e^{(\alpha \log{\mu})^2} - 1}.
\end{align*}
Rearranging this expression, we can write down $\alpha$ in terms of $\mu$ and $\mathrm{CoV}(X)$, where
\begin{align*}
    \alpha = \frac{\sqrt{\log{[\mathrm{CoV}(X)^2 + 1}]}}{\log{\mu}}
\end{align*}
This means that we can derive a Log-Normal distribution that obeys the empirical mean and coefficient of variation for transcription factors. In Fig~\ref{fig12}, we simulated noisy synthetic datasets using Log-Normal distributions with $P = 5000$ and $R = 100$ as the mean copy numbers and a range of coefficients of variation from $0.1$ to $10^2$, which covers the levels of fluctuations that are physiologically relevant. In Fig~\ref{figS12}, we show the distributions of copy numbers given the three levels of fluctuations that we particularly discuss in Sec~\ref{sec:extrinsic}. Note that when the coefficient of variation is set to 100, the copy number of RNAPs can reach as high as $10^9$ and the copy numbers of repressors can reach as high as $10^7$, both of which are very unrealistic levels of copy numbers.

\section{Extrinsic noise under low signal} \label{S13_Appendix}

\begin{figure}[h!]
    \centering
    \includegraphics[width=0.6\textwidth]{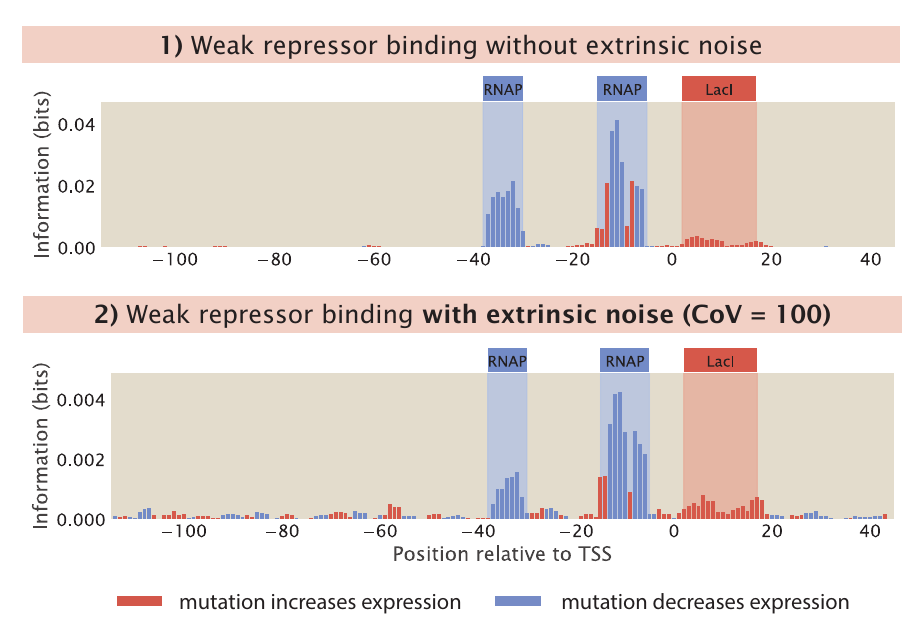}
    \caption{{\bf Effects of extrinsic noise on information footprints with weak repressor binding.}
    For both footprints, the repressor binding energy $\Delta \varepsilon_\mathrm{rd} = -11\, k_BT$. In the top information footprint, no extrinsic noise is introduced. In the bottom information footprint, the Log-Normal distribution from which RNAP and repressor copy numbers are drawn has a coefficient of variation of 100.}
    \label{figS13}
\end{figure}

In Sec~\ref{sec:extrinsic}, we explored the effect of extrinsic noise while keeping all other parameter at their standard values. That is to say, we expect that the fluctuations in copy numbers should not drastically change the level of signal in the footprints. However, one reasonable suspicion is that when the signal from binding events is sufficiently low, even low levels of noise may affect our interpretation of information footprints and expression shift matrices. To examine whether this is the case, we built footprints with lowered binding energy for the repressor while allowing copy numbers to fluctuate. As shown in Fig~\ref{figS13}, even when the signal is low and the coefficient of variation is set to as high as 100, we can still identify signal at the repressor binding site. Therefore, it seems that this is not an issue unless the level of noise is unrealistically high.

\section{Extrinsic noise for different architectures} \label{S14_Appendix}

\begin{figure}[h!]
    \centering
    \includegraphics[width=0.8\textwidth]{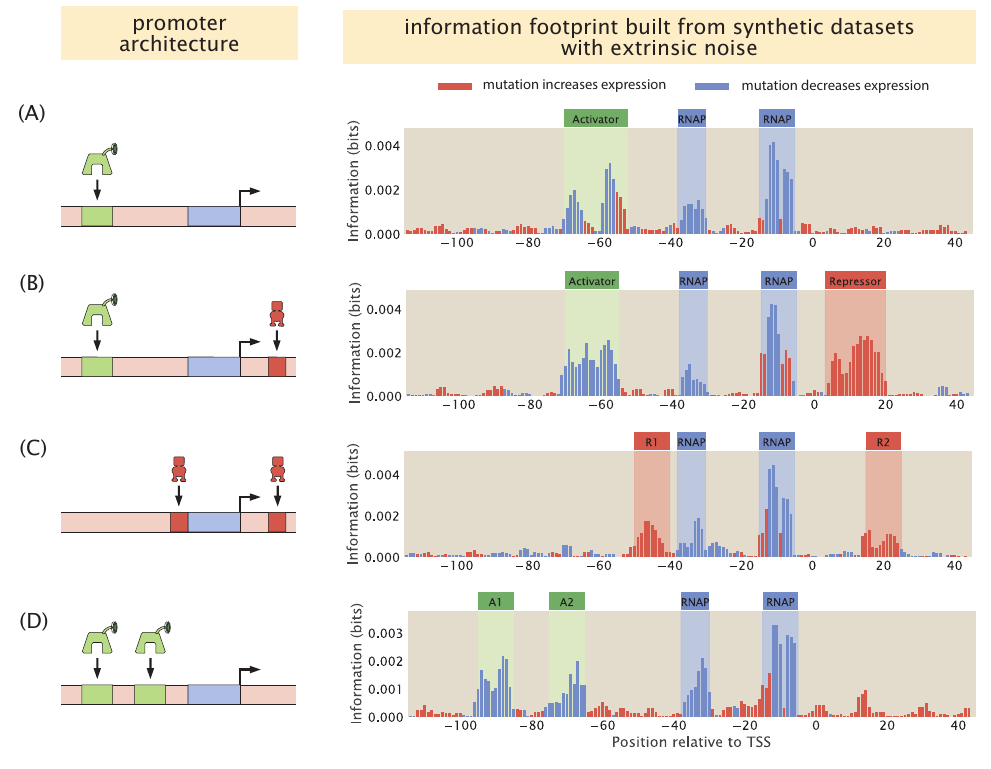}
    \caption{{\bf Information footprints built from synthetic datasets with extrinsic noise under common regulatory architectures.}
    In all four plots, the coefficient of variation for RNAP and transcription copy numbers is set to 10. Panel (A) to (D) show the footprints from the simple activation architecture, repression-activation architecture, double repression architecture, and the double activation architecture, respectively.}
    \label{figS14}
\end{figure}

To see if our investigation of extrinsic noise is generalizable, we test the cases where there are copy number fluctuations under the other four common regulatory architectures (simple activation, double repression, double activation, and repression-activation). As shown in Fig~\ref{figS14}, the signal-to-noise ratio remains high regardless of the architecture even when copy numbers are allowed to fluctuate to 10 times above or below the average copy number.

\section{Statistical weights for the simple activation promoter based on spanning trees} \label{S15_Appendix}

\begin{figure}[h!]
    \centering
    \includegraphics[width=\textwidth]{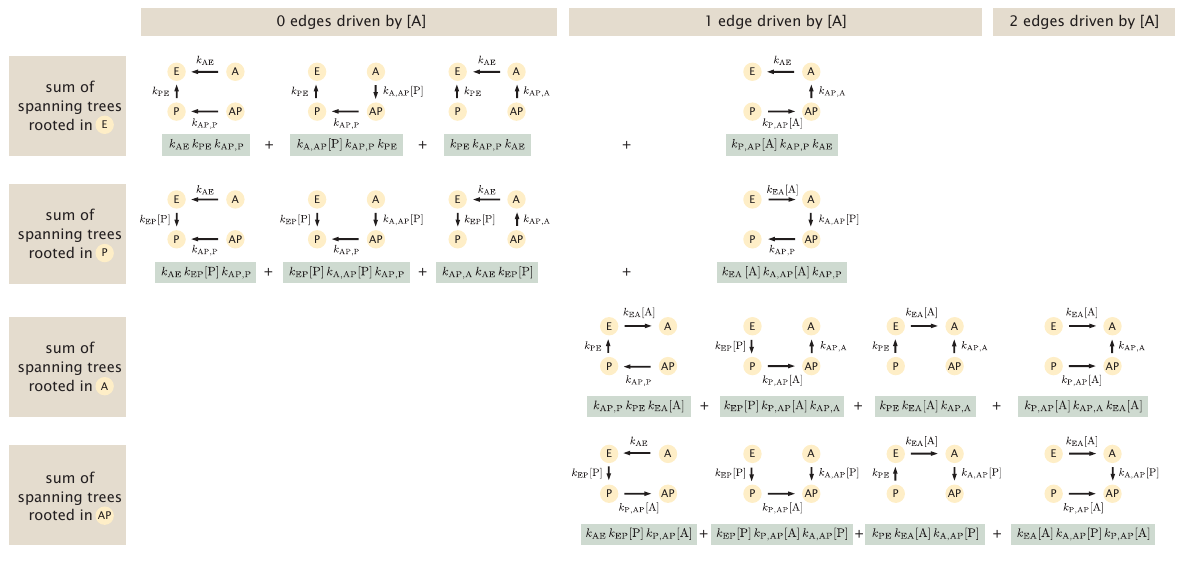}
    \caption{{\bf Deriving statistical weights of promoter states using spanning trees.}
    Each row corresponds to a different root for the spanning trees. The columns are grouped based on the number of edges in the spanning tree that depend upon the concentration of the activator. The figure is adapted from Mahdavi, Salmon et al.~\cite{Mahdavi2023-vr}.}
    \label{figS15}
\end{figure}

To derive the statistical weights of all states using the graph-theoretic approach introduced in Sec~\ref{sec:nonequilibrium}, we make use of the Matrix Tree Theorem, which states that at steady state, the probability of state $i$ is proportional to the sum of products of rate constants across all spanning trees rooted in the vertex representing state $i$. By definition, a spanning tree rooted in vertex $i$ is a subgraph that (1) contains all vertices in the original graph and (2) has all edges incoming in vertex $i$. The graph describing a simple activation promoter is shown in Fig~\ref{fig15}. As shown in Fig~\ref{figS15}, we can enumerate all spanning trees for each of the four vertices. This gives us the following statistical weights for each of the promoter states
\begin{align}
    \rho_\mathrm{E} &= k_\mathrm{AE}\,k_\mathrm{PE}\,k_\mathrm{AP,P}
    + k_\mathrm{A,AP}\mathrm{[P]}\,k_\mathrm{AP,P}\,k_\mathrm{PE}
    + k_\mathrm{PE}\,k_\mathrm{AP,P}\,k_\mathrm{AE}
    + k_\mathrm{P,AP}\mathrm{[A]}\,k_\mathrm{AP,P}\,k_\mathrm{AE} \\
    \rho_\mathrm{P} &= k_\mathrm{AE}\,k_\mathrm{EP}\mathrm{[P]}\,k_\mathrm{AP,P}
    + k_\mathrm{EP}\mathrm{[P]}\,k_\mathrm{A,AP}\mathrm{[P]}\,k_\mathrm{AP,P}
    + k_\mathrm{AP,A}\,k_\mathrm{AE}\,k_\mathrm{EP}\mathrm{[P]}
    + k_\mathrm{EA}\,\mathrm{[A]}\,k_\mathrm{A,AP}\mathrm{[A]}\,k_\mathrm{AP,P}\\
    \rho_\mathrm{A} &= k_\mathrm{AP,P}\,k_\mathrm{PE}\,k_\mathrm{EA}\mathrm{[A]}
    + k_\mathrm{EP}\mathrm{[P]}\,k_\mathrm{P,AP}\mathrm{[A]}\,k_\mathrm{AP,A}
    + k_\mathrm{PE}\,k_\mathrm{EA}\mathrm{[A]}\,k_\mathrm{AP,A} 
    + k_\mathrm{P,AP}\mathrm{[A]}\,k_\mathrm{AP,A}\,k_\mathrm{EA}\mathrm{[A]}\\
    \rho_\mathrm{AP} &= k_\mathrm{AE}\,k_\mathrm{EP}\mathrm{[P]}\,k_\mathrm{P,AP}\mathrm{[A]}
    + k_\mathrm{EP}\mathrm{[P]}\,k_\mathrm{P,AP}\mathrm{[A]}\,k_\mathrm{A,AP}\mathrm{[P]}
    + k_\mathrm{PE}\,k_\mathrm{EA}\mathrm{[A]}\,k_\mathrm{A,AP}\mathrm{[P]}
    + k_\mathrm{EA}\mathrm{[A]}\,k_\mathrm{A,AP}\mathrm{[P]}\,k_\mathrm{P,AP}\mathrm{[A]}
\end{align}
Finally, we can calculate the probability of each state by taking the weight of each state and dividing by the sum of all weights
\begin{align}
    p_\mathrm{E} &= \frac{\rho_\mathrm{E}}{\rho_\mathrm{E} + \rho_\mathrm{P} + \rho_\mathrm{A} + \rho_\mathrm{AP}} \\
    p_\mathrm{P} &= \frac{\rho_\mathrm{P}}{\rho_\mathrm{E} + \rho_\mathrm{P} + \rho_\mathrm{A} + \rho_\mathrm{AP}} \\
    p_\mathrm{A} &= \frac{\rho_\mathrm{A}}{\rho_\mathrm{E} + \rho_\mathrm{P} + \rho_\mathrm{A} + \rho_\mathrm{AP}} \\
    p_\mathrm{AP} &= \frac{\rho_\mathrm{AP}}{\rho_\mathrm{E} + \rho_\mathrm{P} + \rho_\mathrm{A} + \rho_\mathrm{AP}}
\end{align}
In particular, since $A$ and $AP$ are the transcriptionally active states, the probability that the promoter is on is given by $p_\mathrm{active} = p_\mathrm{A} + p_\mathrm{AP}$.

\newpage

\printbibliography

\end{refsection}
